\documentclass[a4paper,fleqn,usenatbib]{mnras}
\pdfoutput=1
\usepackage{mathptmx}
\usepackage{txfonts}

\usepackage[T1]{fontenc}
\usepackage{ae,aecompl}

\usepackage{graphicx}
\usepackage[space]{grffile}
\usepackage{longtable}
\usepackage{gensymb}
\usepackage{url}
\usepackage{array}
\newcounter{rowno}
\usepackage[utf8]{inputenc}
\usepackage[english]{babel}

\def\mone{{\rm M}_1}
\def\mtwo{{\rm M}_2}

\def\aone{{\rm a}_1}
\def\atwo{{\rm a}_2}
\def\vk{{\rm v}_{\rm k}}
\def\af{{\rm a}_{\rm f}}
\def\orbl{{\rm L_{\rm orb}}}

\def\xeff{\chi_{\rm eff}}
\def\mcrit{{\rm M}_{\rm crit}}

\newcommand{\Ms}{\ensuremath{M_{\odot}}}

\newcommand{\eg}{{\it e.g.}}
\newcommand{\cf}{{\it c.f.~}}
\newcommand{\ie}{{\it i.e.}}

\newcommand{\beq}{\begin{equation}}
\newcommand{\eeq}{\end{equation}}
\newcommand{\mtot}{\ensuremath{M_{\rm tot}}}
\newcommand{\mzams}{\ensuremath{M_{\rm ZAMS}}}

\newcommand{\mchirp}{\ensuremath{M_{\rm chirp}}}

\newcommand{\kmps}{\ensuremath{{\rm~km~s}^{-1}}}

\newcommand{\tinsp}{\ensuremath{t_{\rm insp}}}
\newcommand{\thub}{\ensuremath{t_{\rm Hubble}}}

\newcommand{\mcl}{\ensuremath{M_{cl}}}
\newcommand{\rh}{\ensuremath{r_h}}

\newcommand{\tmrg}{\ensuremath{t_{\rm mrg}}}

\newcommand{\tauinsp}{\ensuremath{\tau_{\rm insp}}}
\newcommand{\tauinspr}{\ensuremath{\tau_{\rm insp}^\prime}}

\newcommand{\tkl}{\ensuremath{T_{\rm KL}}}
\newcommand{\torb}{\ensuremath{T_{\rm orb}}}

\newcommand{\nmrgin}{\ensuremath{N_{\rm mrg,in}}}
\newcommand{\nmrgout}{\ensuremath{N_{\rm mrg,out}}}

\newcommand{\nbhbound}{\ensuremath{N_{\rm BH,bound}}}

\newcommand{\nbseven}{{\tt NBODY7}}
\newcommand{\nbsix}{{\tt NBODY6}}
\newcommand{\nbpp}{{\tt NBODY6++GPU}}

\newcommand{\bse}{{\tt BSE}}

\newcommand{\mobse}{{\tt MOBSE}}
\newcommand{\mocca}{{\tt MOCCA}}
\newcommand{\cmc}{{\tt CMC}}
\newcommand{\startrack}{{\tt StarTrack}}
\newcommand{\chain}{{\tt CHAIN}}
\newcommand{\archain}{{\tt ARCHAIN}}
\newcommand{\mesa}{{\tt MESA}}
\newcommand{\tevol}{\ensuremath{T_{\rm evol}}}
\newcommand{\tlisa}{\ensuremath{T_{\rm LISA}}}
\newcommand{\fbin}{\ensuremath{f_{\rm bin}}}
\newcommand{\nbin}{\ensuremath{N_{\rm bin}}}
\newcommand{\fobin}{\ensuremath{f_{\rm Obin}}}
\newcommand{\fgwp}{\ensuremath{f_{\rm GWp}}}

\newcommand{\fbfac}{\ensuremath{f_{\rm fb}}}
\newcommand{\ptsone}{{\tt pts1}}
\newcommand{\ptstwo}{{\tt pts2}}
\newcommand{\ptsthree}{{\tt pts3}}

\newcommand{\mbh}{\ensuremath{M_{\rm BH}}}

\newcommand{\mecsns}{\ensuremath{m_{\rm ECS,NS}}}
\newcommand{\mrem}{\ensuremath{M_{\rm rem}}}
\newcommand{\mco}{\ensuremath{M_{\rm CO}}}

\newcommand{\vkickns}{{\rm v}_{\rm kick,NS}}
\newcommand{\vkickecs}{{\rm v}_{\rm kick,ECS}}
\newcommand{\vesc}{{\rm v}_{\rm esc}}

\newcommand{\fmrg}{\ensuremath{f_{\rm mrg}}}
\newcommand{\ftz}{\ensuremath{f_{\rm TZ}}}

\title[Stellar-mass black holes in young clusters IV]
{Stellar-mass black holes in young massive and open stellar
clusters IV:
updated stellar-evolutionary and black hole spin
models and comparisons with the LIGO-Virgo O1/O2 merger-event data}

\author[S. Banerjee]{
Sambaran Banerjee$^{1,2}$\thanks{E-mail: sambaran@astro.uni-bonn.de (SB)}
\\
$^{1}$Helmholtz-Instituts f\"ur Strahlen- und Kernphysik (HISKP),
Nussallee 14-16, D-53115 Bonn, Germany\\
$^{2}$Argelander-Institut f\"ur Astronomie (AIfA),
Auf dem H\"ugel 71, D-53121, Bonn, Germany
}

\pubyear{2020}

\begin{document}
\label{firstpage}
\pagerange{\pageref{firstpage}--\pageref{lastpage}} 
\maketitle

\begin{abstract}
I present a set of long-term, direct, relativistic many-body computations of model dense stellar clusters with
up-to-date stellar-evolutionary, supernova (SN), and remnant natal-kick models, including pair instability and
pulsation pair instability supernova (PSN and PPSN), using an updated version of
$\nbseven$ N-body simulation program. 
The N-body model also includes stellar evolution-based natal spins of BHs and
treatments of binary black hole (BBH) mergers based on numerical relativity.
These, for the first time in a direct N-body simulation,
allow for second-generation BBH mergers. The set of 65 evolutionary models have initial masses $10^4\Ms-10^5\Ms$,
sizes 1 pc-3 pc, metallicity $0.0001-0.02$, with the massive stars in primordial binaries and they represent
young massive clusters (YMC) and moderately massive open clusters (OC). Such models produce dynamically-paired BBH mergers
that agree well with the observed masses, mass ratios, effective spin parameters, and final spins of the LVC O1/O2 merger
events, provided BHs are born with low or no spin but spin up after
undergoing a BBH merger or matter accretion onto it. In particular, the distinctly higher mass, effective spin parameter,
and final spin of GW170729 merger event is naturally reproduced, as also the mass asymmetry of the O3 event GW190412.
The computed models produce massive, $\sim100\Ms$ BBH mergers with primary mass within the `PSN gap'
and also yield mergers involving remnants in the `mass gap'.
They also suggest that YMCs and OCs produce persistent, Local-Universe GW sources detectable by LISA.
Such clusters are also capable of producing eccentric LIGO-Virgo mergers.
\end{abstract}


\section{Introduction}\label{intro}

The LIGO-Virgo collaboration (hereafter LVC) has so far published 10 binary black hole
(hereafter BBH) and one binary neutron star (hereafter BNS) merger events,
in their first gravitational wave transient catalogue \citep[][GWTC-1]{Abbott_GWTC1},
through ground-based,
interferometric detection of gravitational waves \citep[][hereafter GW]{2016PhRvL.116f1102A}
during their first (O1)
and second (O2) observing runs. In their recently-concluded third observing run
(O3; \url{https://gracedb.ligo.org/superevents/public/O3/}),
56 additional candidates of compact-binary mergers are detected,
a few of those being of BNS and even neutron star-black hole (hereafter NSBH) mergers.
Additionally, a few runs are labelled as ``mass gap'' in the sense that 
one or both of the merging members lie in the potential gap between the masses of
neutron stars (hereafter NS) and stellar-remnant black holes (hereafter BH) that certain supernova 
(hereafter SN) models \citep[\eg,][]{Fryer_2012} predict.
The handful of events from O1/O2 already suggest that detection of GW transients
from compact binary merger events not only is interesting by its own right
\citep{2016PhRvL.116f1102A,2016ApJ...818L..22A} but also has the potential
to provide unprecedented information regarding masses, spins, and their
boundaries \citep{Abbott_GWTC1_prop}, of stellar-remnant BHs and
NSs. Such information would provide strongest
constraints on the formation mechanisms of compact stellar remnants and
of the environment in which their parent stars form and evolve
\citep{Belczynski_2020,Olejak_2020}.

It is as well of wide interest and diverse implications \citep[\eg,][]{Abadie_2010,Mandel_2017}
to consider how and under which conditions NSs and BHs would pair up in tight-enough
binaries so that they can spiral in by emitting GW radiation and
merge within the Hubble time.
Recent numerical studies based on analytical \citep{Henon_1975},
direct N-body integration \citep{Aarseth_2003},
and Monte Carlo approach \citep{Henon_1971,Joshi_2000,Hypki_2013}
show that the retention of BHs in
dense stellar clusters of wide mass range,
beginning from low-/medium-mass young and open clusters
\citep[\eg,][]{Banerjee_2010,Ziosi_2014,Mapelli_2016,Park_2017,Banerjee_2017,Banerjee_2017b,Banerjee_2018,Rastello_2019,DiCarlo_2019,Kumamoto_2019}
through globular clusters
\citep[\eg,][]{Sippel_2013,Morscher_2013,Breen_2013,ArcaSedda_2016,Rodriguez_2016,Rodriguez_2018,Hurley_2016,Wang_2016,Askar_2016,Chatterjee_2017a,Chatterjee_2017b,Fragione_2018b,Antonini_2020,Kremer_2020}
to galactic nuclear clusters
\citep[\eg,][]{Antonini_2016,Antonini_2019,ArcaSedda_2017,ArcaSedda_2020,Hoang_2017,Hoang_2019},
comprise environments where BHs can pair up
through close dynamical interactions, which, furthermore,
lead to general-relativistic (hereafter GR)
coalescences of these BBHs. Being much more massive than the rest of the stars,
the BHs, which remain gravitationally bound to a cluster after
their birth, spatially segregate and remain highly concentrated in the cluster's
innermost (and densest) region \citep[\eg,][]{Banerjee_2010,Morscher_2015}
due to dynamical friction \citep{Chandrasekhar_1943,Spitzer_1987} from the stellar background.
This is essentially an early core collapse of the cluster
leading to its post-core-collapse behaviour
\citep{Henon_1975,Spitzer_1987,Heggie_2003}, \ie, energy generation
in the ``collapsed'' BH core leading to an overall expansion of
the cluster with time \citep{Breen_2013,Antonini_2020}.
Inside this core, BHs undergo close binary-single and
binary-binary encounters giving rise to compact subsystems (triples, quadruples, or
even higher-order multiples) whose resonant evolution can lead to GR inspiral
and merger of their innermost binaries
\citep{Leigh_2013,Samsing_2014,Geller_2015a,Banerjee_2018,Samsing_2018,Zevin_2019},
through the binaries' eccentricity pumping.
The breakup of such subsystems or simply close, flyby encounters may also
lead to a sufficient boost in eccentricity of a BBH such that it
merges either promptly, in between two close encounters \citep[\eg,][]{Kremer_2019}
or within a Hubble time
if it gets ejected from the cluster as a result of the interaction
\citep[\eg,][]{Rodriguez_2015,Park_2017,Kumamoto_2019}.
Note that such GR mergers can also happen in hierarchical systems, containing
NSs and BHs, in a galactic field that derive from field massive-stellar multiplets
\citep[\eg,][]{Toonen_2016,Antonini_2017,Fragione_2019,Fragione_2020}.

Alternatively, compact-binary mergers can happen in a galactic field
either via common-envelope (hereafter CE) evolution of massive-stellar binaries
with hydrogen-rich envelope \citep{Dominik_2012,Belczynski_2016,Stevenson_2017,Giacobbo_2018,Baibhav_2019}
or in close, tidally-interacting,``over-contact'' binaries composed of
chemically-homogeneous members \citep{DeMink_2009,DeMink_2016,Marchant_2016}.
In such studies, BBH inspiral detection rate of $\sim10-\sim1000{\rm~yr}^{-1}$
has been estimated for the full-sensitivity LIGO, \ie, from similar to up to 2 orders
of magnitudes higher detection rate than what is estimated for dynamically-formed BBHs
\citep{Banerjee_2010,Rodriguez_2016,Askar_2016,Banerjee_2017}.

Recently, studies by \citet[][hereafter Paper I]{Banerjee_2017},
\citet[][Paper II]{Banerjee_2017b}, and \citet[][Paper III]{Banerjee_2018}
have shown the importance of triple-/higher-order-dynamical
interactions, including post-Newtonian (hereafter PN; \citealt{Blanchet_2014}) terms, in
triggering in-cluster\footnote{Like in Papers I-III, ``in-cluster coalescence'' in this work
will refer to any GR coalescence taking place inside a cluster, while being gravitationally bound to the cluster.}
GR BBH coalescences, particularly inside
relatively low-velocity-dispersion systems
such as open clusters and lower-mass globular clusters (hereafter GC).
These studies have utilized the direct (star-by-star) N-body integration code
$\nbseven$ \citep{Aarseth_1999,Nitadori_2012,Aarseth_2012} that couples the
fourth-order Hermite orbit integration technique with
advanced subsystem identification and their regularization \citep{Aarseth_2003},
PN treatment for binaries, triples, and higher-order subsystems containing
NS or BH by adopting the $\archain$ algorithm \citep{Mikkola_1999,Mikkola_2008},
and semi-analytical, recipe-based population synthesis of single and binary stars
by adopting the $\bse$ program \citep{Hurley_2000,Hurley_2002}.
Note that $\nbseven$ uses $\archain$ to treat triples and higher-order subsystems 
even if they do not contain NS or BH. For close star-star binaries or
those containing a white dwarf (hereafter WD) or close hyperbolic
passages, KS regularization \citep{Kustaanheimo_1965} is
applied along with energy loss due to tidal interaction and
gravitational radiation (PN-2.5 term; \citealt{Peters_1964}).

In the $\nbseven$ computations presented in Paper I, II, and III, a somewhat old
SN remnant mass scheme, as in \citet{Belczynski_2008}, is applied along
with stellar wind mass loss recipes of \citet{Belczynski_2010}, as implemented
in the currently official, public version of $\nbseven$ (in its $\bse$ sector).
In particular, the remnant scheme did not take into account the BH mass ceiling
and the ``upper mass gap'' due to pulsation pair-instability supernova and
pair-instability supernova respectively \citep[hereafter PPSN/PSN;][]{Langer_2007,Woosley_2017,Mapelli_2020}.
Also, the possibility of having a ``lower mass gap'' between NS and BH masses 
due to ``rapid'' core-collapse SN \citep{Fryer_2012} was not incorporated.
The schemes adopted for assigning natal kicks of NSs and BHs, that are crucial
for their retention inside clusters, were rather basic.
As GW observations continue to provide mass measurements
of an increasing number BHs and NSs \citep{Abbott_GWTC1},
signatures of such theoretically-predicted features are being widely
investigated and discussed
\citep[\eg,][]{Fishbach_2017,Fishbach_2017a,Fishbach_2020,Abbott_GWTC1,Chatziioannou_2019,Kimball_2020,Giacobbo_2018,Spera_2019,Rodriguez_2019,Olejak_2020}.
Full-fledged, long-term, direct, relativistic N-body (or many-body) computations of dense stellar clusters with
up-to-date stellar-evolutionary and remnant-formation and retention recipes is, therefore,
of urgent need and highly desired.

In this work, such a set of 65 computations is presented for the first time using
an updated version of $\nbseven$ that explicitly incorporates up-to-date
stellar remnant formation models. It additionally includes schemes for
assigning spins of stellar-remnant (first-generation\footnote{In this work,
BHs that are direct descendants of stars will be referred to as ``first generation'',
their merger product with other first-generation BHs as ``second generation'', and so on.})
BHs based on detailed stellar-evolutionary
models, run-time tracking of the BHs' spins and assigning final spins and GW recoil velocities
to in-cluster BBH mergers based on numerical-relativity (hereafter NR) results.
Note that one or more of these aspects have been incorporated in recent
Monte Carlo
(that use $\cmc$ or $\mocca$ codes; \citealt{Joshi_2000,Fregeau_2007,Morscher_2015,Hypki_2013,Giersz_2013})
and direct N-body (that use $\nbpp$ code; \citealt{Spurzem_2008,Wang_2015})
studies \citep{Morawski_2018,Rodriguez_2019,DiCarlo_2019,Kremer_2020}.
The new BH-spin aspects of the updated $\nbseven$ now automatically allows the formation
and tracking of second-generation BHs and BBH mergers involving them
as in latest $\cmc$-based Monte Carlo studies \citep{Rodriguez_2018,Rodriguez_2019}.
This work introduces the new $\nbseven$ computations, focusing on the compact-remnant merger
outcomes from them and comparing them with LVC O1/O2 merger-event data.

In Sec.~\ref{newnb}, elements of the updated $\nbseven$ are described.
Sec.~\ref{comp} describes the cluster model characteristics and the set of direct N-body computations.
Sec.~\ref{result} describes the compact-remnant
merger outcomes from these new N-body simulations and compares them with the O1/O2 events. 
Sec.~\ref{summary} summarizes the results and indicates future prospects.

\begin{figure*} 
\includegraphics[width=8.7cm,angle=0]{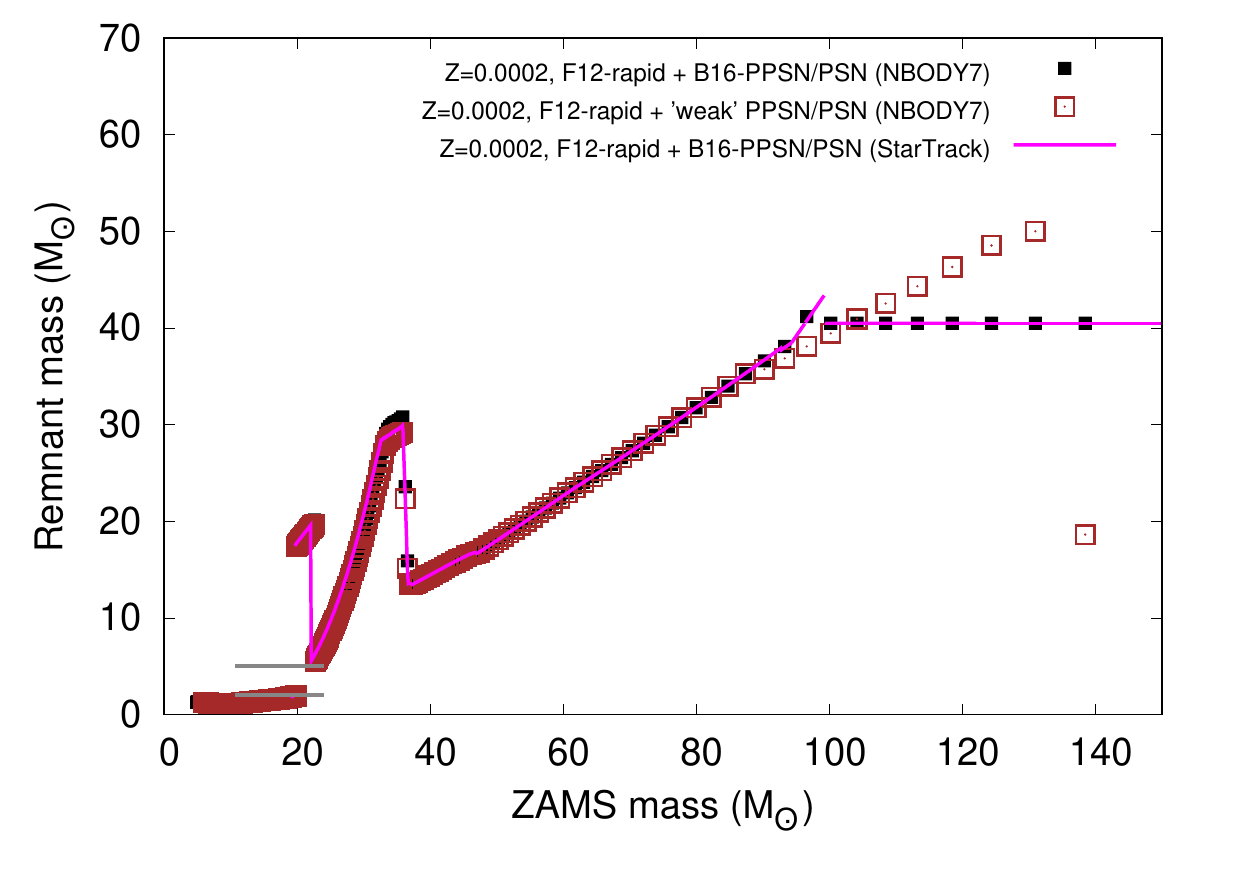}
\includegraphics[width=8.7cm,angle=0]{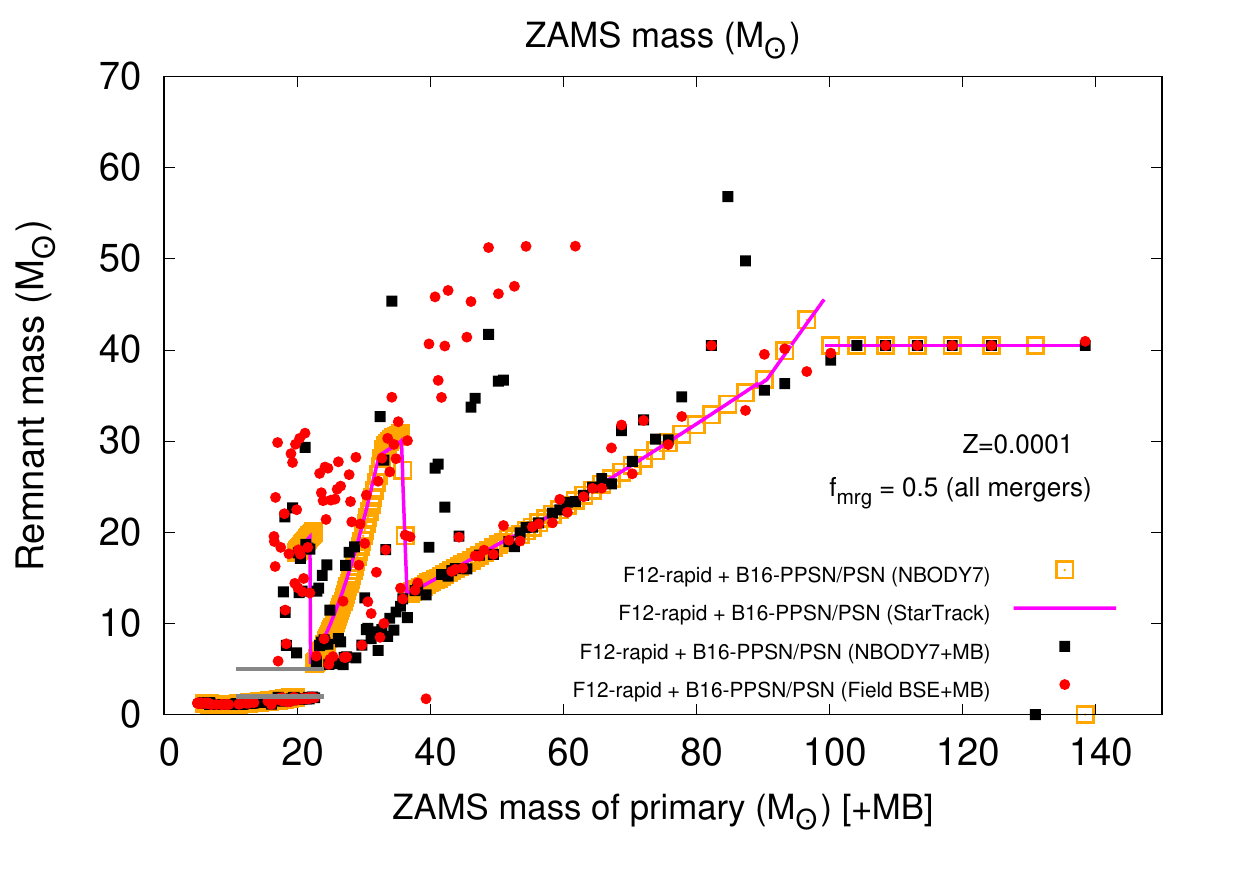}
\caption{{\bf Left panel:} examples of ZAMS mass-remnant mass relations as obtained from updated
$\nbseven$ used in this work. The outcomes for F12-rapid+B16-PPSN/PSN (filled, black squares) and
F12-rapid+weak-PPSN/PSN (empty, brown squares) remnant-mass schemes (Sec.~\ref{newrem})
are shown for the metallicity $Z=0.0002$. For F12-rapid+B16-PPSN/PSN remnant model, the comparison
with the corresponding relation from $\startrack$ (solid, magenta line) is demonstrated.
In these N-body models (initially of total cluster mass $\mcl(0)=5.0\times10^4\Ms$
and half-mass radius $\rh(0)=2.0$ pc), all stars are initially single
whose ZAMS masses range from $0.08\Ms-150.0\Ms$ and which are distributed according to the standard IMF.
{\bf Right panel:} ZAMS mass-remnant mass relations with updated $\nbseven$ from a model
cluster of the same $\mcl(0)$ and $\rh(0)$ ($Z=0.0001$; F12-rapid+B16-PPSN/PSN;
standard IMF over $0.08\Ms-150.0\Ms$), where all stars with $\mzams\geq16\Ms$
are in primordial binaries (see Sec.~\ref{comp}; filled, black squares). Models involving such
a massive primordial-binary population are indicated by `+MB' in the legends and axis labels. 
For progenitor stars that have undergone a star-star merger before the remnant formation,
the ZAMS mass of the primary (the more massive of the members participating in the star-star merger,
at the time of the merger) is plotted along the abscissa. In all star-star mergers,
$\fmrg=0.5$ of the secondary's mass is assumed to be lost in the merger process
(see Sec.~\ref{mrgr}). The corresponding single-star $\nbseven$ outcomes (empty, orange squares)
and $\startrack$ outcomes (solid, magenta line) are shown for comparison.
Also shown for comparison is the corresponding outcome (filled, red circles)
when the initial population of massive binaries from the N-body model
is evolved individually using the updated standalone $\bse$ of Ba20.
Likewise the N-body model with primordial binaries, the primary mass is plotted along
the abscissa if the outcome of the binary evolution is a single remnant ($\fmrg=0.5$ is assumed).
}
\label{fig:inifnl}
\end{figure*}

\begin{figure*}
\includegraphics[width=8.7cm,angle=0]{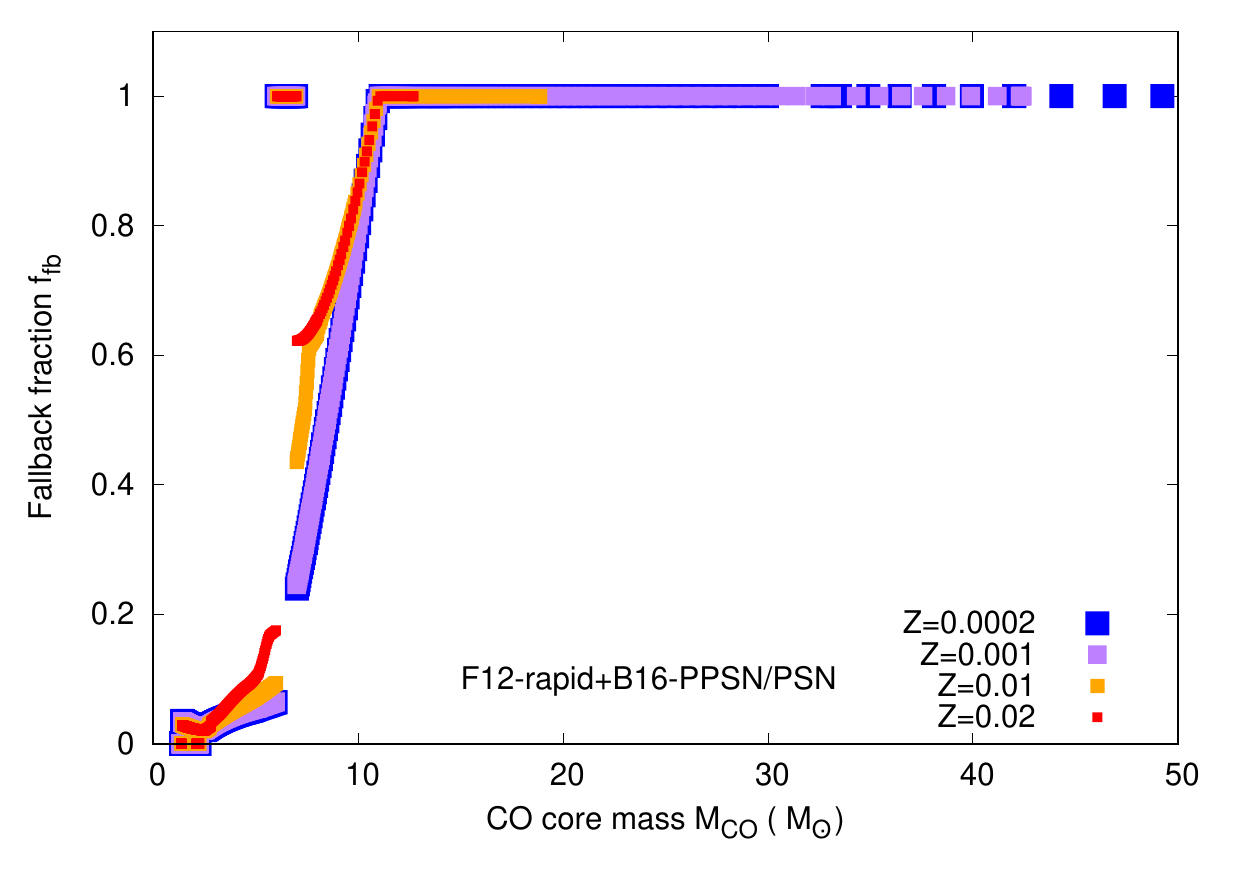}
\includegraphics[width=8.7cm,angle=0]{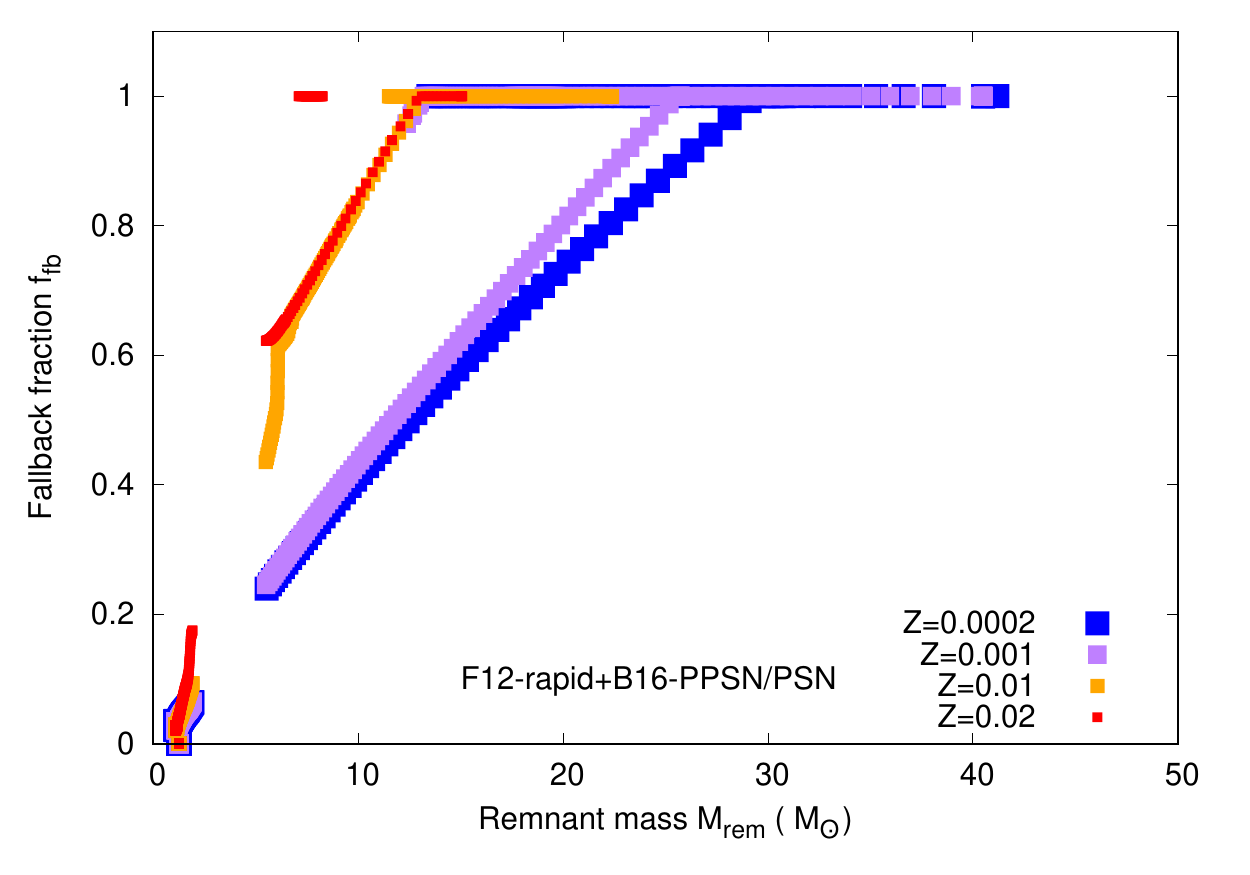}
\caption{Supernova fallback fraction (alternatively, fallback factor), $\fbfac$, as
a function of the progenitor star's carbon-oxygen
core mass, $\mco$ (left panel), and remnant mass, $\mrem$ (right panel).
The data points are obtained directly from $\nbseven$ computations in this work,
of model clusters composed of only single stars initially whose ZAMS masses range from
$0.08\Ms-150.0\Ms$ and which are distributed according to the standard IMF.
The $\fbfac-\mco$ and $\fbfac-\mbh$ dependencies are shown at the four metallicities $Z=0.0002$,
0.001, 0.01, and 0.02 (legend) for the remnant-mass model F12-rapid+B16-PPSN/PSN (Sec.~\ref{newrem}).
$\fbfac=1.0$ corresponds to direct-collapse BHs.}
\label{fig:fbfrac}
\end{figure*}

\section{The updated $\nbseven$}\label{newnb}

Both the standalone $\bse$ code \citep{Hurley_2000,Hurley_2002} and its
version that is integrated with $\nbseven$ have recently been updated in parallel with
up-to-date stellar wind mass loss and remnant-formation
recipes. The updated versions are thoroughly tested to agree well with the remnant
outcomes of $\startrack$ population-synthesis code \citep{Belczynski_2008,Belczynski_2016}. 
Furthermore, material fallback during a core-collapse SN is considered
explicitly in determining natal kicks of NSs and BHs.
The reader is advised to consult \citet[][hereafter Ba20]{Banerjee_2020} where
all these new implementations and $\bse$-$\startrack$ comparisons 
are elaborated. Below, these updates to $\nbseven$ are summarized and further
new ingredients are described. At present,
the updated $\nbseven$ is private (which will be publicized at a later occasion)
but the parallel standalone $\bse$ is available publicly with
Ba20.

\subsection{Stellar wind}\label{newwind}

The stellar wind mass loss follows the recipe of \citet[][hereafter B10]{Belczynski_2010}
as described in Sec.~2.1 of Ba20. The main difference with respect to the
currently public version of standalone $\bse$ is the application of
\citet{Vink_2001} wind model, that exhibits a line-driven
bi-stability jump feature at $\approx25000$K surface temperature,
for massive, hot stars (of surface temperature $>12500$K).
Other new ingredients are the use of the luminous-blue-variable (LBV) wind
of \citet{Humphreys_1994} and, for Helium stars, the Wolf-Rayet (WR) wind of
\citet{Hamann_1998,Vink_2005}. For lower-mass, colder stars,
the original \citet{Hurley_2000} wind model is maintained. The wind mass
loss depends on metallicity, $Z$, of the star through the use of the
bi-stability jump, WR, and \citet{Hurley_2000} wind models. Note that
all these wind ingredients are in principle implemented in the
current public $\nbseven$. However, the
implementation is revised so that the sequencing of the various wind
elements are parallel to that of B10, as elaborated in Ba20.

\subsection{Stellar remnant formation}\label{newrem}

The remnants (NSs and BHs) are formed according to the rapid- or delayed-SN
mass and fallback models of \citet[][hereafter F12]{Fryer_2012} which
are available as options. The options for the existing older remnant models
are also retained. Additionally, there is now the option of having a BH mass ceiling
due to PPSN and the upper mass gap due to PSN, according to the prescriptions
of \citet[][hereafter B16]{Belczynski_2016a}. These new implementations and the
corresponding initial mass-final mass relations are
detailed in Ba20. Optionally, PPSN-derived BHs can also be formed with masses
according to the ``moderate'' and ``weak'' PPSN models of \citet{Leung_2019},
following their implementations in \citet[][hereafter B20]{Belczynski_2020}.
For obtaining the gravitational mass of
the remnant from its baryonic mass, a neutrino mass loss of 10\% is assumed
for BH formation and the mass loss is according to \citet{Lattimer_1989,Timmes_1996}
for NS formation. NS formation through
electron-capture-supernova \citep[ECS;][]{Podsiadlowski_2004} is incorporated, as default
in $\bse$ (which scheme is analogous to the scheme in \citealt{Belczynski_2008}, producing 
the characteristic $\mecsns=1.26\Ms$ ECS-NSs).
In this work, F12-rapid and F12-delayed remnant mass models, with B16-PPSN/PSN
(indicated as, \eg, F12-rapid+B16-PPSN/PSN remnant model), are applied
in most of the model computations, except for a few where the weak PPSN
is applied.

Note that with B16-PPSN/PSN, the (baryonic) mass of the pre-collapse
Helium star in a PPSN is taken to be of $45\Ms$ \citep[B16,][]{Woosley_2017}
which collapses directly
to a $40.5\Ms$ BH, taking into account the 10\% neutrino mass loss.
$45\Ms$ is widely considered as the theoretical upper mass limit of BHs due to PPSN
and such a mass limit is also supported by BH masses measured in
LVC O1/O2 BBH mergers \citep{Abbott_GWTC1,Chatziioannou_2019,Kimball_2020}.
Hence, a $>45\Ms$ BH is generally considered as a BH in the PSN (upper) mass gap.
However, if one applies the weak PPSN model instead of the B16 model
(in B20, B16-PPSN model is referred to as ``strong'' PPSN), PPSN-derived
BHs can reach up to $\approx50\Ms$ (assuming 10\% neutrino mass loss)
for very low metallicities, \ie, BH mass can lie
in the ``classical'' PSN mass gap. This is demonstrated in Fig.~\ref{fig:inifnl} (left panel)
where remnant masses are plotted against zero age main sequence (hereafter ZAMS) masses
directly from $\nbseven$ outputs (metallicity $Z=0.0002$ is assumed). The
characteristic lower mass gap between NSs and BHs,
between $\approx 2.5\Ms - 5.0\Ms$, of F12-rapid remnant scheme is also indicated.
See Ba20 for further examples of initial-final relations for various
remnant mass schemes and metallicities and their comparisons with $\startrack$.

Even with B16-PPSN/PSN, the remnant BH mass can enter the PSN mass gap 
due to star-star mergers at low metallicities producing stars with over-massive Hydrogen envelope, 
when, as observations suggest, the BH-progenitor stars are in tight massive primordial
binaries instead of being single. This possibility is discussed in detail
and demonstrated in Ba20; see also \citet{Spera_2019}. Fig.~\ref{fig:inifnl} (right panel) 
demonstrates this from $\nbseven$ outputs (filled, black squares)
for a model star cluster with massive primordial binaries, where,
with F12-rapid+B16-PPSN/PSN remnant model, BHs up to $\approx60\Ms$ form at $Z=0.0001$.
Also shown in Fig.~\ref{fig:inifnl} (right panel)
is the initial-final outcome when the same primordial-binary population from the
model cluster
is evolved as isolated binaries using the standalone $\bse$ of Ba20 that is
updated in parallel (filled, red circles). Although the overall
pattern of the initial-final points is similar for the isolated and the in-cluster
binary evolutions, individual points do differ. This difference
can be attributed to the stochastic, weak and strong perturbations that
the binaries receive during their evolutions, altering their
orbital parameters, when they are cluster members. However, differences
can also arise due to the way $\bse$'s binary evolution engine is integrated
with the direct N-body evolution engine in $\nbseven$.
(The binary-evolution physics and its $\nbseven$ integration remains unaltered
in the updated standalone $\bse$ and $\nbseven/\bse$.)
A detailed investigation on the differences between dynamically-influenced
and isolated evolution of a massive binary population is beyond the present scope;
see \citet{DiCarlo_2019} in this context.

As detailed in Ba20 (see their Sec.~2.3.1), for numerical convergence with the updated
wind and remnant recipes, both standalone $\bse$ and $\nbseven$ are run with
the $\bse$ time step parameters $\ptsone=0.001$, $\ptstwo=0.01$ (along with $\ptsthree=0.02$
for standalone $\bse$). These choices of the time step parameters
practically do not slow down $\nbseven$ or isolated-binary runs.

\begin{figure*}
\includegraphics[width=8.5cm,angle=0]{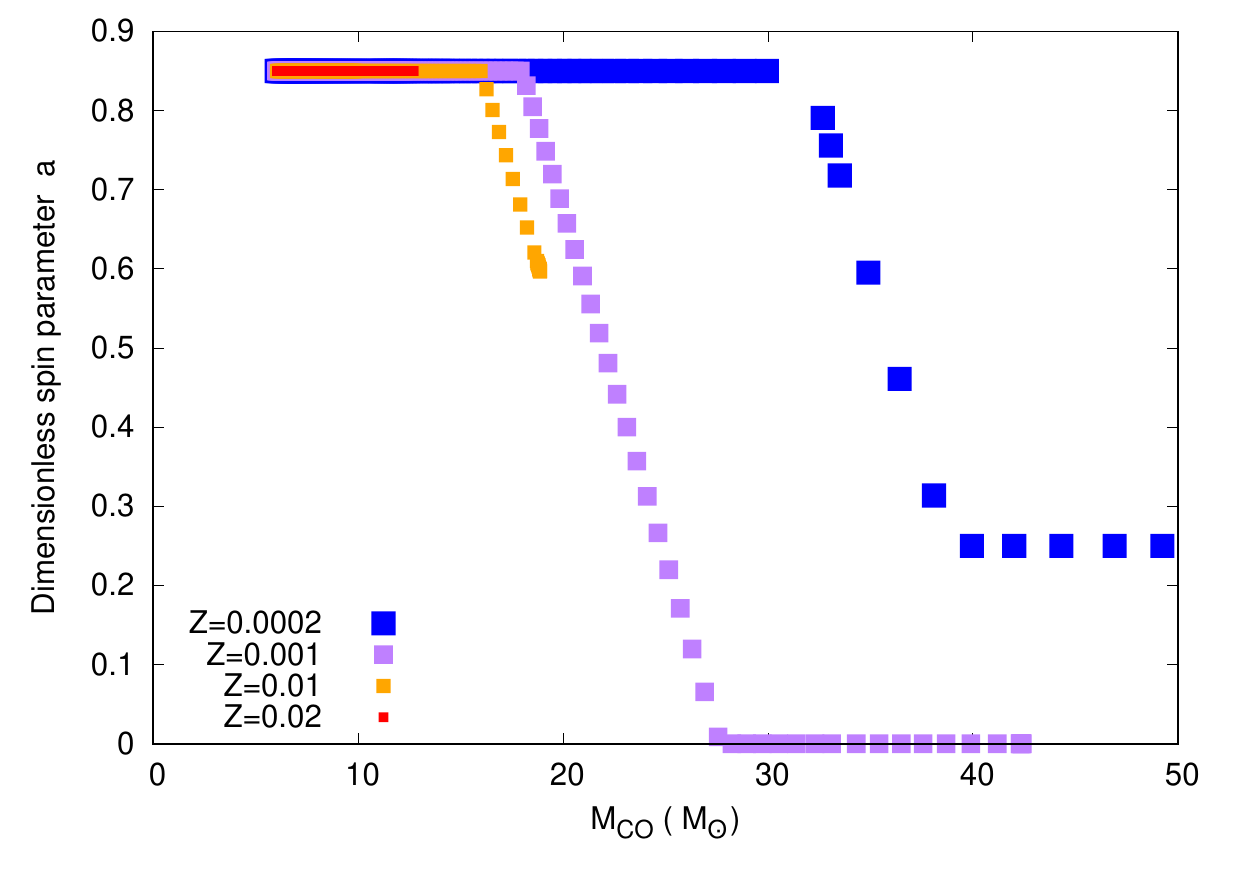}
\includegraphics[width=8.5cm,angle=0]{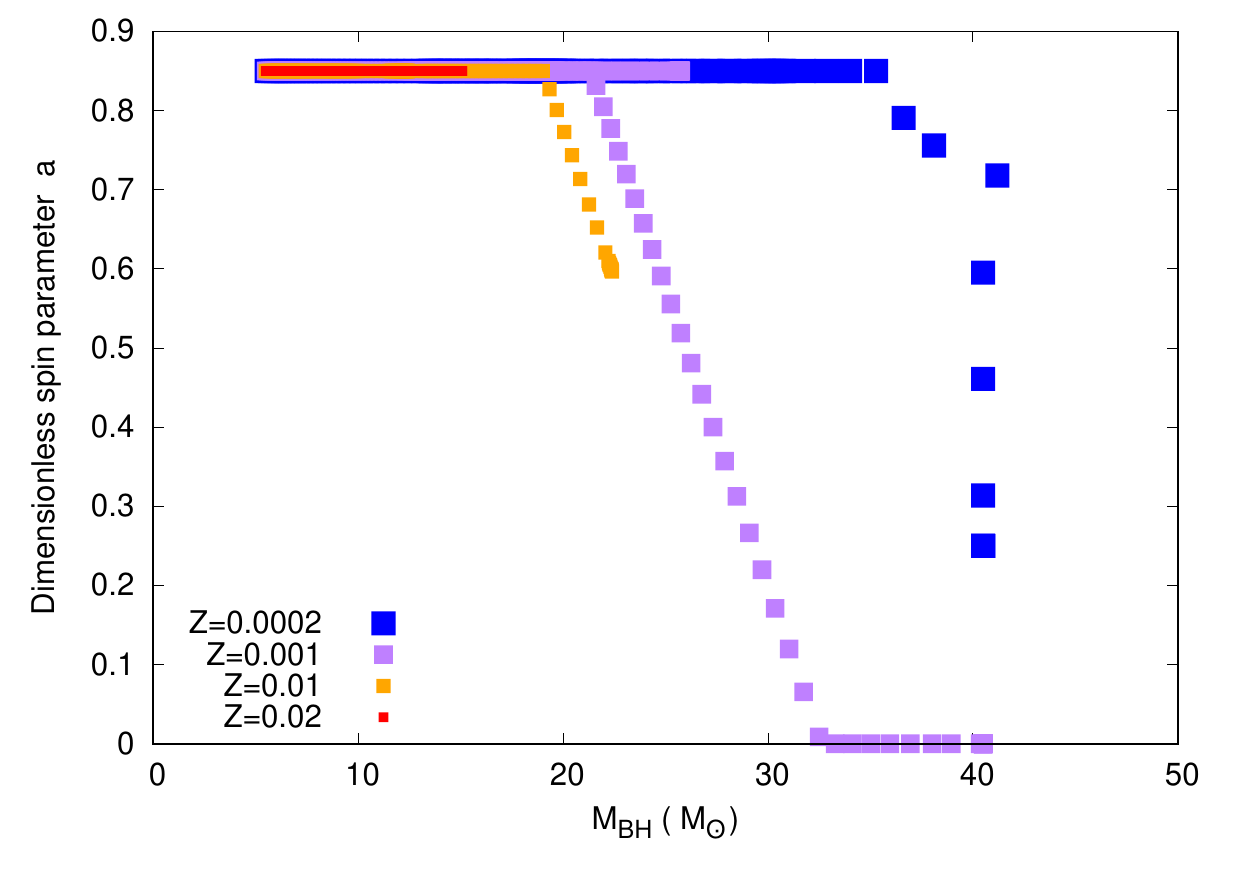}\\
\includegraphics[width=8.5cm,angle=0]{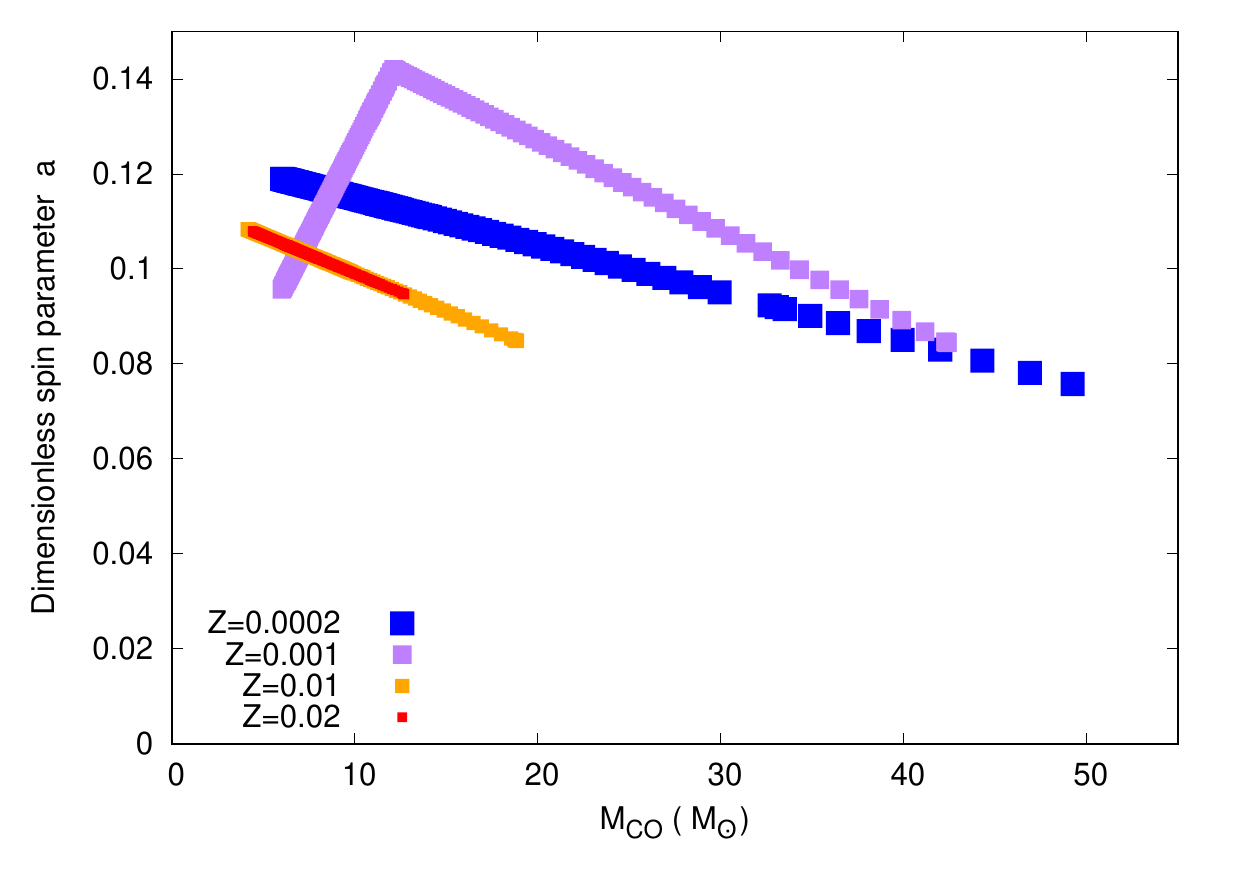}
\includegraphics[width=8.5cm,angle=0]{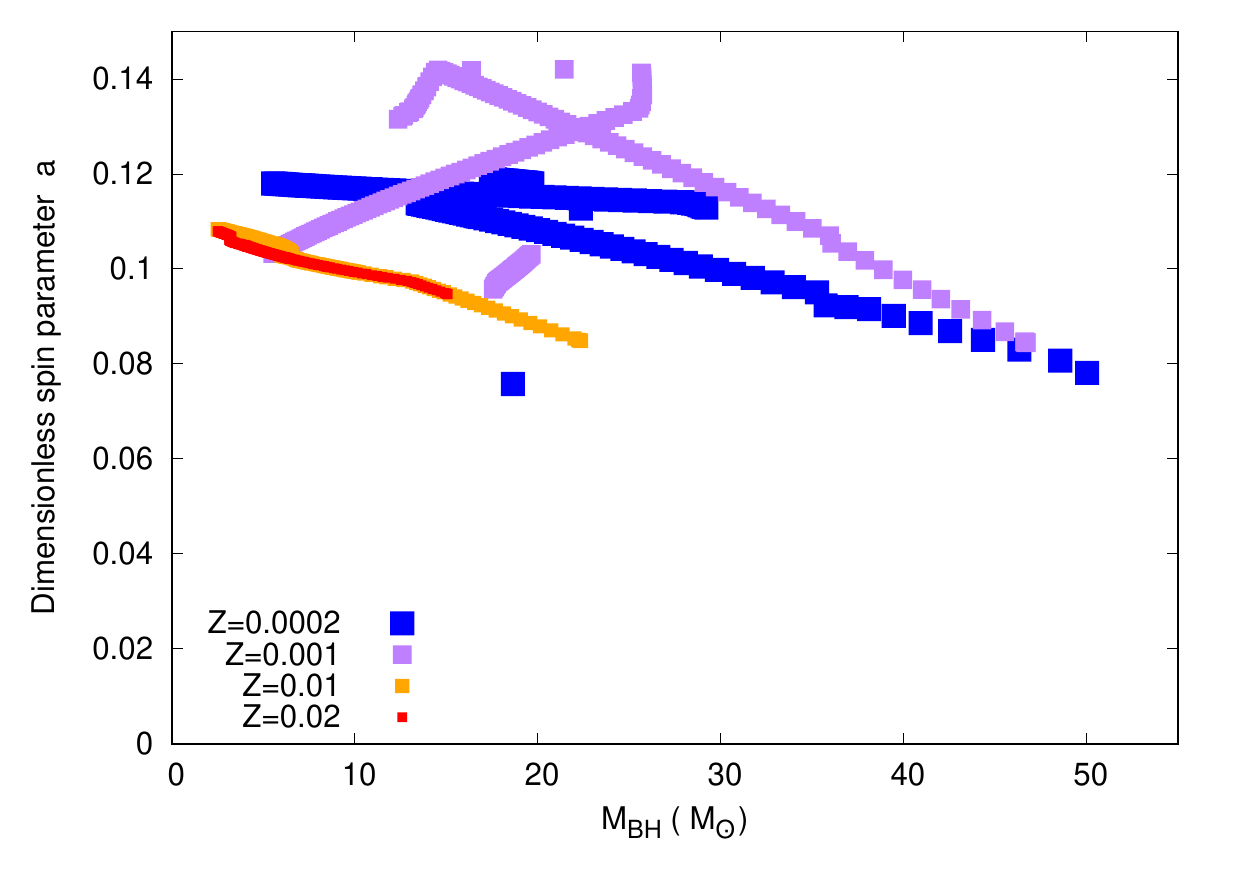}\\
\includegraphics[width=8.5cm,angle=0]{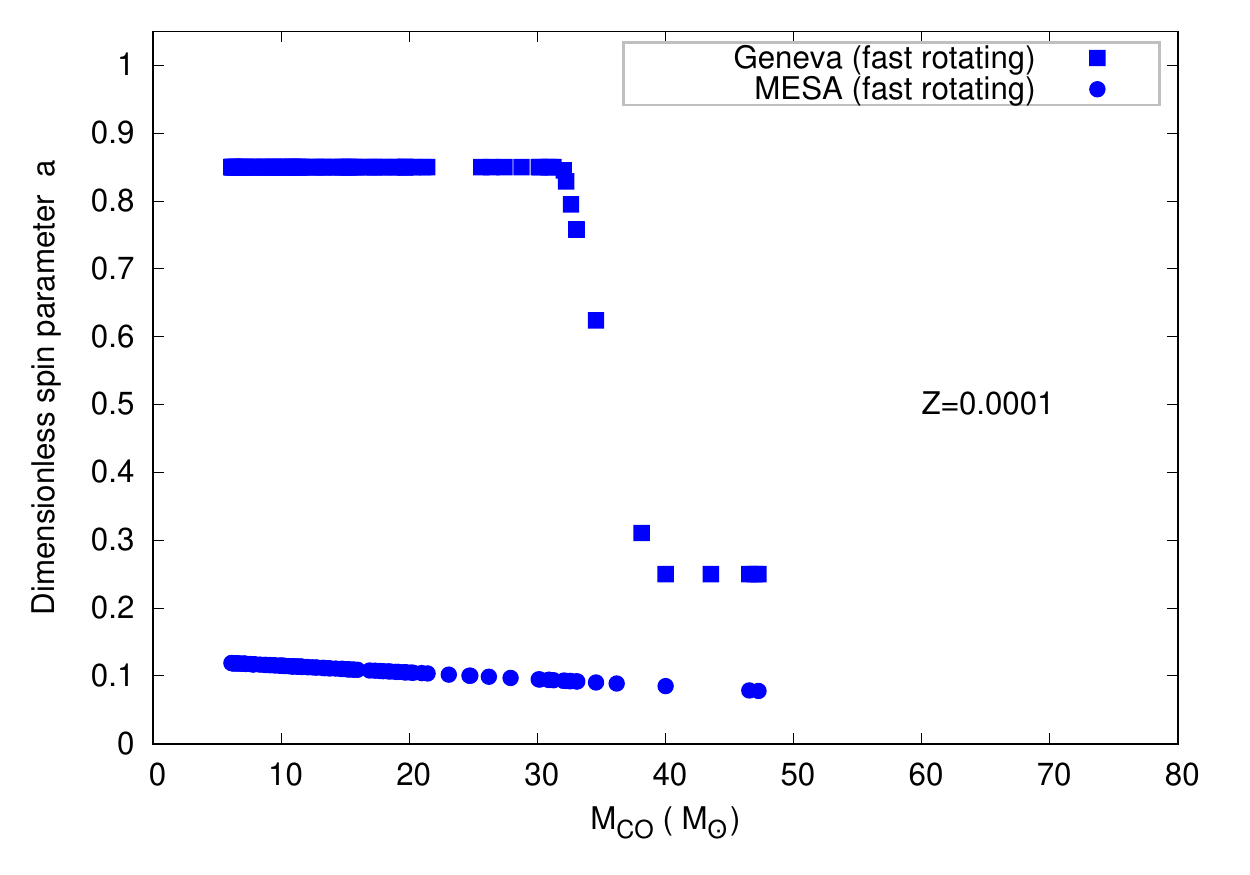}
\includegraphics[width=8.5cm,angle=0]{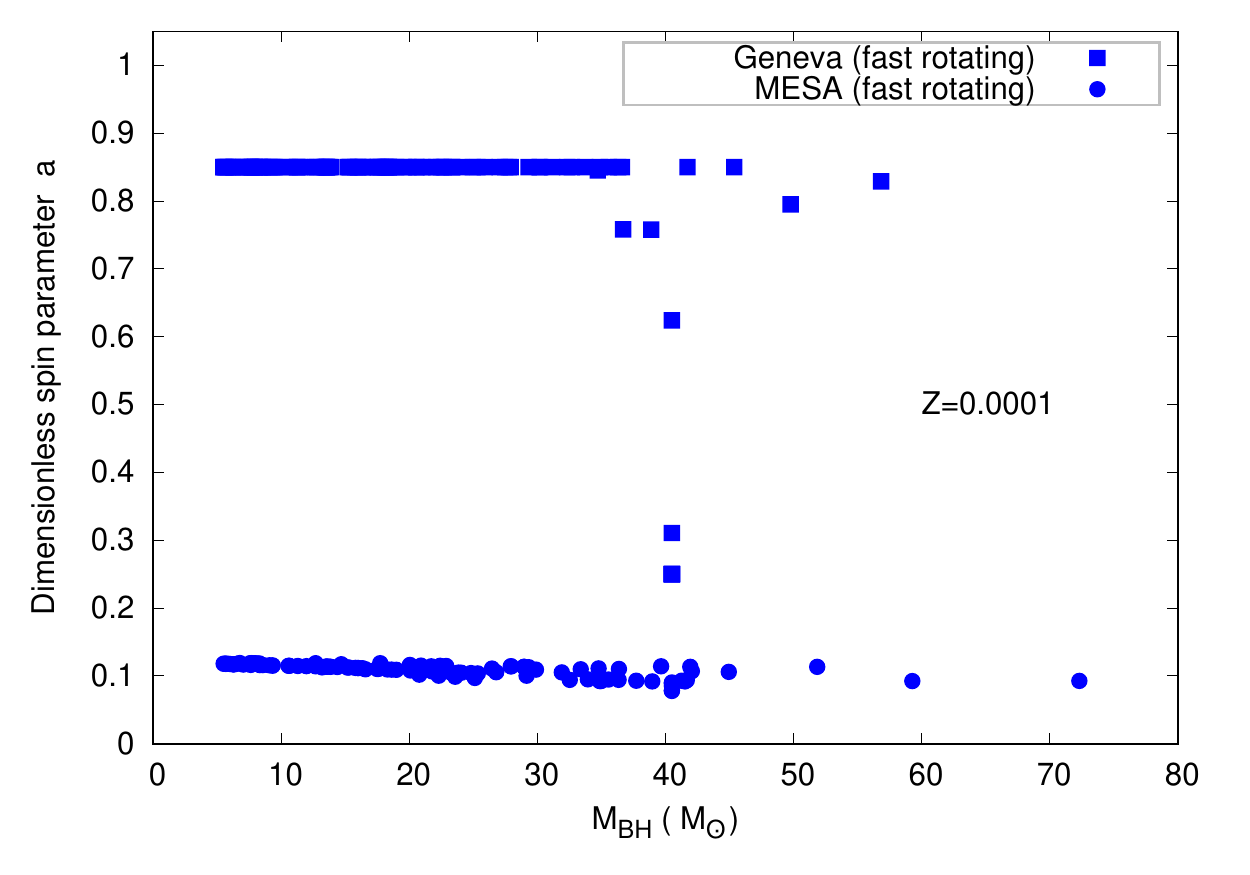}
\caption{Magnitude of dimensionless spin parameter, $a$, of stellar-remnant BHs at birth
(\ie, of BHs that have \emph{not} undergone any mass accretion or GR coalescence
after their formation) as a function of the progenitor star's carbon-oxygen
core mass, $\mco$ (left column), and the BH mass, $\mbh$ (right column).
The data points are obtained directly from $\nbseven$ computations in this work.
{\bf Top panels:}
the N-body models corresponding to these panels employ the ``Geneva model'' of
\citet[][B20]{Belczynski_2020} for BH spin (Sec.~\ref{bhspin})
and comprise only single stars initially, whose ZAMS masses range from
$0.08\Ms-150.0\Ms$ and which are distributed according to the standard IMF.
These models use the F12-rapid+B16-PPSN/PSN remnant mass prescription (Sec.~\ref{newrem}). The models
are for four metallicities, $Z=0.0002$, 0.001, 0.01, and 0.02 as indicated in the
legends.
{\bf Middle panels:}
the N-body models corresponding to these panels employ the ``MESA model'' of B20
for BH spin (Sec.~\ref{bhspin}).
The other model characteristics are the same as those in the top panels except
that the ``weak'' PPSN mass prescription (Sec.~\ref{newrem}) of \citet{Leung_2019}
is utilized (resulting in the non-monotonic behaviour w.r.t. $\mbh$ which, here, extends
up to $\approx50\Ms$ as opposed to the models in the top panels where $\mbh$ is
capped at $40.5\Ms$ due to the use of B16-PPSN/PSN).
{\bf Bottom panels:}
these $\mco-a$ and $\mbh-a$ relations, employing the Geneva and MESA BH-spin
prescriptions (see legend), are from model computations with $Z=0.0001$ where all stars with $\mzams\geq16\Ms$
are in primordial binaries (see Sec.~\ref{comp}).
The other model characteristics are the same
as those in the top panels. Here, as a result of star-star mergers occurring in the
massive binaries, $\mbh$ exceeds the widely-accepted PPSN upper limit
of $45\Ms$. Such ``mass-gap BH''s possess high (low) spins when Geneva (MESA) models are applied.
}
\label{fig:bhspin}
\end{figure*}

\subsection{Remnant natal kick}\label{newkick}

The remnant natal kick is based on observed kick distribution of
NSs (for single NSs in the Galactic field, the one-dimensional
kick velocity dispersion is $\vkickns\approx165\kmps$; \citealt{Hobbs_2005})
that is reduced linearly by the fraction of material fallback in
the SN. This ``momentum conserving'' formulation, as given by Eqn.~1 of Ba20,
is identical to the natal kick treatment in other widely-used population synthesis
programs such as $\startrack$ \citep{Belczynski_2008}, {\tt TrES} \citep{Toonen_2016},
and $\mobse$ \citep{Giacobbo_2018}.
The amount and fraction of the SN material fallback are provided by the
chosen remnant-mass scheme (see Sec.~\ref{newrem}). Fig.~\ref{fig:fbfrac}
shows the fallback fraction, $\fbfac$, as a function of carbon-oxygen
core mass, $\mco$, and remnant mass, $\mrem$, for
F12-rapid+B16-PPSN/PSN remnant mass model (Sec.~\ref{newrem}),
at different metallicities.

Apart from this standard momentum-conserving
recipe, two of its variants can be opted for, namely, models for 
``convection-asymmetry-driven'' kick \citep{Scheck_2004,Scheck_2008,Fryer_2007} 
and ``collapse-asymmetry-driven'' kick \citep{Burrows_1996,Fryer_2004,Meakin_2006,Meakin_2007}.
These alternative recipes are given by Eqns.~2 and 3 of Ba20. A recipe
for ``neutrino-driven'' \citep{Fuller_2003,Fryer_2006}
kick is also available as an option (Eqn.~4 of Ba20).  
The effects of these various natal kick models on the retention of NSs and BHs
in clusters, right after their birth, is demonstrated
and discussed in detail in Ba20.

To summarize, taking into account the slow down of natal kicks due to fallback,
the momentum-conserving and convection-asymmetry-driven kicks lead to similar BH and NS retention
in all types of clusters. The collapse-asymmetry-driven kick, including fallback reduction,
would retain most of the BHs in
young massive clusters (hereafter YMC; \citealt{PortegiesZwart_2010}),
open clusters (hereafter OC), and globular clusters (hereafter GC)
which have central escape speeds, $\vesc$, ranging from a few $\kmps$ 
(for low-mass OCs), through $\sim10\kmps$ (for YMCs,
massive OCs, and GCs; \citealt{PortegiesZwart_2010}),
up to 100 $\kmps$ (for massive GCs; \citealt{Georgiev_2009,Baumgardt_2018}).
The relatively slower NSs in the collapse-asymmetry-driven kick model
would lead to a higher NS retention in GCs \citep[\eg,][]{Goswami_2014,Kremer_2020}
but still similar retention as
in the momentum-conserving and convection-asymmetry-driven cases, in YMCs and OCs.
Note that in the above three kick models, the majority of the NSs that retain after birth are
products of ECS, assuming that ECS produces low
kicks $\sim$ few $\kmps$ \citep{Podsiadlowski_2004,Gessner_2018}. BHs of 
$\gtrsim10\Ms$ would also retain in clusters since they receive small or
zero natal kicks as, for them, $\fbfac\approx1$ (see Fig.~\ref{fig:fbfrac}). 
$\fbfac=1$ implies a failed SN, leading to the formation of a BH via
direct collapse.

Finally, since neutrino-driven kick is independent
of fallback, natal kicks of all core-collapse-SN produced NSs and BHs would be much higher
and they would escape from YMCs, OCs, and GCs. ECS-produced NSs would be
the only retaining remnants in the case of neutrino-driven natal kick mechanism.
See Sec.~3.2.4, Fig.~9, and Fig.~11 of Ba20 for further discussions.
The retention fractions would be moderately affected if the BH and NS progenitor
stars are in primordial binaries as demonstrated in Sec.~4 and Table~1 of Ba20.
Note that irrespective of the natal kick mechanism, a large fraction of NSs and BHs
will always retain in nuclear star clusters (hereafter NSCs) owing to their
very high $\vesc$ between $300-500\kmps$ \citep{Schoedel_2014,Georgiev_2009,Georgiev_2016}.

Based on these considerations, all computations in this work uses either momentum-conserving or
collapse-asymmetry-driven natal kick prescription. The kick-velocity components of core-collapse-SN
NSs (without fallback) have a Gaussian distribution with the \citet{Hobbs_2005} dispersion
of $\vkickns=265\kmps$. The counterpart for ECS-NSs is assigned a much lower
value of $\vkickecs=3\kmps$ \citep{Gessner_2018}. In this work, no natal kick
is applied to WDs; $\sim$ few $\kmps$ natal kicks of WDs would not significantly influence
the outcomes of the present models except for the least massive ones (see Sec.~\ref{comp}; Table~\ref{tab_nbrun}),
where WD kicks would have led to a somewhat higher mass loss from the cluster
over its long-term evolution \citep{Fellhauer_2003}.

\subsection{Natal spins of black holes}\label{bhspin}

The O-type parent stars of BHs typically possess significant spin angular momentum,
as observations of young, massive stars suggest \citep{Ramirez_2013,Ramirez_2015}.
However, this doesn't necessarily translate
into high spins of their remnant BHs. The spin angular momentum of a BH is often expressed
through its dimensionless counterpart, namely, the Kerr vector/parameter
(dimensionless spin vector/parameter; \citealt{Kerr_1963}) $\vec a$, which is defined as
\beq
\vec a = \frac{c\vec S_{\rm BH}}{G\mbh^2}.
\label{eq:adef}
\eeq
Here $\vec S_{\rm BH}$ is the total angular momentum vector of a Kerr BH of (non-spinning) mass
$\mbh$.

In the absence of any angular momentum supply on to the evolved parent star (\eg, due to
mass accretion or tidal interaction from a binary companion), the remnant BH can potentially
have a low spin if the angular momentum of the inner stellar core is carried away along
with the stellar wind. Therefore,
the magnitude of $\vec a$ ($a$) depends on (i) how efficient is the transport of
angular momentum from the core of the pre-SN star to its envelope and (ii) how
high is the wind mass loss rate. In this work, the BH spin estimates of B20
are adopted for BHs \emph{at their birth}
\ie, for those BHs that are non-recycled, having \emph{not} undergoing
any mass accretion or GR coalescence after their formation
(see Secs.~\ref{grkick} and \ref{mrgr} for the treatment of spins of second/higher-generation and
mass-accreted BHs). B20 estimate
BH spins from detailed evolutionary models of fast-rotating single stars.

Fig.~\ref{fig:bhspin} (top row) shows the outcomes, as a function of metallicity $Z$,
of the present adoption of
the BH natal spin model of B20 that is based on rotating stellar models using
the Geneva stellar-evolution code
(\citealt{Eggenberger_2008,Ekstrom_2012}; hereafter ``Geneva'' BH-spin model).
Since the Geneva code does not include magnetic field,
core-to-envelope angular momentum transport is purely convective
and, therefore, inefficient leading to most
of the BHs forming with a high spin, $a=0.85$. Only the strong wind
of the most massive stars are effective in eliminating the core's angular momentum
so that the most massive BHs form with
$0.25\geq a \geq 0.0$ as see in Fig.~\ref{fig:bhspin} (top row; \cf Fig.~1 of B20).
Note that the $\mco-a$ relation (and hence the $\mbh-a$ relation) depends on $Z$:
here, the $\mco-a$ function is taken piecewise over $Z$-ranges in the same
way as in \citet[][see their Table~1]{Morawski_2018}.

Fig.~\ref{fig:bhspin} (middle row) shows the $\mco-a$ and $\mbh-a$ relations
in $\nbseven$ for the BH natal spin model of B20 that is based on
rotating stellar models using the $\mesa$ stellar-evolution code
(\citealt{Paxton_2011,Paxton_2015}; hereafter ``MESA'' BH-spin model). Unlike the Geneva code,
$\mesa$ includes magnetic field that makes the outwards angular momentum
transport from the core much more efficient by forming a
Tayler-Spruit magnetic dynamo \citep{Spruit_2002,Fuller_2019}.
This causes BHs of all masses to form with a small residual spin, $a\sim0.1$,
as seen in Fig.~\ref{fig:bhspin} (middle row; \cf Fig.~2 of B20).
Likewise the Geneva BH-spin model, the piecewise $Z$-dependence of
the MESA $\mco-a$ relation is divided over the same $Z$-ranges
as in \citet[][]{Morawski_2018}.

When a primordial massive-binary population as in the
current models (see Sec.~\ref{comp}) is present, star-star mergers can form
BHs in the PSN mass gap (see Sec.~\ref{gapbh}; Ba20). In conjunction with the Geneva (MESA)   
BH-spin model, such BHs will also form with high (low) spins as demonstrated
in Fig.~\ref{fig:bhspin} (bottom row).

Note that according to the recent work by \citet{Fuller_2019}, the Tayler-Spruit
magnetic dynamo can essentially extract all of the angular momentum of the
proto-remnant core, leading to nearly non-spinning BHs. To consider this possibility,
the option of assigning $a=0$ irrespective of $\mbh$ is kept in $\nbseven$
(hereafter Fuller BH-spin model).

Note that, for the time being, the above natal spin models are applied only to BHs.
The NS spins are still treated as defaulted in $\nbseven/\bse$\footnote{The
new implementations described in Secs.~\ref{newwind}, \ref{newrem}, \ref{newkick},
and \ref{bhspin} are also adopted in and independently
tested to work correctly in a version of $\nbpp$ (in preparation).}.

\subsection{General relativistic merger recoil and final spin}\label{grkick}

The final merged BH out of a BBH merger would receive a recoil kick, $\vec\vk$,
due to asymmetric radiation of GW. The magnitude and direction (w.r.t. the BBH orbit)
of the recoil
depends on the BBH's mass ratio and on the magnitudes and directions
w.r.t. the orbital angular momentum, $\vec\orbl$, of the component BHs' $\vec a$s
\citep{Pretorius_2005,Campanelli_2007,Hughes_2009}.
If the merging BHs' spins are zero, $\vec\vk$ will be aligned
along the line joining the BHs just before the 
merger. Its magnitude, $\vk$, is zero (small) for equal-mass (extreme-mass-ratio) 
components and maximizes to $\approx170\kmps$
at the mass ratio of $\approx1/2.9$ \citep{Baker_2007,Baker_2008}.

The GW emission is
particularly asymmetric if the merging BHs are spinning. In that case,
depending on the spins' magnitudes and orientations, 
$\vec\vk$ will as well have an in-orbital-plane component perpendicular to the mass axis 
and a component perpendicular to the orbital plane. For (near-) maximally-spinning
BHs, this off-plane component typically dominates and can well 
exceed $500\kmps$; for certain configurations, it 
can reach $\approx3000\kmps$ \citep{Campanelli_2007,Baker_2008,vanMeter_2010,Lousto_2013}.

For typical configurations, the orbital angular momentum is the primary
contributor to the final (dimensionless) spin, $\vec\af$, of the merged BH \citep{Rezzolla_2008}.
If the merging BHs' spins are zero, then the only source of angular momentum
in the BBH system is $\vec\orbl$. Accordingly, $\vec\af$, will be aligned with $\vec\orbl$,
and, for equal-mass components, will have the magnitude $\af\approx0.7$ \citep{Pretorius_2005}.
With finite, misaligned spins of the merging BHs,
$\vec\af$ will be misaligned relative to $\vec\orbl$ and its magnitude will be 
augmented (suppressed) w.r.t the non-spinning-merger value if the spins are 
pro-aligned (anti-aligned). Since, for most configurations,
$\vec\orbl$ dominates the BBH system's angular momentum budget, the misalignment
is, typically, $<10\degree$. 

As a first attempt to incorporate on-the-fly NR-based treatments of a BBH merger that
takes place while the BBH is bound to the cluster, $\nbseven$
utilizes the NR-based fitting formulae of \citet{vanMeter_2010}
for evaluating the components of $\vec\vk$. These formulae 
incorporate cases where the BHs' spins are inclined w.r.t. $\vec\orbl$
and hence would undergo spin-orbit precession during the in-spiral 
and merger phases. These formulae agree with NR outcomes within $5\%$.
As for $\vec\af$, the NR-based fitting functions of \citet{Rezzolla_2008}
are applied. When the members of the BBH do not derive from the
same primordial binary (\ie, the BBH is dynamically assembled),
the orientations of the BHs' spins are chosen randomly and isotropically (their
dimensionless magnitudes being according to the chosen BH-spin model;
see Sec.~\ref{bhspin}) to evaluate $\vec\vk$ and $\vec\af$.
If both BH members retain the membership of the same parent primordial binary,
then, considering the parent stars' close birth locations and potential interactions
thereafter, the BHs' random relative spin orientations are restricted within $\leq90\degree$.
The same procedure is followed for computing $\vec\vk$ and $\vec\af$ of
BBH mergers that take place within a Hubble time after getting ejected from the cluster, using
a standalone version of the above mentioned NR formulae.

For all BNS and NSBH mergers, $\vec\vk=0$ is assigned and $\vec\af$ is
evaluated in the same way as for BBH mergers, as a preliminary treatment.
In both cases, the merger product is assumed to be a BH of mass equal to
the total binary mass.

For all in-cluster and ejected GR mergers, the ``effective spin parameter'' $\xeff$,
which is a measure of the spin-orbit alignment of the merging system \citep{Ajith_2011,Abbott_GW170104},
is evaluated according to the expression
\beq
\xeff = \frac{\mone\aone\cos\theta_1 + \mtwo\atwo\cos\theta_2}{\mone+\mtwo}.
\label{eq:xdef}
\eeq
Here, $\mone$, $\mtwo$ are, respectively, the masses of the merging members with
dimensionless spins $\vec\aone$, $\vec\atwo$ that project with angles
$\theta_1$, $\theta_2$ on $\vec\orbl$.

After assigning $\vec\vk$, the {\tt escape} algorithm of $\nbseven$ \citep{Aarseth_2003}
determines whether the merged BH should remain bound in the cluster or
escape, in the usual way. If the BH remains bound, it is treated alike
other cluster members. The $\af$ of the BH is also remembered and recycled if
the BH engages itself in further GR mergers. How $\vec\vk$ is assigned and spins of
first and higher-generation BHs are tracked are detailed in Sec.~\ref{nbspin}.

\subsection{Star-star and star-remnant mergers}\label{mrgr}

In a dynamically-active dense stellar system, mergers among stellar entities
can take place due to either binary evolution or dynamical interactions (in-orbit
collisions, due to binary-single and binary-binary encounters boosting the binaries'
eccentricities, and hyperbolic collisions). In both standalone and $\nbseven/\bse$,
a complete mixing is assumed in composing the merged star \citep{Hurley_2002}.
In this work, all star-star mergers are assigned a mass loss, as indicated in
theoretical studies \citep[\eg,][]{deMink_2013}. The amount of mass loss is taken to be equal to
a fixed fraction, $\fmrg$, of the secondary's mass 
(the less massive of the merging stars, at the time of the merger).
The merged product is a rejuvenated star of type and age according to
the merger-product-type scheme of \citet{Hurley_2002}. The total
mass of the merged product is equal to the total stellar mass just before the merger
minus the merger mass loss.

Furthermore, in a BH-star merger (forming a BH-Thorne-Zytkow-object; hereafter BH-TZO),
a fixed fraction, $\ftz$, of the star's mass is assumed to be accreted on to the BH, increasing
the BH's final mass.
The post-stellar-accretion BH is assumed to be maximally spinning, \ie, has $a=1$
irrespective of the BH's natal spin, due to the associated accretion of angular momentum.
Observations of high mass X-ray binaries indeed
suggest near-maximal spins of accreting BHs \citep{Miller_2015}.
(This treatment is different from that of a BBH merger product whose spin is assigned based on NR;
see Sec.~\ref{grkick}.) As shown in \citet{Banerjee_2020b}, a significant BH-TZO accretion
can lead to BHs well within the PSN mass gap, at all metallicities. 
Finally, if a BH forms during a tidal-interaction (symbiotic), mass-transfer,
or CE episode with a stellar binary companion,
then also $a=1$ is assigned to it, as supported by observations and theoretical
modelling \citep[\eg,][]{Qin_2019}.

No matter-accretion is assumed on to the NS, in an NS-star merger (forming
an NS-TZO; \citealt{TZ_1975}), whose final outcome is just the NS (the stellar material of the TZO
is assumed to be completely irradiated away by the NS).

See Secs.~\ref{nbspin} and \ref{nbmrg} for the details of the implementations
of these aspects in updated $\nbseven$.

\subsection{Black holes in the ``upper'' and ``lower'' mass gaps}\label{gapbh}

Combining with the discussions from Secs.~\ref{newrem}, \ref{newkick}, \ref{bhspin}, \ref{grkick}, \ref{mrgr} a BH can be
in the PSN (upper) mass gap, in the classical sense, due to
(i) weak PPSN, (ii) a star-star merger, (iii) a BBH merger,
(iv) a BH-TZO accretion. In a dynamically-active environment of a dense stellar cluster,
all the four processes can, in principle, take place and can even combine. Among these,
a PSN-gap BH from channel (i) or (ii) is a first-generation BH, that from channel
(iii) is a second-generation BH, and that from channel (iv) is a first (second)
generation BH if the accreting BH is of first (second) generation
\footnote{The outcome of a BH-TZO accretion can also be referred to
as a ``recycled'' BH but, in this work, we simply stick to referring to BH generations.}.
A PSN-gap BH out of channel (i) or (ii) will have its spin as per the adopted
BH natal spin model (Sec.~\ref{bhspin}) and that from channel (iv)
will necessarily be maximally spinning, as assumed here (Sec.~\ref{mrgr}).
For most BBH mergers, channel (iii) will also produce highly spinning BHs ($a\geq0.5$).
However, depending on the spin-orbit configuration of the merging BHs, the
second-generation (PSN-gap) BH can also be of moderate or low spin ($a<0.5$),
as discussed in \citet{Belczynski_2020c}.

If the PSN-gap BH is retained in the cluster, it can, in principle, participate in (further)
GR mergers and be detectable by ground- and space-based
GW detectors such as LIGO-Virgo and LISA
\citep[laser interferometer space antenna][]{eLISA}. Among the above four channels,
channel (iii) will typically eject the PSN-gap BH from any cluster when one or both of
the merging, first-generation BHs are highly spinning (but see \citealt{Belczynski_2020c}).
When they are equal (close) in mass and are (nearly) non-spinning (Fuller BH-spin model
in Sec.~\ref{bhspin}), they would retain in most OCs and more massive systems. 
For small BH spins (MESA BH-spin model in Sec.~\ref{bhspin})
or for unequal-mass non-spinning BHs, the recoiled BH would typically
retain in NSCs \citep{Miller_2009,Antonini_2019} but escape from
OCs, YMCs, and GCs. These inferences are based on the GR merger recoils
as discussed in Sec.~\ref{grkick}.
The other channels will retain the PSN-gap BH in any cluster
(all PPSN BHs and core-collapse-SN BHs of $\gtrsim10\Ms$
are direct collapse BHs, in the present remnant schemes; see Sec.~\ref{newkick}).

Note that if a seed BH grows in mass via extreme-mass-ratio inspirals (hereafter EMRI) with many small BHs,  
as proposed in, \eg, \citet{Miller_2002,Hughes_2003},
then a cluster could sustain many generations of BBH mergers, due to the small GR recoils in
EMRIs. That way, BHs can enter the PSN gap with low spins. 
In the present computed model clusters (Sec.~\ref{comp}), BHs derive from a continuous stellar initial
mass function (hereafter IMF) and comprise a (nearly) continuous mass distribution.
In that case, seed BH growth will be ineffective.
(The present models also do not undergo an early runaway stellar merger episode as
in, \eg, \citealt{Portegies_2007,Fujii_2013} that would produce a seed BH.)

As for the lower mass gap between NSs and BHs (Sec.~\ref{newrem}), BHs can lie in the
gap only in the case of F12-delayed remnant mass scheme that does
not produce this gap (Sec.~\ref{newrem}). This is due to the current
assumptions regarding NS-star mergers (Sec.~\ref{mrgr}). Note that
such low-mass BHs are formed with small SN fallback (Sec.~\ref{newrem}), so that
they receive large natal kicks in the momentum-conserving
kick scenario (Sec.~\ref{newkick}). Therefore, to retain
such low-mass BHs in OCs, YMCs, and GCs, it is also necessary to
adopt the collapse-asymmetry-driven kick model (Sec.~\ref{newkick}) instead.

With F12-rapid remnant
model, for which this lower gap is inherent (Sec.~\ref{newrem}; Fig.~\ref{fig:inifnl}),
BNS mergers can produce BHs in the gap that would then retain in the
clusters (Sec.~\ref{grkick}) and potentially participate in further GR mergers. However,
in-cluster BNS mergers are rare in the present computed models
(see Sec.~\ref{comp}, Table~\ref{tab_nbrun}).

\section{Direct, post-Newtonian many-body computations with updated stellar-evolutionary and
black hole spin models}\label{comp}

Using the updated $\nbseven$ as described in Sec.~\ref{newnb}, 
a set of 65 long-term N-body computations of model stellar clusters
are performed with varied remnant-mass, BH
natal spin, metallicity, and initial cluster parameters. The initial
cluster parameters and model choices are given in Table~\ref{tab_nbrun}.
These runs are performed over about an year whilst the
code developments in $\nbseven$ which is why, at present,
the evolutionary model grid in Table~\ref{tab_nbrun}
is somewhat heterogeneous. However, the present set is still
composed of representative examples of the various
remnant and natal spin schemes and their observable outcomes (see Sec.~\ref{result}).

The initial model clusters have \citet{Plummer_1911} profiles with masses
$1.0\times10^4\Ms \leq \mcl(0) \leq 1.0\times10^5\Ms$, half-mass radii
$1.0{\rm~pc} \leq \rh(0) \leq 3.0{\rm~pc}$, \emph{overall} (see below)
primordial-binary fractions $0.0 \leq \fbin(0) \leq 0.1$, 
and have metallicities $0.0001\leq Z\leq0.02$.
All the computed models are taken to be initially virialized \citep{Spitzer_1987} and
unsegregated and they are subjected to a solar-neighbourhood-like
external galactic field.
All initial models are composed of
ZAMS stars. All models are evolved until 11.0 Gyr unless
the cluster is dissolved earlier\footnote{For a few
models, the run had to be concluded prematurely
as it could not be recovered after a crash or could
be run only with impractically small regular time steps,
after arriving at its core-collapsed state \citep{Spitzer_1987}.
This is, of course, not the case with all core-collapsed models.}. 

Note that the values of $\fbin(0)>0.00$ quoted in Table~\ref{tab_nbrun}
represent the \emph{overall} initial binary fraction in the cluster
for the entire stellar mass distribution, the latter being
taken to be the standard IMF \citep{Kroupa_2001} and in
the ZAMS mass range $0.08\Ms-150.0\Ms$.
However, as in Papers II \& III, the initial binary fraction of the
O-type stars (ZAMS mass $\geq\mcrit$; $\mcrit=16.0\Ms$), which are paired among themselves,
is taken to be $\fobin(0)=100$\%, to be consistent with the observed high binary fraction
among O-stars in young clusters and associations
\citep[\eg,][]{Sana_2011,Sana_2013}. The initial binary fraction among the non-O-type binaries
would still be $\approx\fbin(0)$ since, for the adopted standard stellar IMF, the
O-stars comprise only a small fraction of the total stellar population.
The O-star binaries are taken to initially follow
the observed orbital-period distribution of \citet{Sana_2011}
and a uniform mass-ratio distribution.
The orbital periods of the non-O-star primordial binaries are taken to follow the period
distribution of \citet{Duq_1991} and their mass-ratio distribution is also taken to be uniform.
The initial eccentricities of the O-star
binaries follow the \citet{Sana_2011} eccentricity distribution and those
for the rest of the binaries are drawn from the thermal eccentricity distribution
\citep{Spitzer_1987}. As explained in Paper II, such a scheme of including
primordial binaries provides a reasonable compromise between the economy of computing
and consistencies with observations.

When $\fbin(0)=0.00$ is quoted in Table~\ref{tab_nbrun},
an initial model with all single ZAMS star (standard IMF) is implied.
Although having an observationally-motivated primordial-binary population
is more realistic, large-N runs, as in here, with primordial binaries
is significantly more time-consuming (and tedious). Therefore, to obtain reasonable
statistics, initially single-star models are also performed. 
In a few primordial-binary runs, $\mcrit=5.0\Ms$ is taken instead of $16.0\Ms$ as test
cases; it is mentioned in the table's footnote when so. Unless stated otherwise
in Table~\ref{tab_nbrun}, $\fmrg=\ftz=0.5$ (Sec.~\ref{mrgr}) is taken.
In several runs, a reduced $\fmrg$ and an enhanced $\ftz$ are
assumed to facilitate BHs enter the PSN mass gap (Sec.~\ref{gapbh}) and explore
its consequences on observable GR mergers (Sec.~\ref{result}).

The star cluster models computed here represent young massive and open clusters that
continue to form and dissolve throughout a gas-rich galaxy, as in, \eg,
the Milky Way and Local Group galaxies. Note that
the 1 -3 pc  initial size of the present models is consistent with
the typical sizes observed in gas-free YMCs
in the Milky Way and neighbouring
galaxies \citep{PortegiesZwart_2010,Banerjee_2017c}.
It is, therefore, implicit in the present models that these clusters
have survived their assembling and violent-relaxation phases
and have expanded to parsec scale sizes from sub-parsec sizes
(due to, \eg, gas expulsion),
as observed in newly-formed, gas-embedded and partially-embedded
clusters and associations; see \citet{Banerjee_2018b} and references
therein. The long-term evolution explored here would essentially wash
out any imprint \citep{Heggie_2003} of the complex birth history of the
clusters. The present models bridge between lower-mass embedded-cluster-type
initial conditions explored in works such as \citet{Kumamoto_2019,DiCarlo_2019}
and GC-progenitor systems as in \citet{Askar_2018,Kremer_2020}.

\begin{figure*}
\includegraphics[width=14.0cm,angle=0]{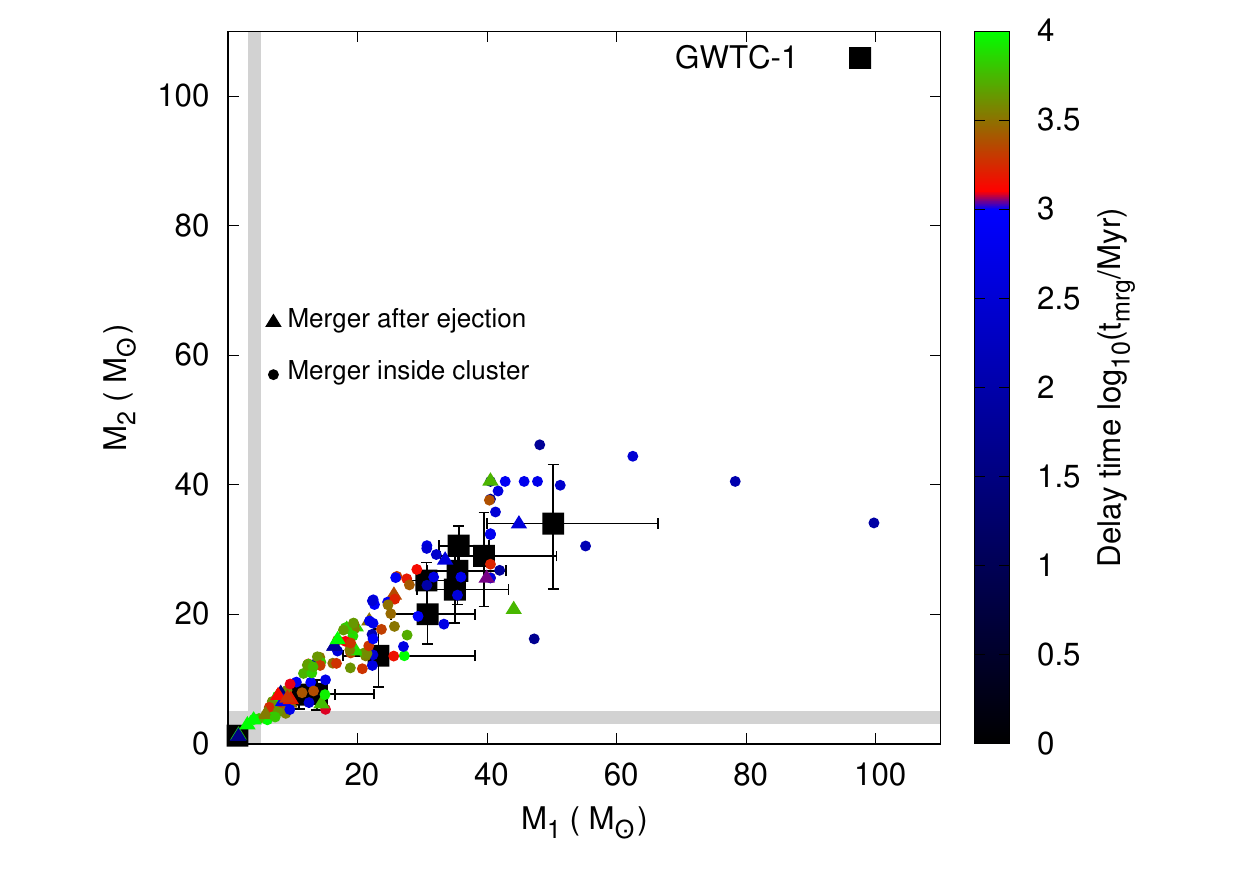}
\caption{Primary mass, $\mone$, versus secondary mass, $\mtwo$ ($\mone\geq\mtwo$), of GR compact-binary mergers
from all the computed models in Table~\ref{tab_nbrun} (filled circles and triangles). They are colour-coded
according to their delay times (colour bar). The circles represent mergers inside a model cluster
while being bound to it and the triangles represent those happening in compact-binary systems after
they get ejected from the cluster. The filled, black squares with error bars (90\% credible intervals)
represent the observed merger-event data from LVC O1/O2
\citep[GWTC-1,][]{Abbott_GWTC1}. The ``lower mass gap'' over $3.0\Ms\leq\mone,\mtwo\leq5.0\Ms$
is indicated with the grey shades.}
\label{fig:m1m2}
\end{figure*}

\section{Results}\label{result}

In the present paper, focus is given on how the GR-merger outcomes from the
present N-body computations (Sec.~\ref{comp}; Table~\ref{tab_nbrun}) agree
with the measured parameters of those observed in LVC O1/O2 \citep{Abbott_GWTC1}
and how they comply with the O3 candidates
(\url{https://gracedb.ligo.org/superevents/public/O3/}).

\subsection{Merger masses}\label{mrgmass}

Fig.~\ref{fig:m1m2} plots the primary and secondary masses, $\mone$, $\mtwo$ ($\mone\geq\mtwo$),
of all the GR mergers from the computed models in Table~\ref{tab_nbrun}.
The vast majority these mergers are in-cluster BBH mergers as is the case with somewhat
older prescriptions and treatments, for similar types of clusters, in Papers I, II, \& III.
See the discussions and analyses in these papers.

Fig.~\ref{fig:m1m2} demonstrates that the trend of the $\mone-\mtwo$ combination,
their ranges, and their scatter,
in the mergers from the computed models, agree well with those of the
GR merger events from LVC O1/O2 (taking
into account the latter's 90\% confidence intervals). For $\mone\gtrsim30\Ms$, all BBH mergers
happen at short delay times (the time interval between the beginning of the cluster evolution
and the occurrence of the merger), $\tmrg<1.0$ Gyr, except for a few ejected mergers
(see also Fig.~\ref{fig:delay_all}).
In this work, an ejected double-compact binary is labelled as a ``merger'' if
its GR coalescence time, $\tauinsp$, is less than the Hubble time, $\thub=13.7$ Gyr.
The adopted $\tauinsp$ for this purpose is given by \citep{Peters_1964}
\beq
\tauinsp\approx\frac{5}{64}\frac{c^5a_{\rm ej}^4(1-e_{\rm ej}^2)^{7/2}}
	{G^3 \mone \mtwo (\mone+\mtwo)}
	\left(1+\frac{73}{24}e_{\rm ej}^2+\frac{37}{96}e_{\rm ej}^4\right)^{-1},
\label{eq:taumrg}
\eeq
$a_{\rm ej}$ and $e_{\rm ej}$ being the binary's semi-major-axis and eccentricity, respectively, at the time
of its ejection.

A handful of mergers are ejected BNS mergers (left lower corner of Fig.~\ref{fig:m1m2}). All
these BNS mergers are ``isolated'' mergers in the sense that the merging NS members have maintained
their original primordial-binary membership. These binaries got ejected as eccentric BNSs
due to either the later-born NS member's natal kick or a dynamical encounter. The ``dynamical heating'' effect
of the BH core (see Sec.~\ref{intro}) prevents mass segregation of lighter members including NSs,
inhibiting strong dynamical interactions and dynamical pairing among themselves until
all BHs are nearly depleted \citep[Paper II;][]{Fragione_2018,Ye_2019}.
The present set of models do not yield dynamically-paired BNS mergers or NSBH mergers
\citep[but see][]{Rastello_2020}.
In contrast, as in Papers I-III and other studies
involving intermediate-mass and massive stellar clusters (see Sec.~\ref{intro}
and references therein), most of the BBH mergers are dynamically assembled and
take place inside the clusters (like
the BNS mergers, there are only a few BBH mergers where the primordial binary membership
is maintained until the merger). 

\subsubsection{Mergers in the ``upper'' and ``lower'' mass gaps}\label{gapmrg}

Fig.~\ref{fig:m1m2} shows that in the present models, BBH mergers take place with the primary
being in the PSN mass gap (see Sec.~\ref{newrem}), \ie, of $\mone>45\Ms$. All these
mergers happen inside clusters and at delay times $\tmrg<1.0$ Gyr (see also Fig.~\ref{fig:delay_all}).
Sec.~\ref{gapbh}
discusses the various ways in which a BH can appear in the PSN gap. Among the
four most prominent PSN-gap mergers in Fig.~\ref{fig:m1m2} ($\mone\gtrsim55\Ms$),
only the one with $\mone\approx80\Ms$ is an in-cluster second generation merger;
this particular sequential BBH-merger event is demonstrated
in Example 1 of Sec.~\ref{nbspin} (outcome from model 49 of Table.~\ref{tab_nbrun}).
The rest are mass-gained BHs through significant BH-TZO accretion
($\ftz\geq0.5$; see Table.~\ref{tab_nbrun}), which is the most common form of participation
in PSN-gap BBH mergers, in the present models. The $\mone\approx100\Ms$ BBH-merger primary
has originated due to a $\ftz=0.95$ BH-TZO accretion in a low $Z$ model (model 61 of Table.~\ref{tab_nbrun}). 
All the PSN-gap mergers took place in models with GC-like metallicities, of $Z\leq0.001$,
where the BHs' birth mass and mass gain are the largest.

Note that $\mone-\mtwo$s of the BBH mergers from the computed models
show a linear trend up to $\mone\approx40\Ms$,
with mass ratios typically $\mtwo/\mone>0.5$ (but see Sec.~\ref{asymrg}; see also Fig.~\ref{fig:delay_all})
and agreeing well with the O1/O2 merger-event data. This preferred
pairing of similar masses as in the O1/O2 events \citep{Fishbach_2020b}
comes out naturally of the present computations and is a consequence of
dynamical pairing. As already explained in Paper II, despite
pairing of unrelated members (\eg, members not belonging to the same primordial binary
or that have formed well separated from each other), commonly
termed as `random pairing', the most massive members segregate to the innermost
part of the host cluster and, hence, interact preferentially
among each other. In fact, this selection effect operates at
all levels: among the segregated members, the most massive ones would
engage themselves in close encounters most frequently due to
their stronger gravitational focusing \citep{Spitzer_1987} and
such encounters (\eg, three-body, binary-single, or binary-binary encounters)
would most likely result in pairing up the most massive participants since such an
outcome would be energetically favourable \citep{Heggie_2003}.
Since the BH mass distribution up to $\approx40\Ms$
is \emph{continuous} (see Fig.~\ref{fig:inifnl}; Figs.~8 and 12 of Ba20),
the pairing is automatically biased
towards similar masses, typically resulting in $0.5<\mtwo/\mone<1.0$.

However, this trend deviates sharply for larger $\mone$,
in the PSN gap, as seen in Fig.~\ref{fig:m1m2}.
(Since most of the computed models assume B16-PPSN/PSN model,
the PSN gap actually begins from $\approx 40\Ms$, in the present model set; see Sec.~\ref{newrem};
for the sake of discussions, PSN gap will still refer to $>45\Ms$.) Within the PSN gap,
the mass ratio tends to decline from 0.5 with $\mone$. This is, again, a consequence of random, dynamical
pairing of the merging BBHs and of the fact that
mass-gained PSN-gap BHs are, in most models, \emph{distinctively} the most massive 
members. The computed data points, nevertheless, well encompass the
most massive and marginally-PSN-gap \citep[but see][]{Chatziioannou_2019,Kimball_2020} 
BBH merger event GW170729 (the most massive O1/O2
$\mone-\mtwo$ pair in Fig.~\ref{fig:m1m2}).

For the typical $\sim10\kmps$ escape speeds from the present models, the only way a
second-generation BH retains in a cluster is
via mergers of two similarly-massive, first-generation BHs, that did not  
undergo any previous matter-interaction (see Sec.~\ref{mrgr}),
with the Fuller BH-spin model (\ie, BHs having practically zero spin at birth; see Sec.~\ref{bhspin}).
In all other cases, one or both of the merging BHs
will have spins of $a>0.1$ and the merger recoil speed will eject the merged BH from the cluster
(see Secs.~\ref{grkick} and \ref{gapbh})\footnote{An exception is the application
of the Geneva BH-spin model with $Z=0.001$ which would also make the most massive BHs non-spinning;
see Fig.~\ref{fig:bhspin} (top panel); see also \citealt{Morawski_2018}.}.
Even if the second-generation BH retains in
a cluster, it doesn't necessarily participate in further (in-cluster or ejected) mergers and may get ejected
via dynamical encounters, as seen in the present models that employ the Fuller BH spin.
Nevertheless, such models do tend to produce a larger number of in-cluster BBH mergers;
see Table~\ref{tab_nbrun}. The probability of a retained second-generation BH to
actually participate in further mergers would increase in higher-escape-velocity
systems like massive GCs, as seen in, \eg, \citet{Rodriguez_2018,Rodriguez_2019}.
Also, owing to their $\sim100\kmps$ escape speed, the chances of retaining second
and even higher generation BHs and of their participation in
further mergers increases in NSCs, as explored in, \eg, \citet{Miller_2009,Antonini_2019,ArcaSedda_2020b}.

Fig.~\ref{fig:m1m2} also demonstrates BBH mergers in the lower NS-BH mass gap (the shaded strips;
see Secs.~\ref{newrem} and \ref{gapbh}), happening both inside clusters and
after ejection. As explained in Sec.~\ref{gapbh}, under the present assumptions
and adopted models, only the combination of F12-delayed remnant mass model (Sec.~\ref{newrem})
with collapse-asymmetry-driven natal kick (Sec.~\ref{newkick}) would retain such lower mass gap BHs in
the clusters, after their birth. The lower mass gap mergers in Fig.~\ref{fig:m1m2}  
are from such models (see Table~\ref{tab_nbrun}), exclusively. Note that
these mergers happen at long delay times of $1{\rm~Gyr} \lesssim \tmrg \lesssim 10{\rm~Gyr}$. 
This is expected since such low-mass BHs have to wait until their
more massive counterparts are dynamically depleted from the cluster, before
they can effectively participate in dynamical encounters, as also found in
previous studies \citep[\eg,][]{Morscher_2015,Chatterjee_2017b}.
Note, further, that the clusters with F12-rapid remnant model, that have
actually produced such lower mass gap mergers, are of higher $Z$ (0.01 or 0.02),
which produce and retain the least number of massive, $\gtrsim10\Ms$ BHs. 
Interestingly, LVC has indeed detected a few of such (lower) ``mass gap'' mergers
in their O3.
Note that although in all model clusters the BH population depletes with
time due to dynamical ejections (plus BBH mergers), BHs continue
to be present in them until late evolutionary times, as demonstrated in
Fig.~\ref{fig:nbh}. This happens due to the fact that the energetic
dynamical interactions, that form BBHs and eject BHs, also lead to
heating and expansion of the clusters which effect, in turn, moderates
the BHs' dynamical activities
(see, \eg, \citealt{Breen_2013,Morscher_2015,Banerjee_2017}).

\subsubsection{Mass-asymmetric mergers: on GW190412}\label{asymrg}

In the light of the above discussions, it would be worth highlighting
the BBH merger event GW190412 of the LVC O3,
published very recently by the collaboration \citep{LSC_GW190412}.
With respect to all the LVC GW merger events so far, the most striking aspect
of GW190412 is its highly asymmetric mass ratio of $\mtwo/\mone=0.28^{+0.13}_{-0.07}$.
Although relatively rare, similarly low mass ratio BBH mergers, with the
primary below the PSN gap, do take place in the present computed models.
In Fig.~\ref{fig:m1m2} (see also Fig.~\ref{fig:delay_all}), they are the few lower mass ratio outliers
of the overall O1/O2/computed-merger trend. Apart from these, the BBH
mergers with $\mone$ within the PSN gap are also of $\mtwo/\mone\lesssim0.5$,
as discussed in Sec.~\ref{gapmrg}. Overall, 3\% of all GR mergers  
from the present models are highly asymmetric in mass, in the sense
of having $\mtwo/\mone<0.5$, with the most asymmetric one being
of $\mtwo/\mone\approx0.3$ (primary below the PSN gap). All but one of such mergers are 
in-cluster mergers; since the majority of the mergers are in-cluster (see above),
this is also reflected in their mass-asymmetric subset (the asymmetric merger
fraction is 4\% w.r.t. the in-cluster mergers).

Despite the ``dynamical selection'' effect discussed in Sec.~\ref{gapmrg},
asymmetric outliers are statistically possible in random pairing. The most
concentrated ones among the present model set produce the asymmetric mergers;
those with $\mcl(0)\geq7.5\times10^5\Ms$ and with $\mcl(0)=5\times10^4\Ms$,
$\rh(0)=1.0$ pc (Table~\ref{tab_nbrun}). Higher stellar density allows
central mass segregation of BHs down to lower-mass BHs
(Sec.~\ref{intro} and references therein) promoting close interactions and pairings among
more dissimilar masses. If the GR inspiral (Sec.~\ref{insp}) begins
early enough in the cluster evolution that many exchange interactions do \emph{not}
equalize the in-cluster BBH component masses \citep[\eg,][]{Antonini_2020},
highly mass-asymmetric mergers are possible, as the present computations
demonstrate. Indeed, all the mass-asymmetric BBH mergers, in the present
computations, take place within $\tmrg<1$ Gyr
(see Figs.~\ref{fig:m1m2}, \ref{fig:delay_all}; the ejected one
is ejected from the parent cluster at $\approx500$ Myr cluster-evolutionary time).

This is further supported by the very recent work of \citet{DiCarlo_2020},
who have indeed obtained a larger fraction of such asymmetric, GW190412-like BBH mergers
in their computations of a large number of cluster models with $\rh(0)$ 0.2 pc-1.5 pc
and $\mcl(0)$ $10^3\Ms-3\times10^4\Ms$. Such models have shorter
two-body relaxation and hence BH mass-segregation times
\citep{Spitzer_1987,Banerjee_2010,Breen_2013,Antonini_2020} and, hence,
are expected to produce more of such asymmetric BBH mergers.
The shorter mass-segregation times in their models also induce  
runaway stellar mergers, enabling the formation and participation in mergers
of PSN-gap BHs which are also, typically, asymmetric
(\citealt{DiCarlo_2019}; see also Sec.~\ref{gapmrg}).

Therefore, YMCs and OCs serve as sites that naturally enable mass-asymmetric BBH mergers.
Apart from star clusters, other astrophysical environments that enable close interactions
and mergers among unequal-mass BHs are active galactic nucleus (AGN) gas discs \citep[\eg,][]{Secunda_2019}
and field triple and higher-order systems
\citep[\eg,][]{Silsbee_2017,Antonini_2017,Fragione_2019b}.

\begin{figure*}
\includegraphics[width=9.0cm,angle=0]{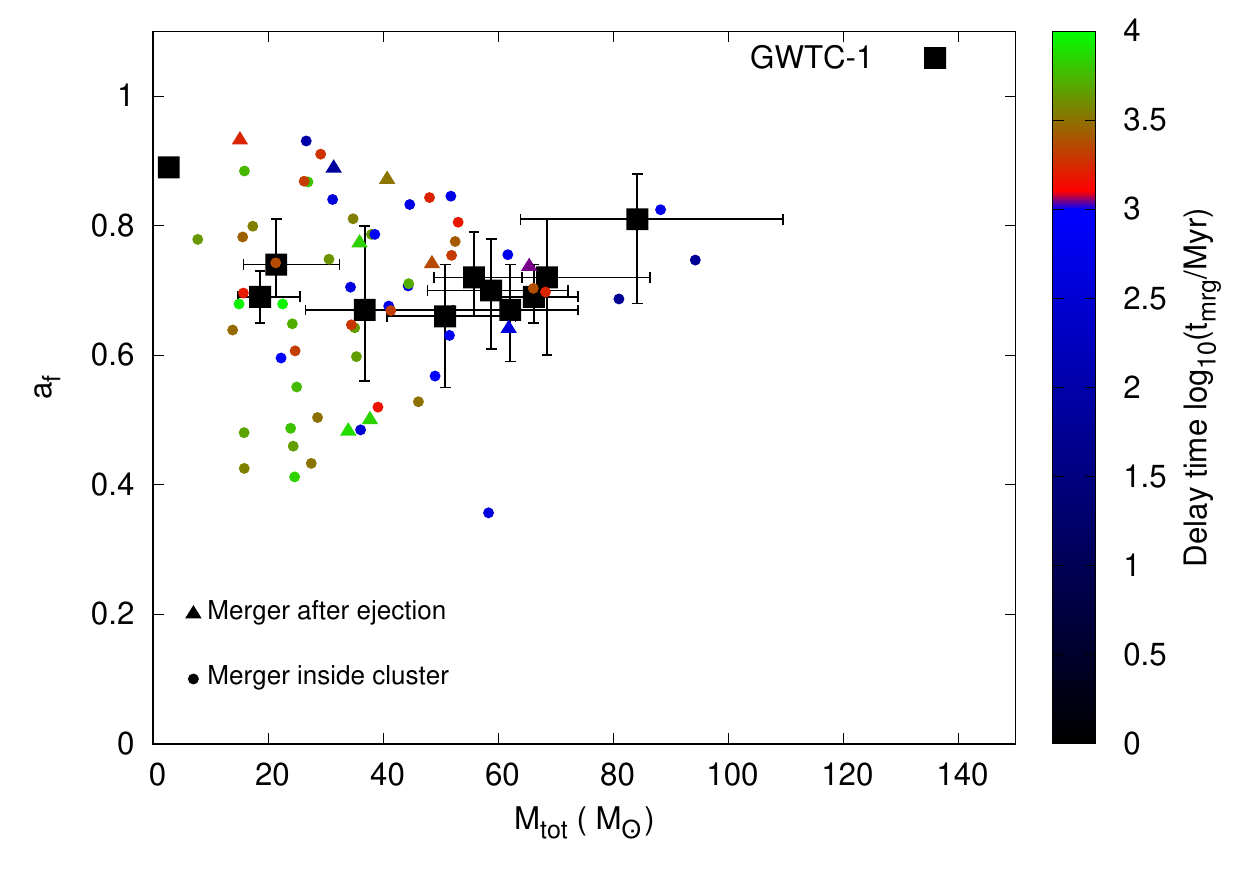}
\hspace{-0.5cm}
\includegraphics[width=9.0cm,angle=0]{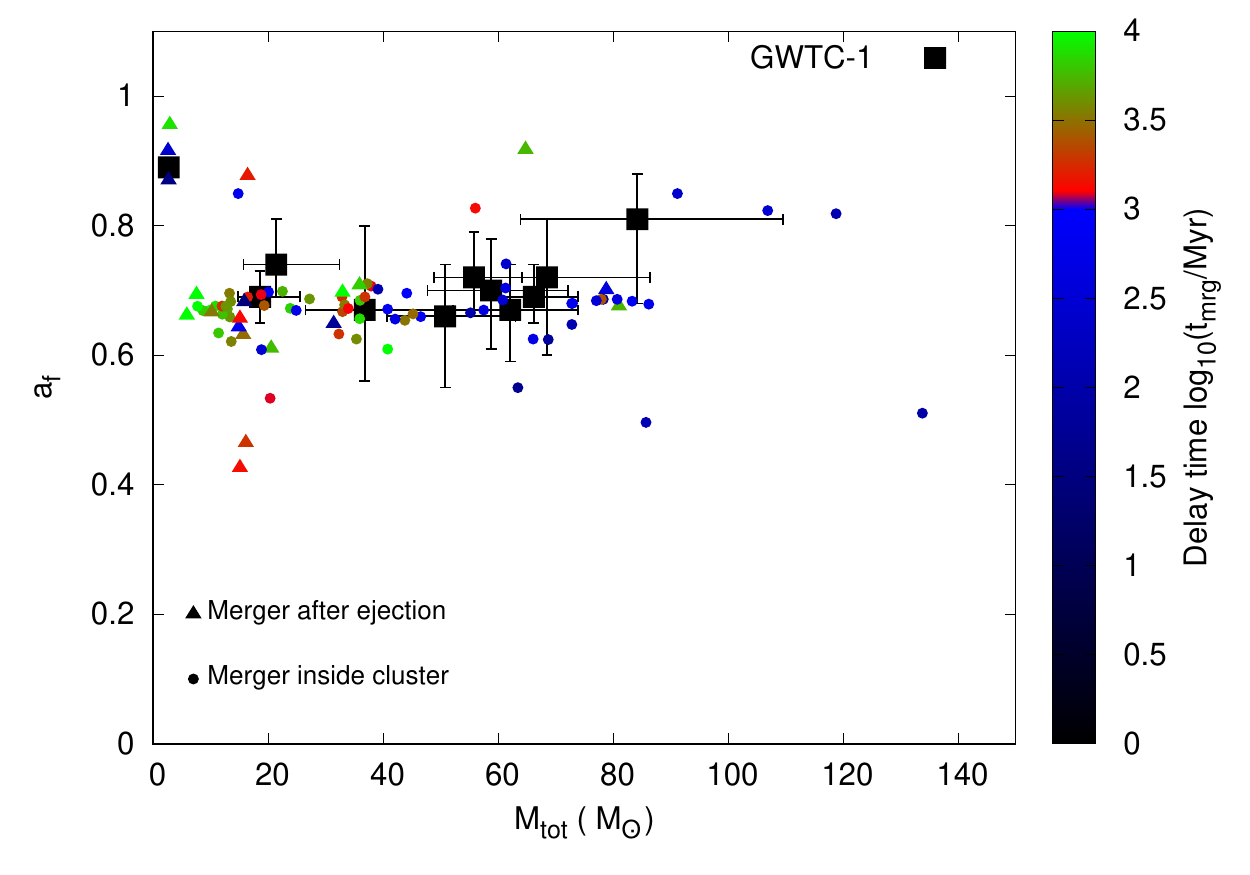}
\caption{Total mass, $\mtot$, versus final spin, $\af$, of GR mergers from those computed models
in Table~\ref{tab_nbrun} that employ the high BH natal spin, \ie, Geneva BH-spin model (left panel)
and that employ the low BH natal spin, \ie, MESA and Fuller BH-spin models (right panel). The legend
for the points is the same as in Fig.~\ref{fig:m1m2}. The $\af$ values for both the in-cluster
and ejected mergers are computed as described in Sec.~\ref{grkick}.}
\label{fig:mtot_af}
\end{figure*}

\begin{figure*}
\includegraphics[width=9.0cm,angle=0]{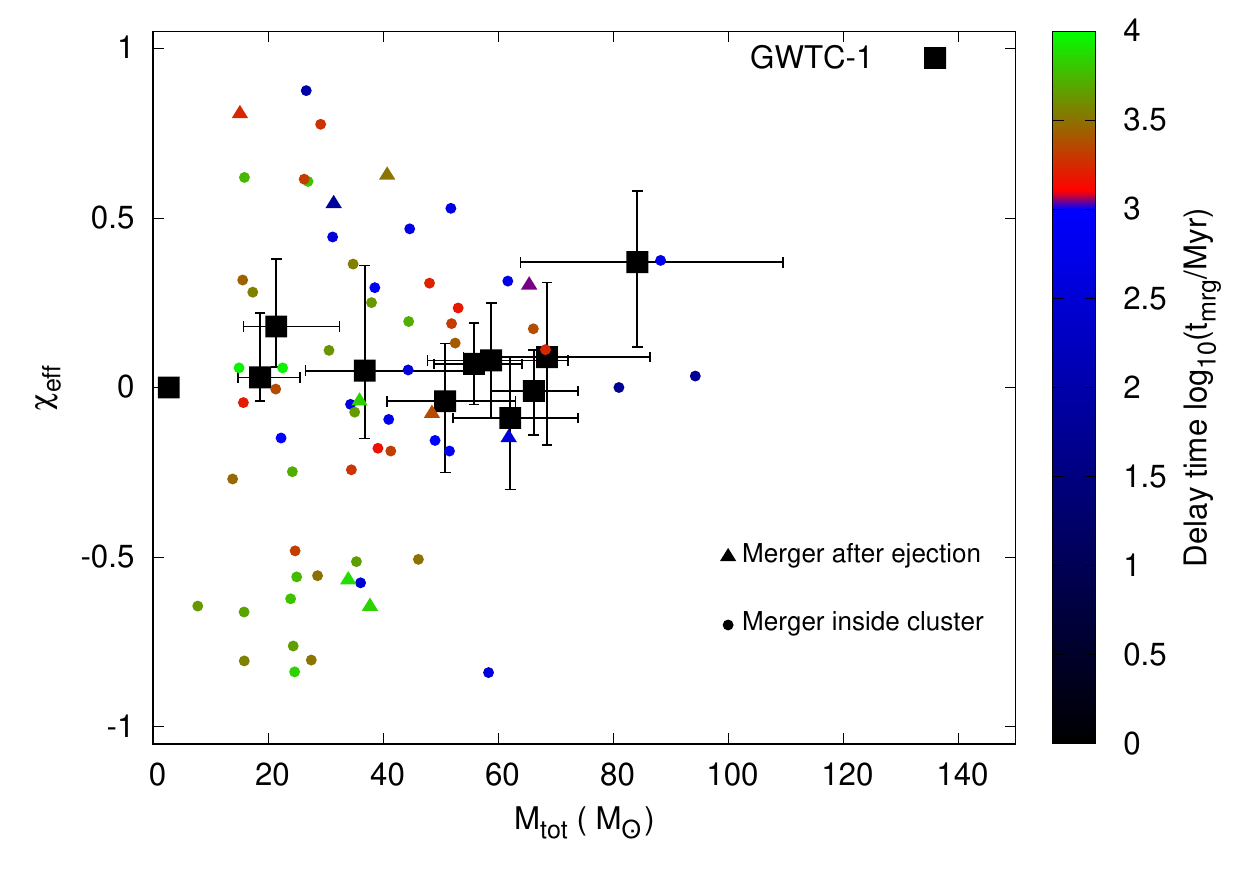}
\hspace{-0.5cm}
\includegraphics[width=9.0cm,angle=0]{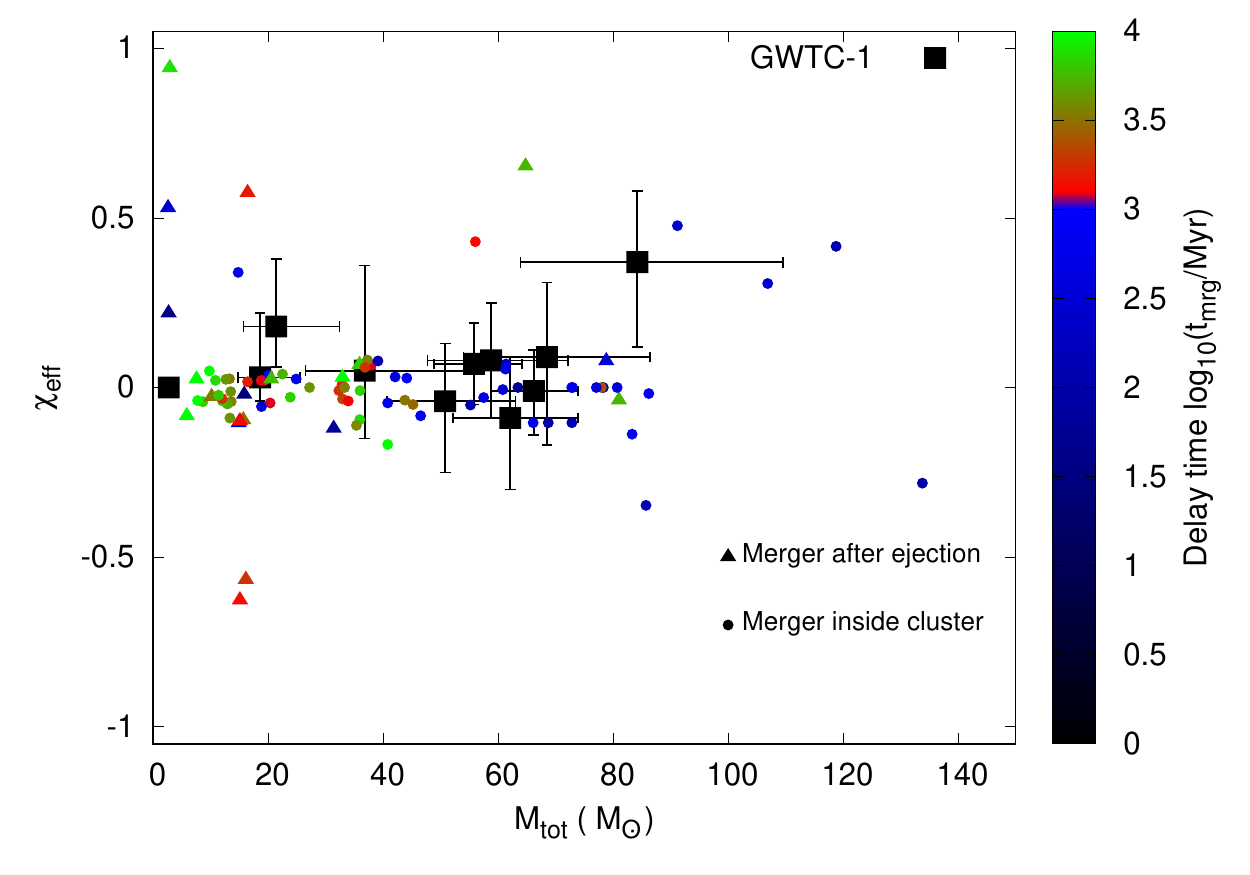}
\caption{Total mass, $\mtot$, versus effective spin parameter, $\xeff$, of GR mergers from those computed models
in Table~\ref{tab_nbrun} that employ the high BH natal spin, \ie, Geneva BH-spin model (left panel)
and that employ the low BH natal spin, \ie, MESA and Fuller BH-spin models (right panel). The legend
for the points is the same as in Fig.~\ref{fig:m1m2}. The $\xeff$ values for both the in-cluster
and ejected mergers are computed as described in Sec.~\ref{grkick}.}
\label{fig:mtot_xeff}
\end{figure*}

\begin{figure*}
\includegraphics[width=8.7cm,angle=0]{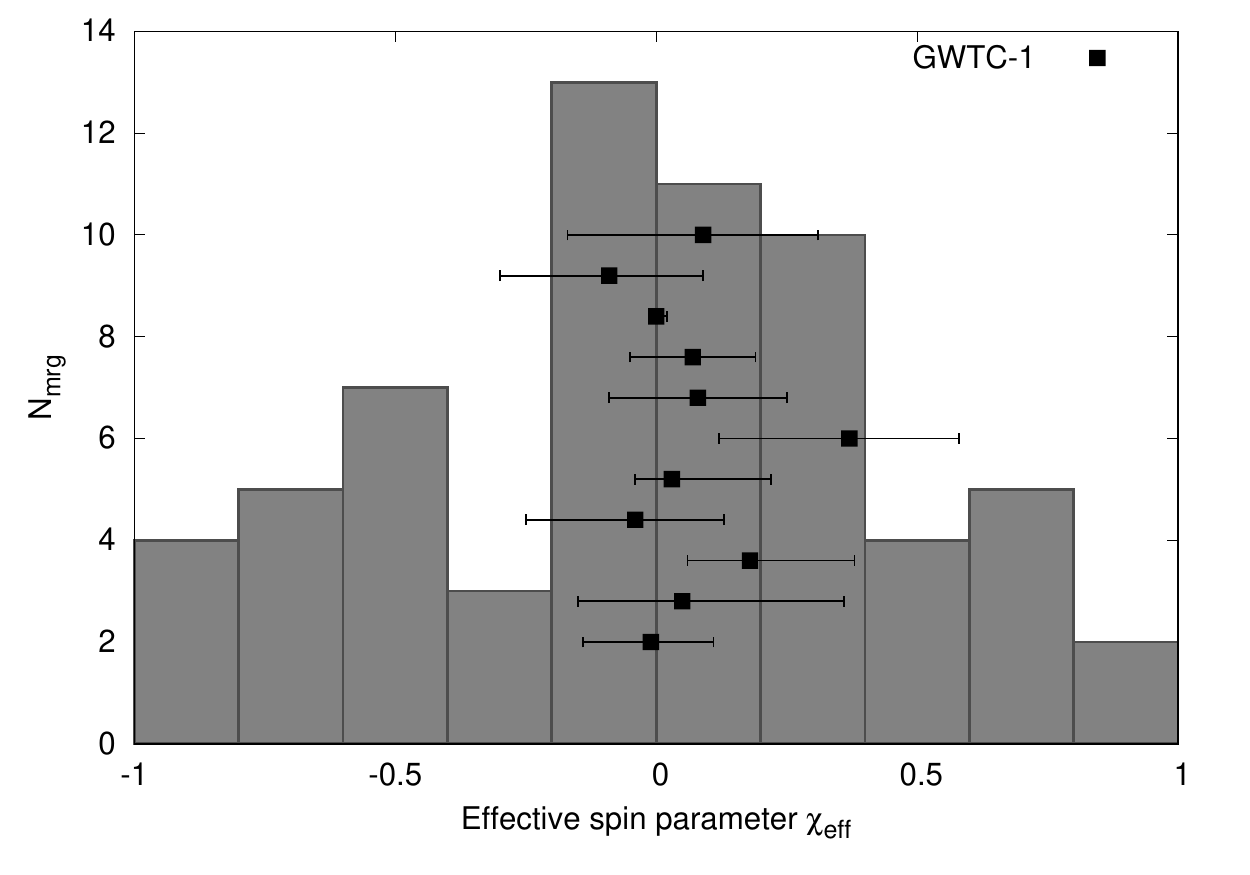}
\includegraphics[width=8.7cm,angle=0]{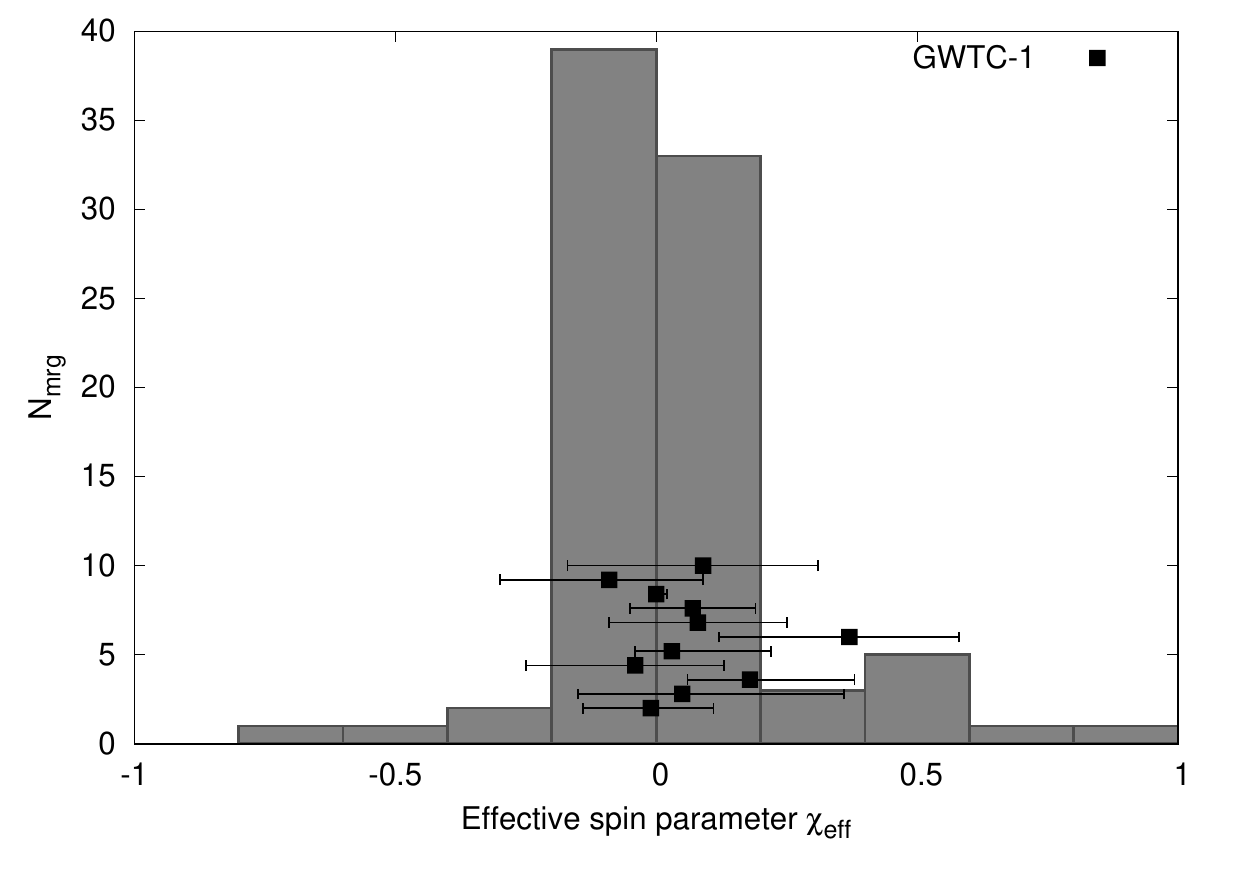}
\caption{Distribution of effective spin parameter, $\xeff$ (shaded histogram),
of GR mergers from those computed models
in Table~\ref{tab_nbrun} that employ the high BH natal spin, \ie, Geneva BH-spin model (left panel)
and that employ the low BH natal spin, \ie, MESA and Fuller BH-spin models (right panel).
The $\xeff$ values for in-cluster and ejected mergers are computed as described in Sec.~\ref{grkick}.
The filled, black squares with error bars (90\% credible intervals) represent the observed $\xeff$ values
in LVC O1/O2 merger events \citep[GWTC-1,][]{Abbott_GWTC1}. For clarity, the LVC O1/O2
data points are displaced along the Y-axis in the increasing order of their date of discovery.}
\label{fig:xeff_dist}
\end{figure*}

\subsection{Merger final spins and effective spin parameters: on GW170729}\label{mrgspin}

Figs.~\ref{fig:mtot_af} and \ref{fig:mtot_xeff} show the final spins, $\af$s, and the
effective spin parameters, $\xeff$ (Sec.~\ref{grkick}), of all the mergers from the computed models in
Table~\ref{tab_nbrun}. In these figures, the outcomes from the models employing high BH natal
spins, \ie, Geneva BH-spin scheme, and low BH natal spins, \ie,
MESA and Fuller BH-spin schemes, are distinguished (left and right panels, respectively).

As one expects, high BH natal spin would result in significant scatter of $\af$ and $\xeff$
values around their equal-mass, non-spinning merger values, $\af\approx0.7$ and $\xeff=0.0$,
due to the random mass pairing and spin orientations of the merging members,
in dynamical mergers (Sec.~\ref{grkick}). This is demonstrated in
Figs.~\ref{fig:mtot_af} and \ref{fig:mtot_xeff} (their left panels). The notable tapering
of the scatter with increasing $\mtot$ is due to the fact that even in Geneva
BH-spin model, depending on the BH-progenitor star's metallicity, the most massive
BHs still receive low natal spins (Sec.~\ref{bhspin}; top panel of Fig.~\ref{fig:bhspin}).
This leads to the convergence to the non-spinning $\af$ and $\xeff$ values with
increasing $\mtot$. Beyond
$\mtot\approx70\Ms$, mass-gained BHs become the abundant participants of BBH mergers
(Sec.~\ref{gapmrg}), which BHs have high or maximal spins (Secs.~\ref{mrgr}, \ref{gapbh}).
Therefore, the scatter of $\af$ and $\xeff$ resumes for $\mtot\gtrsim70\Ms$
(left panels of Figs.~\ref{fig:mtot_af}, \ref{fig:mtot_xeff}). Overall,
the scatter in $\af$ and $\xeff$, for the computed mergers with Geneva BH-spin model,
well exceeds that in the O1/O2 merger events (left panels of Figs.~\ref{fig:mtot_af},
\ref{fig:mtot_xeff}, and \ref{fig:xeff_dist}).

On the other hand, BBH mergers from the models with low BH natal spin (MESA and Fuller BH-spin model;
Sec.~\ref{bhspin}; Fig.~\ref{fig:bhspin}) yield $\xeff$ and $\af$ values that agree well
with the observed values of these quantities from O1/O2 events and as well with their overall
trend and scatter (except for GW170729, see below, the O1/O2 values are similar to the non-spinning values).
This is demonstrated in Figs.~\ref{fig:mtot_af},
\ref{fig:mtot_xeff}, and \ref{fig:xeff_dist} (their right panels). Nonetheless, as discussed
above, highly-spinning, mass-gained BHs cause the model mergers to fluctuate between spin-orbit-aligned and
spin-orbit-anti-aligned values of $\af$ and $\xeff$ for $\mtot\gtrsim70\Ms$, as seen
in Figs.~\ref{fig:mtot_af} and \ref{fig:mtot_xeff} (their right panels). In particular,
the distinctively high (positive) value of $\af$ ($\xeff$) for GW170729 could be reproduced
(taking into account their 90\% confidence intervals).

For $\mtot\lesssim70\Ms$, a few mergers still show large scatter in MESA and Fuller BH-spin
models. These are either lower-mass, mass-gained BHs or those mergers where the
primordial-binary membership is maintained and, therefore, a partial spin-orbit-alignment
is imposed (Sec.~\ref{grkick}). These include the ejected BNS mergers (Sec.~\ref{mrgmass}).

Overall, the mergers from the computed models can explain well the masses, final spins, and
spin-orbit alignments of the BNS and BBH mergers observed by LVC in their O1/O2,
including the distinctively large masses, pro spin-orbit alignment, and high final spin of GW170729,
provided the natal spins of the BHs are taken to be $a\lesssim0.1$ (MESA or Fuller BH natal spin model).
The results also imply that beyond $\mtot\approx70\Ms$, both high and low final spins and (correspondingly)
pro- and anti-aligned spin-orbit configurations can generally be expected in BBH mergers
despite low natal spins of BHs. This is also occasionally possible for $\mtot\lesssim70\Ms$. 
Example 3 (from model 61 of
Table~\ref{tab_nbrun}) of Sec.~\ref{nbspin} shows a specific example of a GW170729-like merger
between a mass-gained (through BH-TZO accretion) and an intact BH.

In the low BH natal spin models of the current set, very massive BBH mergers, up to $\mtot\approx140\Ms$,
are obtained that lead to (final) BHs of intermediate mass and moderate
to near-maximal spins ($\af\gtrsim0.5$; see right panel
of Fig.~\ref{fig:mtot_af}). These mergers happen in low-$Z$ models through either second-generation
BBH merger (when Fuller BH-spin is applied) or significant BH-TZO accretion ($\ftz\gtrsim0.7$).
The current set with Geneva BH natal spins assume smaller $\ftz$ (Table~\ref{tab_nbrun})
and second-generation mergers in them is unlikely (see Sec.~\ref{gapmrg}), which is why
such massive BBH mergers are missing in them (left panel of Fig.~\ref{fig:mtot_af}).

\subsection{General-relativistic coalescences inside clusters driven by compact subsystems}\label{inmrg}

As seen in Table~\ref{tab_nbrun}, the majority of the GR mergers
(most of which are dynamically-driven BBH mergers) 
take place inside clusters. While, overall, the total number of mergers increases with increasing
$\mcl(0)$ and decreasing $\rh(0)$, the growth is mainly in in-cluster mergers, over the cluster
mass and size ranges explored here, namely, $10^4\Ms\leq\mcl(0)\leq10^4\Ms$,
$1.0{\rm~pc} \leq \rh(0) \leq 3.0{\rm~pc}$. As explained and demonstrated in Paper III
(see Sec.~1 and 3.1 therein), most of these in-cluster mergers are driven by the short-timescale,
few-body dynamics of dynamically-formed compact subsystems that are made of one or more of the
retained stellar remnants, typically of BHs. 
As shown therein (and found in the present models as well), the in-cluster triples or higher-order
subsystems are strongly perturbed so that it is, typically, the chaotic part of the few-body
evolution  \citep[\eg,][]{Antonini_2016b,Samsing_2018a}
of the subsystem rather than its secular evolution \citep[\eg,][]{Lithwick_2011,Katz_2011}
that drives the eccentricity boost of its 
innermost binary, leading to the latter's GR inspiral.

\begin{figure*}
\includegraphics[width=12.0cm,angle=0]{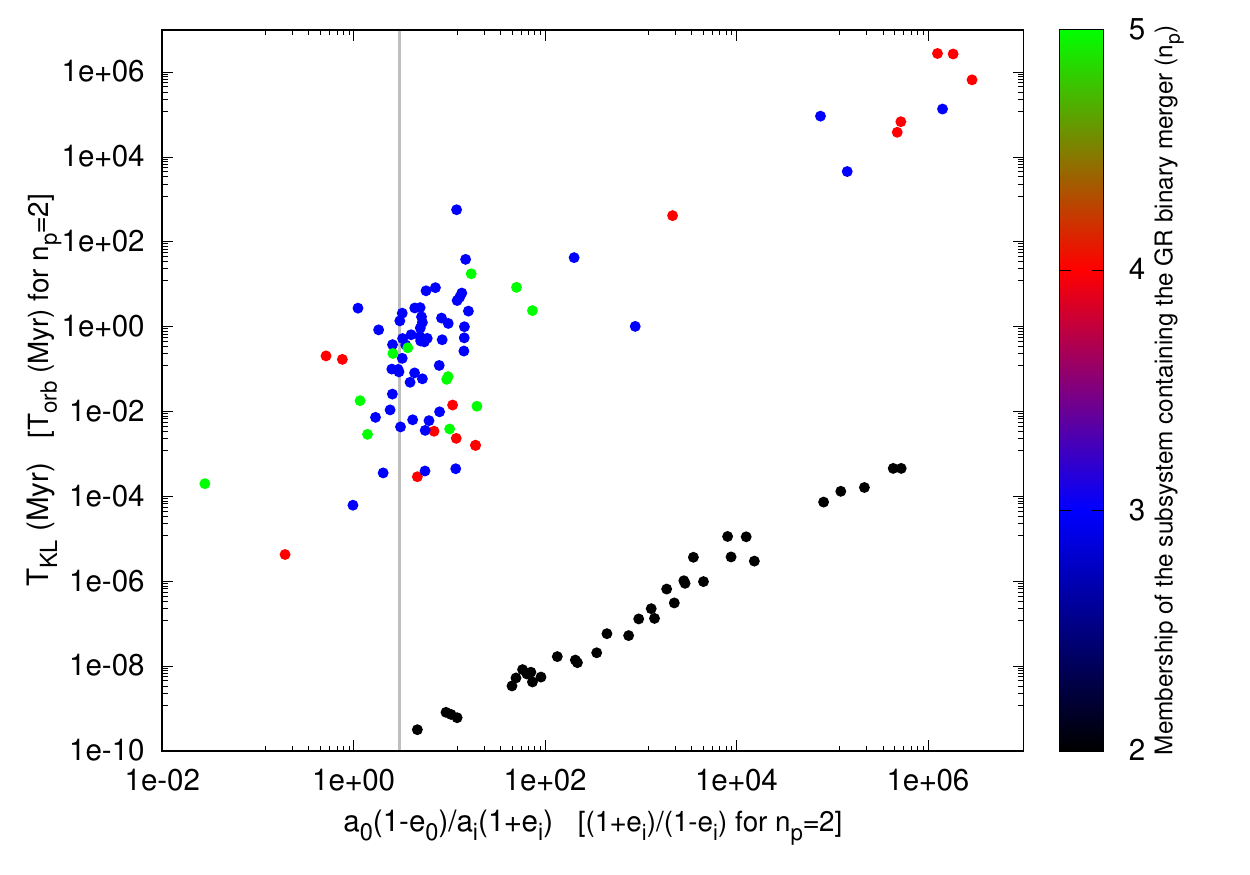}
\caption{The configurations of the subsystems hosting a GR binary merger occurring inside the computed
model clusters in Table~\ref{tab_nbrun} (filled circles). If the merger happens to the innermost binary of a triple
(higher-order) subsystem, then the abscissa is the ratio,
${\mathcal R}$, of the outer periastron to the inner
apoastron of the triple (innermost triple), which measures the extent of hierarchy of this triple.
The ordinate represents the Kozai-Lidov time period (Myr), $\tkl$, of the triple (innermost triple).
The colour coding (colour bar) represents the total membership, $n_p$, of the merger-hosting subsystem.
If the merger happens in an in-cluster binary ($n_p=2$; filled, black circles), then the binary's apoastron
to periastron ratio is plotted along the abscissa and its orbital period (Myr), $\torb$, 
is plotted along the ordinate. These are the configurations of the host subsystems
at the start of the binary's final GR in-spiral.
The grey, vertical line indicates the canonical stability limit, ${\mathcal R}=3$,
for hierarchical triple systems \citep{Mardling_1999}.}
\label{fig:mrgconf}
\end{figure*}

Fig.~\ref{fig:mrgconf} shows the hierarchy of the (innermost) triple configuration, ${\mathcal R}$ (X-axis), against
their respective Kozai-Lidov (hereafter KL) timescale, $\tkl$ (Y-axis), of the in-cluster subsystems hosting a
GR merger, from all the computed models. The triple hierarchy is measured by the ratio of the
triple's outer periastron to inner apoastron, ${\mathcal R}\equiv a_0(1-e_0)/a_i(1+e_i)$,
$a_0$ and $e_0$ being the triple's outer semi-major-axis and eccentricity, respectively, and
$a_i$ and $e_i$ being the inner semi-major-axis and eccentricity, respectively.
The triple's secular KL time period, $\tkl$, is given by \citep{Kiseleva_1998} 
\beq
\tkl=\frac{2P_0^2}{3\pi P_i}(1-e_0^2)^{3/2}\frac{\mone+\mtwo+m_0}{m_0},
\label{eq:kozai}
\eeq
where $P_i$ and $P_0$ are the Keplerian periods of the triple's inner and outer orbits, respectively,
and $\mone$, $\mtwo$, and $m_0$ are the masses of the inner binary's components and the
outer member, respectively.

Fig.~\ref{fig:mrgconf} is essentially similar to Fig.~1 of Paper III but with
a much larger number of data points from the present, much larger set of
computed models. The data points are colour coded with
the total membership, $n_p$ ($n_p=2$ is a binary), of the subsystem containing the GR in-spiralling binary.
Note that $n_p$ simply represents the total number of members that are bound to the
merging binary at the instant the binary is labelled COALESCENCE during the PN
integration of its host subsystem via $\archain$ \citep{Mikkola_1999,Mikkola_2008},
based on the GR merger criteria in $\archain$
\citep{Aarseth_2012}; see Examples 1, 2, and 3 of Sec.~\ref{nbspin}.
These criteria boil down to the arrival of an ($a_i, e_i$) combination
of the innermost PN binary such that it can safely be considered to be spiralling in up to its merger,
unperturbed\footnote{Until this condition is reached, $\archain$ continues to integrate the
subsystem, applying PN treatment to the innermost binary. This seldom results in very long
integration time of the subsystem. All $\archain$ PN sub-integration continues until the
subsystem is resolved via either a GR merger of the innermost binary or the dynamical disintegration
of the subsystem. Note that although spins are assigned to BHs at their formation
and after experiencing matter-interaction (Secs.~\ref{bhspin} \& \ref{mrgr}), BH spins
are not used in the $\archain$ PN integration (both PN members are
treated as non-spinning) and are used only for determining the GW recoil kick and the
final spin (Sec.~\ref{grkick}) upon COALESCENCE declaration. Since the vast majority
of the final inspirals begin when the binary is of sub-Hz peak GW frequency
(Fig.~\ref{fig:inspiral}), spin PN terms would not have a significant effect on most of
the sub-integrations. Although the speed of light can be ``chosen'' in $\archain$,
the physical speed of light has been used in all computations presented here.}.
For each in-cluster merger, $n_p$ is determined by searching for Chain members and Chain
perturbers close to the COALESCENCE, as detailed in Paper III (see Sec.~2.3.1 therein).
At other times, $n_p$ can be different or comprise different members since subsystems
are, typically, strongly perturbed (see above, Paper III).

\begin{figure*}
\includegraphics[width=9.0cm,angle=0]{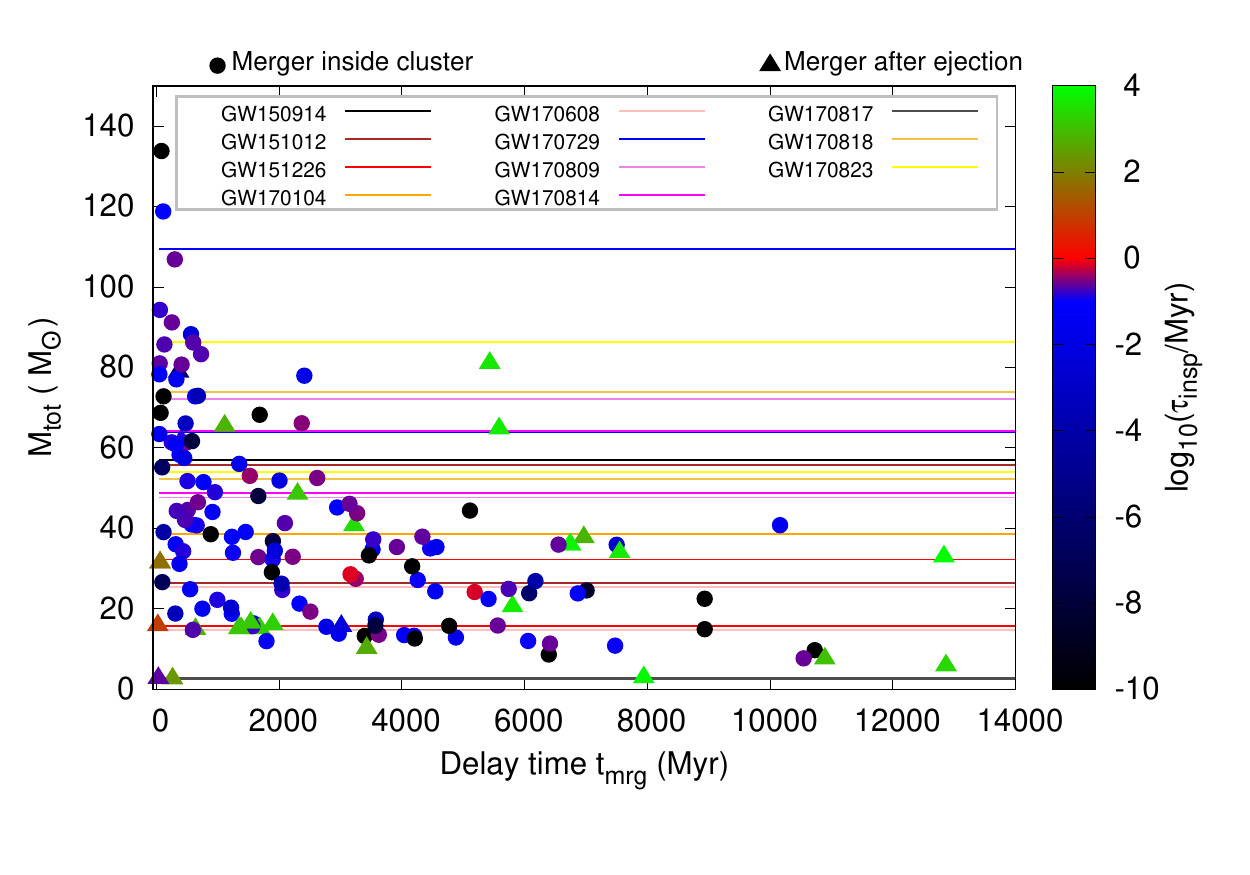}
\hspace{-0.5cm}
\includegraphics[width=9.0cm,angle=0]{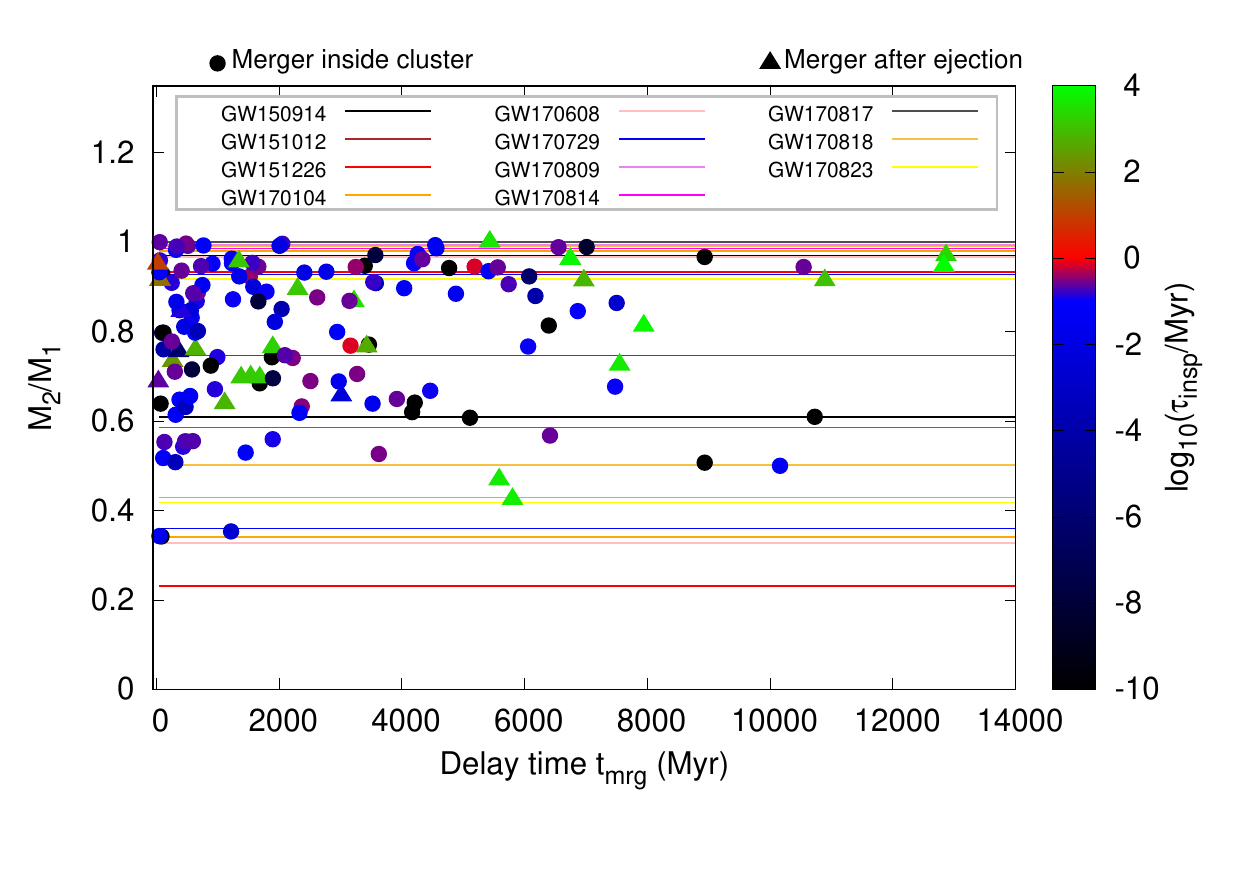}
\caption{Total mass, $\mtot$ (left panel), and secondary-to-primary mass ratio, $\mtwo/\mone$ (right panel),
versus merger delay time, $\tmrg$, of GR compact-binary mergers from all the computed models in
Table~\ref{tab_nbrun} (filled circles and triangles). They are colour-coded
according to their GR in-spiral times, $\tauinsp$ (colour bar).
The circles represent mergers inside a model cluster
while being bound to it and the triangles represent those happening in compact-binary systems after
they get ejected from the cluster. The 90\% credible intervals of the total masses and mass ratios
of the LVC O1/O2 merger events are shown as horizontal lines.
}
\label{fig:delay_all}
\end{figure*}

Note that in Fig.~\ref{fig:mrgconf}, $n_p=2$ implies that the binary has merged without
being a part of a higher-order subsystem: due to a recent strong dynamical interaction, the binary
has shrunk and/or become sufficiently eccentric that it has merged before meeting the next encounter
\footnote{In some works such as \citet{Rodriguez_2018,Kremer_2019}, it is only this type of mergers that
have been labelled ``in-cluster''.}.
In such cases, the apoastron to periastron ratio of the binary at its COALESCENCE stage,  
$(1+e_i)/(1-e_i)$, and the corresponding orbital period,
$\torb=P_i$, are plotted along the X- and Y-axis respectively.

Fig.~\ref{fig:mrgconf} suggests that binary GR mergers inside clusters take place in compact binaries
that are, at the time of the beginning of the final inspiral, either a part of a dynamically-formed, higher-order
subsystem or without any host subsystem, in comparable probabilities (merger inside a subsystem
being of moderately higher probability). When part of a subsystem, a merger happens more often
in triples than in higher-order systems: in the present model set, mergers inside quadruples and those
inside $n_p>4$ systems are comparable in number.
The figure also suggests that the majority of the (innermost) triple systems hosting a merger have
hierarchy ratio ${\mathcal R}\lesssim10$, implying that they are typically of unstable to
hierarchical configuration \citep{Mardling_1999,Mardling_2008}.
For further insight, it would be worth studying the secular evolutionary properties of these systems
(as in, \eg, \citealt{Fragione_2020b})
or integrate them, after extracting them from the clusters, through PN few-body integration
(as in, \eg, \citealt{Antonini_2014,Antonini_2016b,Kimpson_2016}) which will be taken up in a forthcoming study.

\subsection{General-relativistic inspirals}\label{insp}

Fig.~\ref{fig:delay_all} shows the merger delay times, $\tmrg$ (X-axis), against
$\mtot$ (left panel) and $\mtwo/\mone$ (right panel) for all the mergers from the computed models
in Table~\ref{tab_nbrun}. The computed merger data points well encompass 
the 90\% confidence limits of $\mtot$s and $\mtwo/\mone$s of the O1/O2 merger events (horizontal lines),
except for a few $\mtot\gtrsim80\Ms$ mergers with $\tmrg<1.0$ Gyr as consistent with
Fig.~\ref{fig:m1m2} (Sec.~\ref{mrgmass}). The colour coding represents the inspiral time, $\tauinsp$,
corresponding to a merger as estimated from Eqn.~\ref{eq:taumrg} ($a_{\rm ej}$, $e_{\rm ej}$
replaced by $a_i$, $e_i$, see Sec.~\ref{inmrg}, for in-cluster mergers). All in-cluster mergers (filled circles),
consistently, have $\tauinsp << 1$ Myr (see colour coding), which are, therefore,
much shorter than $\sim$ Myr strong-interaction timescale among BHs in such model
clusters with $\lesssim10\kmps$ velocity dispersion \citep{Spitzer_1987,Bacon_1996}.

\begin{figure*}
\includegraphics[width=12.0cm,angle=0]{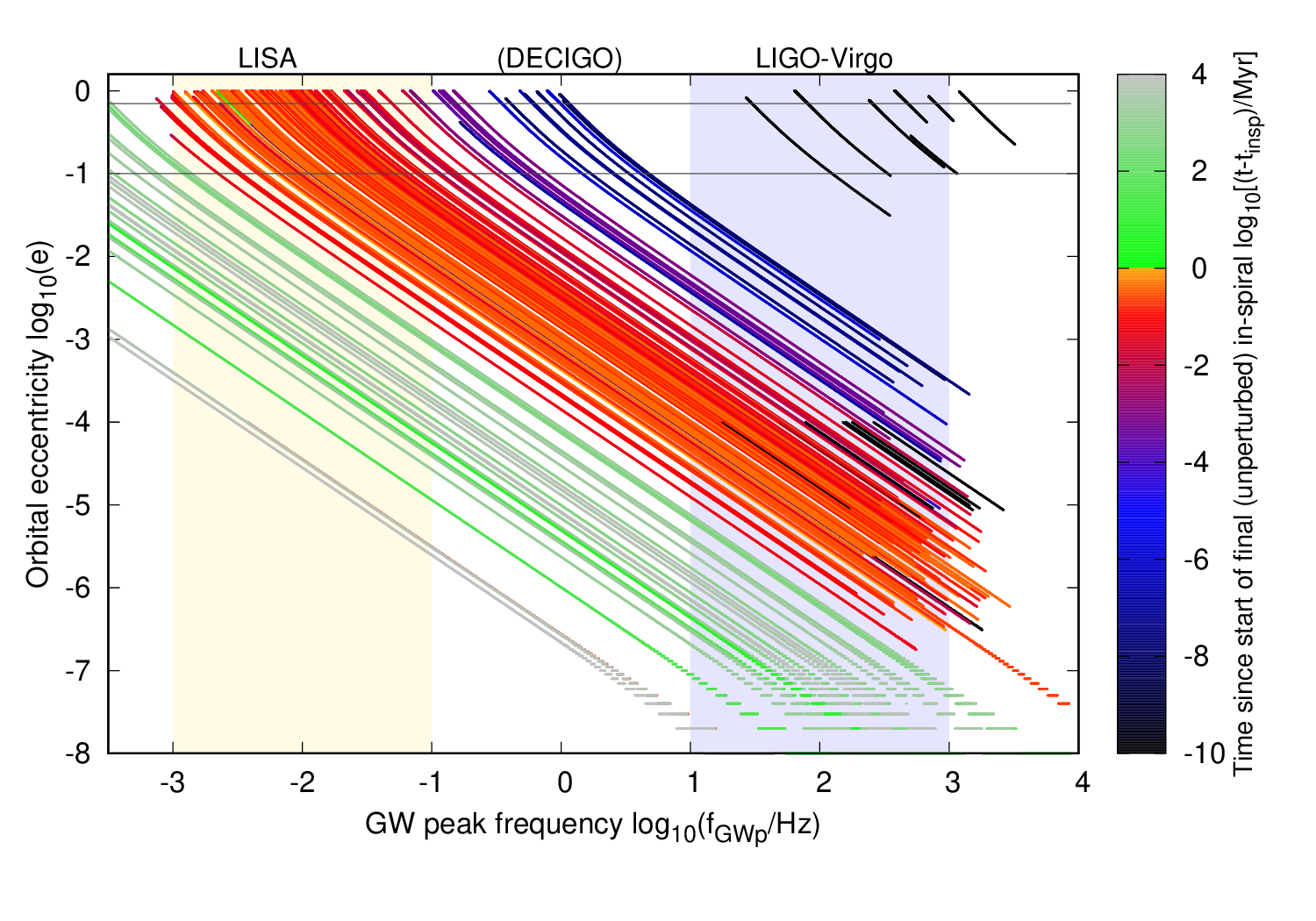}
\caption{Final (unperturbed) orbital inspiral curves of all the GR binary
mergers (both, in-cluster and ejected) from the computed models in Table~\ref{tab_nbrun}
in the $\log_{10}(e)-\log_{10}(\fgwp)$ plane,
$e$ being the binary's eccentricity and $\fgwp$ being the peak GW frequency (see text).
The gold and the blue shades represent the characteristic detection
frequency bands of the LISA and the LIGO-Virgo instruments. The curves are
obtained by integrating the \citet{Peters_1964} orbit-averaged semi-major-axis and eccentricity
decay equations. They are colour-coded according to the time since the
beginning of the final inspiral (colour bar). When $e\lesssim0.7$
(when $\fgwp$ corresponds to the $\approx10$th or a lower harmonic), GW from a binary
is detectable by LISA. LIGO-Virgo will detect an eccentricity in the inspiral
waveform if $e\gtrsim0.1$. These limits in $e$ are indicated by the grey, horizontal
lines.}
\label{fig:inspiral}
\end{figure*}

Fig.~\ref{fig:inspiral} shows the unperturbed orbital evolution
up to coalescence (Sec.~\ref{inmrg}), starting from the
$a_i$, $e_i$ ($a_{\rm ej}$, $e_{\rm ej}$) for in-cluster (ejected) mergers at a
cluster-evolutionary time/delay time $\tinsp$.
The orbits are 
obtained by integrating \citet{Peters_1964} orbit-averaged orbital shrinkage and
circularization equations due to quadrupole GW radiation (PN-2.5 term) and
are plotted in the $\log_{10}(e)-\log_{10}(\fgwp)$ plane, $\fgwp$ being the peak-power
GW frequency. $\fgwp$ is given by \citep{Wen_2003}
\beq
\fgwp=\frac{\sqrt{G(\mone+\mtwo)}}{\pi}
	\frac{(1+e_t)^{1.1954}}{\left[a_t(1-e_t^2)\right]^{1.5}},
\label{eq:gwfreq}
\eeq
$a_t$, $e_t$ being the instantaneous orbital parameters of the
in-spiralling binary. The curves are colour-coded with the time since $\tinsp$.

\begin{figure*}
\includegraphics[width=8.8cm,angle=0]{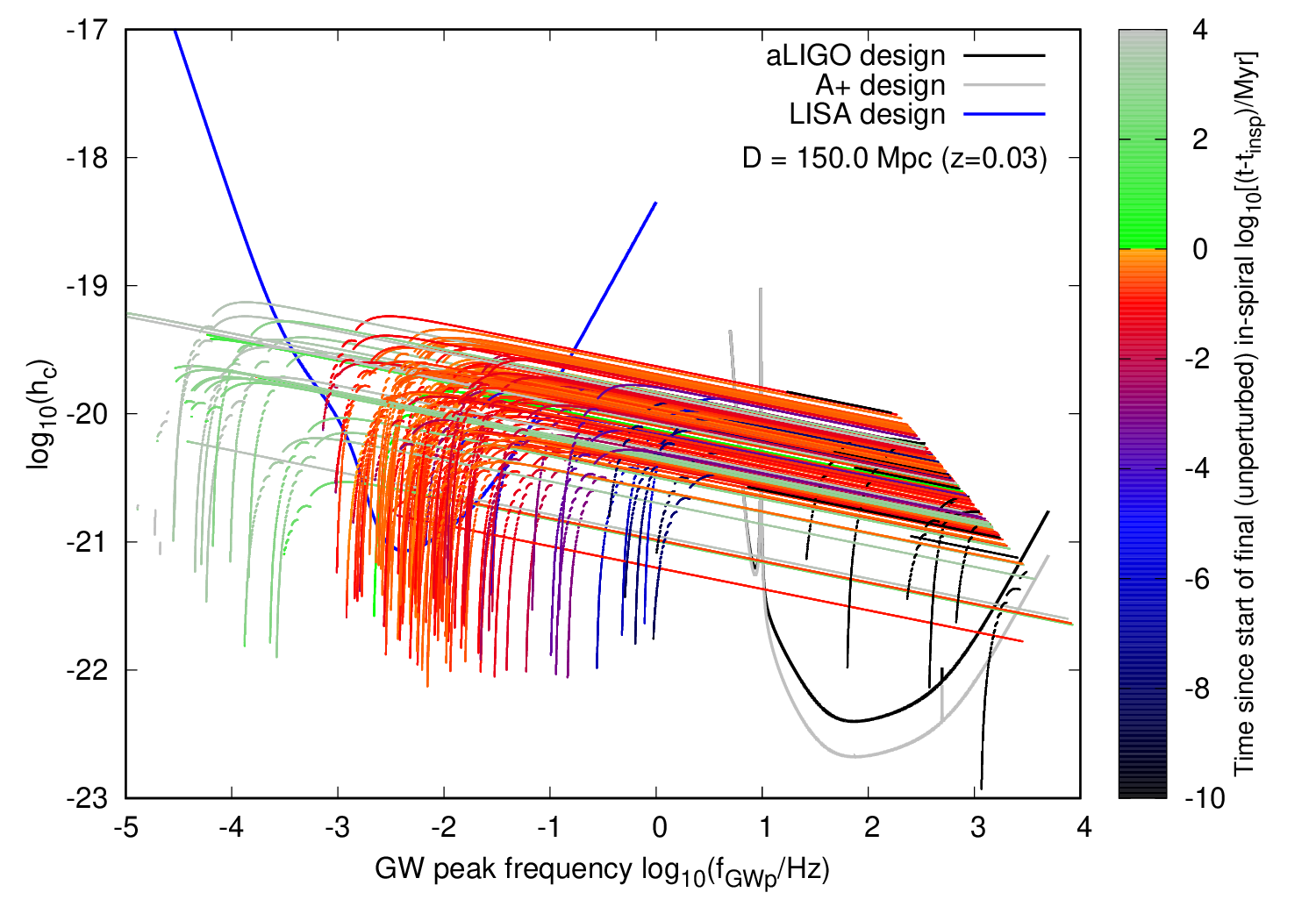}
\includegraphics[width=8.8cm,angle=0]{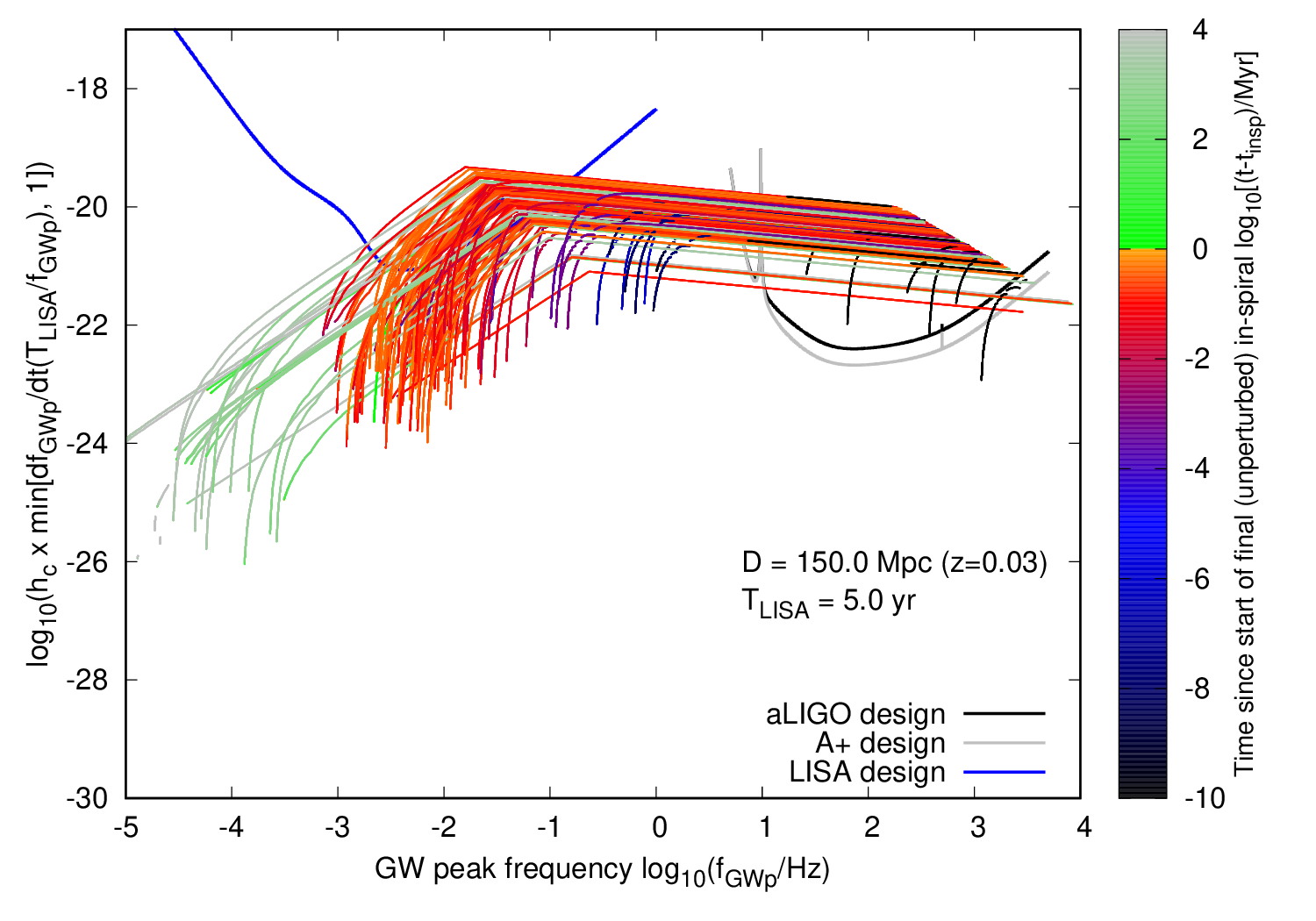}
\caption{{\bf Left panel:} final (unperturbed) orbital inspiral curves of all the GR binary
mergers (both, in-cluster and ejected) from the computed models in Table~\ref{tab_nbrun},
in the $\log_{10}(h_c)-\log_{10}(\fgwp)$ plane. Here, the redshifted, detector-frame peak GW frequency
is plotted along the X-axis and $h_c$, along the Y-axis, is the GW characteristic strain at that
frequency, assuming a comoving distance
of $D=150.0$ Mpc (redshift $z\approx0.03$) of the in-spiraling binary (see text). The same analytic GR orbital evolution
as in Fig.~\ref{fig:inspiral} are utilized to obtain these curves and they bear
the same colour coding. The design sensitivity curves of Advanced LIGO, its proposed
A+ upgrade, and LISA are shown in the same plane (legend).
{\bf Right panel:} same plot as in the left panel, except that $h_c$ (Y-axis) is multiplied with
a correction factor to take into account the GW frequency evolution over a LISA
mission lifetime of $\tlisa=5$ year.}
\label{fig:hcplot}
\end{figure*}

Fig.~\ref{fig:inspiral} (\cf Fig.~7 of Paper II) clearly shows that the
vast majority of the inspirals begin
with $\fgwp$ in the LISA's characteristic frequency band or at an even
lower frequency (especially, the ejected mergers). Then,
after a time of $\tauinspr$\footnote{For highly eccentric binaries,
$\tauinspr$, as obtained by integrating \citet{Peters_1964} orbital-decay equations,
can be up to twice of $\tauinsp$, as obtained from Eqn.~\ref{eq:taumrg}.},
the merger happens after traversing through the decihertz and LIGO-Virgo's
frequency bands. As in Fig.~\ref{fig:delay_all}, $\tauinspr<<1$ Myr for the in-cluster mergers
and $1{\rm~Gyr}\lesssim\tauinspr\lesssim14{\rm~Gyr}$ for the ejected mergers.
Note that $\tauinspr$ reaches down to seconds for the few in-cluster inspirals that begin right at
the LIGO-Virgo band; see Fig.~\ref{fig:inspiral}. These type of mergers
begin with eccentricity $e_i>>0.1$, so that they can be detected as
highly eccentric mergers in the LIGO-Virgo band. All other in-cluster
and ejected mergers enter the LIGO-Virgo band ($\fgwp>10{\rm~Hz}$) with $e<<0.1$ so that
they would appear as circular inspirals by LIGO-Virgo \citep{Huerta_2013}.
In the present model set, such eccentric LIGO-Virgo mergers comprise $\approx5$\%
of the total number of mergers ($\approx6$\% of all in-cluster mergers), which fraction is
comparable to the estimate by \citet{Samsing_2018} and somewhat
lower than that of \citet{Rodriguez_2018}. These studies involve typical GC
models, which are an order of magnitude more massive than the present
YMC and OC-type models, and
would, therefore, contain tighter compact binaries, promoting eccentric
mergers \citep{Samsing_2014}.

Some in-cluster inspirals begin as highly
eccentric in the decihertz frequency range (see Fig.~\ref{fig:inspiral})
and would nearly circularize when they reach the LIGO-Virgo band. These inspirals serve
as the ``missing link'' dynamical mergers \citep{Sesana_2016,Chen_2017} and would be of interest for the future,
space-based decihertz (0.1 Hz - 10 Hz) GW antenna such as DECIGO \citep{Kawamura_2008}.
The rest of the in-cluster mergers begin with $\fgwp$ spread over the LISA band. Although
they also initiate with $e_i>0.9$ and, therefore, would be undetectable by LISA,
they circularize within the LISA band to become of $e\lesssim0.7$, when they are detectable by LISA
\citep{Nishi_2016,Nishi_2017,Chen_2017}. Note that the typical timescale of eccentricity decline
from $e_i$, in the LISA band, is $\tauinspr\sim0.1$ Myr (colour coding in Fig.~\ref{fig:inspiral}).
This means that, depending on the parent cluster's formation epoch, light travel
time from the cluster's distance, and instrument sensitivity, such binaries either are 
already in a state of being observable by LISA as persistent or semi-persistent sources
over its mission lifetime or will remain inaudible by LISA.
It is not possible that such an in-spiralling binary stellar remnant ``becomes''
visible by LISA over its mission lifetime \citep[typically, 5 years;][]{eLISA}.

This is as opposed to all in-cluster and ejected mergers that either initiate in or enter
DECIGO (LIGO-Virgo) band which have $\tauinspr\sim$ year to minute ($\tauinspr\lesssim$ minute).
Irrespective of inspiral timescales, the numbers of persistent, semi-persistent, and
transient sources from YMCs and OCs in the Local Universe at the current cosmic epoch,
in various instrument frequency bands, depend on a
convolution of star formation history, cosmology, and instrument sensitivity.
Such estimates will be taken up in a forthcoming study.

Fig.~\ref{fig:hcplot} (left panel) shows the orbital inspiral curves of Fig.~\ref{fig:inspiral}
in the $\log_{10}(h_c)-\log_{10}(\fgwp)$ plane. In Fig.~\ref{fig:hcplot} (left panel), the X-axis is the
detector-frame peak GW frequency and 
$h_c$ (Y-axis) is the GW characteristic strain on the detector at that frequency, assuming a comoving distance
of $D=150.0$ Mpc (redshift $z\approx0.03$) of the in-spiraling binary (see text). $h_c$
is given by (see \citealt{Kremer_2019} and references therein for a derivation)
\beq
h_c^2 = \frac{2}{3\pi^{4/3}}\frac{G^{5/3}}{c^3}\frac{\mchirp^{5/3}}{D^2}\frac{1}{\fgwp^{1/3}}
	 \left(\frac{2}{n_p}\right)^{2/3}\frac{g(n_p,e)}{F(e)},
\label{eq:hc}
\eeq
where, $\fgwp$ is the source-frame peak GW frequency (as given by Eqn.~\ref{eq:gwfreq} and plotted
along the X-axis of Fig.~\ref{fig:inspiral}), $\mchirp\equiv(\mone\mtwo)^{3/5}/(\mone+\mtwo)^{1/5}$
is the source-frame chirp mass, $n_p$ is the harmonic corresponding to $\fgwp$, $g(n,e)$, with $n=n_p$,
is the relative GW power function as in \citet{Peters_1963}, and $F(e)$ is the eccentricity correction function
as in \citet{Peters_1964} with $(1-e^2)^{-7/2}$ factored in. The source-frame values from the computed
inspiral orbits in Fig.~\ref{fig:inspiral} are directly used in Eqn.~\ref{eq:hc}. In
Fig.~\ref{fig:hcplot} (left panel), the corresponding detector-frame, redshifted frequency, $\fgwp/(1+z)$, is
plotted along the X-axis; since this quantity is very close to $\fgwp$ for the chosen
$D$, a new symbol is not used in Fig.~\ref{fig:hcplot} (left panel). Note that $n_p$ decreases with
the orbital evolution such that $n_p\lesssim10$ for $e\lesssim0.7$ and $n_p=2$ for $e=0$.
The design sensitivity (noise floor) curves of Advanced LIGO (LIGO document number LIGO-T1800044-v5),
its proposed A+ upgrade (LIGO document number LIGO-T1800042-v5), and LISA \citep{eLISA,Robson_2019}
are shown in the same plane.

Fig.~\ref{fig:hcplot} (right panel) repeats the plot of the left panel with
\beq
h_c \times \min\left( \sqrt{\frac{d\fgwp}{dt}\frac{\tlisa}{\fgwp}}, {\rm~~}1 \right)
\label{eq:hcorr}
\eeq
along the Y-axis. The multiplicative factor takes into account the lesser
detectability in the LISA band the slower the evolution of $\fgwp$ is
\citep{Sesana_2005,Willems_2007,Kremer_2019} over a LISA mission lifetime of $\tlisa=5$ yr.
The factor diminishes the characteristic strain over the LISA and decihertz frequency bands; that
over the LIGO-Virgo band remains unaffected.  
Fig.~\ref{fig:inspiral} and \ref{fig:hcplot} suggest that BBH mergers from YMCs and open clusters
within the $\sim100$ Mpc Local Universe
would be detectable by LISA (LIGO-Virgo) with S/N ratio $\sim$ few to 10s ($\sim$ 10s - 100s).

\subsection{Caveats}\label{caveat}

There are several caveats in the current model computations that deserve improvements
in the near future.
An important improvement would be to implement a more consistent scheme for recycling
BH spins due to matter accretion onto it. At present, such BHs are simply assigned
$a=1$ (see Secs.~\ref{mrgr} and \ref{nbspin}). Future models would also incorporate
GW recoil kick and final spins in BBH mergers based on more recent NR studies such as
those of \citet{Lousto_2013,Hofmann_2016}. Although these improvements are
unlikely to alter the main aspects of the present computations and their outcomes, they would, nevertheless,
enrich $\nbseven$ with even more updated and consistent prescriptions and treatments.
The ongoing grid of N-body simulations, the current of which is presented here (Table~\ref{tab_nbrun}),
is somewhat heterogeneous, especially, in $\rh(0)$, BH-spin model, and natal-kick models.
They, nevertheless, encompass all the ingredients and options of the current model.
In the near future, the set of computed models will be expanded, also taking
into account external galactic field in dwarf-galaxy-like environments.
The plethora of merger events from the recently-concluded O3 of LVC will suggest
which ingredients/options to focus on or whether further physical aspects need to be
included.

\section{Summary and outlook}\label{summary}

In this section, the work presented in the previous sections are summarized 
and future prospects are indicated.

\begin{itemize}

\item This work presents a set of 65 long-term, direct, post-Newtonian many-body $\nbseven$ computations
of model dense stellar clusters with up-to-date stellar wind (B10 wind; Sec.~\ref{newwind}), SN
remnant-mass (F12-rapid and delayed SN; Sec.~\ref{newrem}; Fig.~\ref{fig:inifnl}),
and SN natal-kick (momentum-conserving and
collapse-asymmetry-driven, including slow down due to SN material fallback;
Sec.~\ref{newkick}; Fig.~\ref{fig:fbfrac}) prescriptions,
including PPSN and PSN (B16 and weak PPSN/PSN prescriptions; Secs.~\ref{newrem} and \ref{gapbh}; Fig.~\ref{fig:inifnl}).
They also incorporate models of natal spins of stellar-remnant
(\ie, first-generation) BHs (Sec.~\ref{bhspin}; Fig.~\ref{fig:bhspin}),
runtime tracking of NR-based GW recoils and final spins of BBH mergers
(Secs.~\ref{grkick} and \ref{nbspin}),
and preliminary treatments of star-star mergers, star-remnant mergers, and spin-up of BHs due to
matter accretion (Secs.~\ref{mrgr}, \ref{nbspin}, and \ref{nbmrg}).
These ingredients and treatments allow BBH mergers inside clusters involving second-generation and
mass-gained BHs. The model clusters, initially,
have Plummer density profiles with masses $1.0\times10^4\Ms \leq \mcl(0) \leq 1.0\times10^5\Ms$,
half-mass radii $1.0{\rm~pc} \leq \rh(0) \leq 3.0{\rm~pc}$, and ZAMS stars with masses ranging
over $0.08\Ms-150.0\Ms$ that are distributed according to the standard IMF (Sec.~\ref{comp}, Table~\ref{tab_nbrun}).
They range in metallicity over $0.0001\leq Z \leq 0.02$, are initially unsegregated,
are subjected to an external galactic field,
and about half of them have their O-type stars paired among themselves with an observationally-motivated
distribution of primordial binaries (Sec.~\ref{comp}, Table~\ref{tab_nbrun}).
The structure and stellar content of such star cluster evolutionary
models are consistent with those observed in YMCs
and moderately-massive OCs, that bridge low-mass OCs and GCs.

\item These computed models produce GR mergers of BBHs that, primarily, take place while being bound
to the cluster (\ie, are in-cluster mergers) and either being a part of a
triple or higher-order subsystem or by itself following a close encounter
(Sec.~\ref{inmrg}; Fig.~\ref{fig:mrgconf}). The vast majority of the
BBH mergers, in these models, take place following dynamical pairing, \ie,
pairing among BHs whose parent stars were not members
of the same primordial binary or of any binary (Sec.~\ref{mrgmass}).
The models also produce a few BNSs (that maintain the primordial-binary membership)
that merge after getting ejected from their
parent clusters, likewise for a few ejected BBH mergers (Sec.~\ref{mrgmass}). 

\item These mergers, collectively, agree well with the observed masses, mass ratios,
effective spin parameters, and final spins
of the LVC O1/O2 merger events and also with the overall trends and
90\% confidence limits of these quantities in O1/O2, provided
first-generation BHs are born with low or no spin (MESA or Fuller BH-spin model; Sec.~\ref{bhspin})
but spin up after undergoing a (first-generation) BBH merger (Secs.~\ref{grkick}, \ref{nbspin})
or matter accretion onto it (Secs.~\ref{mrgr}, \ref{nbspin}, \ref{nbmrg});
see Secs.~\ref{mrgmass}, \ref{mrgspin}; Figs.~\ref{fig:m1m2}, \ref{fig:mtot_af},
\ref{fig:mtot_xeff}, \ref{fig:xeff_dist}, \ref{fig:delay_all}.
In particular, the distinctly higher mass, effective spin parameter, and final spin of
GW170729 merger event is naturally reproduced. 

\item The computed models also produce massive, $\mtot\sim100\Ms$ BBH mergers with primaries within the 
``PSN gap'' (\ie, with $\mone>45\Ms$; Sec.~\ref{gapmrg}; Fig.~\ref{fig:m1m2}).
Irrespective of the BHs' natal spins, such PSN-gap mergers would show spin signatures, \ie,
typically have pro- or anti-aligned effective spin parameters
leading to high or moderate final spins (Figs.~\ref{fig:mtot_af}, \ref{fig:mtot_xeff}).
If the SN natal kick is driven by collapse asymmetry (Sec.~\ref{newkick}),
the cluster models with the F12-delayed remnant scheme (Sec.~\ref{newrem}) and metallicity $Z\geq0.01$
also yield GR mergers involving remnants (BHs)
with masses within the (lower, NS-BH) ``mass gap'' (Sec.~\ref{gapmrg}; Fig.~\ref{fig:m1m2}).
Candidates of such ``mass-gap'' mergers have been detected during the O3 of LVC.  
Furthermore, mass-asymmetric BBH mergers
with distinctly low mass ratio, similar to that of the LVC O3 event GW190412, is also naturally
obtained in the present models (Sec.~\ref{asymrg}).

\item The computed models also produce BBH inspirals that have eccentricities $>0.1$ in
the LIGO-Virgo frequency band so that their eccentricity is
detectable (Sec.~\ref{insp}; Fig.~\ref{fig:inspiral}). In the present computations, $\approx5$\% of all
GR mergers are such eccentric BBH mergers, which fraction is comparable to that estimated in recent
studies of more massive GC models.

\item These computations produce persistent and semi-persistent
GW sources detectable by LISA from within the $\sim100$ Mpc Local Universe (Sec.~\ref{insp}; Fig.~\ref{fig:hcplot}).

\item The computed evolutionary model set, therefore, shows that with state-of-the-art
stellar-evolutionary and remnant-formation ingredients, YMCs and medium-mass OCs are able to produce
dynamical GR mergers that are well consistent with the observed characteristics of the O1/O2 events. Like
GCs and NSCs, such clusters are also capable of producing second-generation BBH mergers,
PSN-gap mergers, mass-gap mergers, and LIGO-Virgo-detectable eccentric mergers.
Such clusters also assemble highly mass-asymmetric mergers.
The mergers from such clusters are distributed over a wide range of delay times (Figs.~\ref{fig:m1m2},
\ref{fig:delay_all}) and they take place in clusters with a wide range
of metallicity. However, this is compensated by the fact that such clusters form
over a wide range of cosmic epochs and environments, so that it is, in principle, possible
for all such events, from YMCs and OCs, to occur at the current epoch and be observable (Sec.~\ref{insp};
see also \citealt{Kumamoto_2020}).

\end{itemize}

The current model set and their physical ingredients would provide a variety of
opportunities for comparing models with observations and make estimates and
predictions. This study has focused on the GR-merger aspects but, of course,
many other aspects are yet to be studied.
A Monte Carlo approach is being designed to estimate the number of persistent GW
sources and the detection rate of transient GW events from YMCs and OCs, based
on the present (or a somewhat more expanded) model set, taking into account
star formation history, cosmology, and instrument sensitivities
(\citealt{Banerjee_2020d}; Banerjee, S., in preparation).
Another important task is to explicitly compare the size, structure, and kinematics of
the current model clusters, under various remnant-mass and natal-kick scenarios,
with those in individual, well-observed YMCs and OCs, which will be taken up in
the near future \citep[see also][Paper I,II]{2008MNRAS.386...65M}.
Rooms for improvements on the current model ingredients are discussed in Sec.~\ref{caveat}.

\section*{Acknowledgements}

SB is thankful to the anonymous referee for constructive comments and useful suggestions
that have helped to improve the manuscript.
SB acknowledges the support from the Deutsche Forschungsgemeinschaft (DFG; German Research Foundation)
through the individual research grant ``The dynamics of stellar-mass black holes in
dense stellar systems and their role in gravitational-wave generation'' (BA 4281/6-1; PI: S. Banerjee).
SB acknowledges relevant discussions with Sverre Aarseth, Chris Belczynski, Chris Fryer, Mirek Giersz,
Rainer Spurzem, Peter Berczik, Bhusan Kayastha, Sukanta Bose, Parameswaran Ajith, Deirdre Shoemaker, Pablo Laguna,
Achamavedu Gopakumar, Sourav Chatterjee, Harald Pfeiffer, and Philipp Podsiadlowski.
SB acknowledges the generous support and efficient system maintenance of the
computing team at the AIfA and HISKP. SB acknowledges support from
the Silk Road Project while visiting the National Astronomical Observatories of China (NAOC), Beijing,
that has facilitated discussions and independent testings with $\nbpp$ (thanks to Peter Berczik).
A few of the N-body computations presented
here have been partially performed on the KEPLER cluster of the Astronomisches Rechen-Institut (ARI), Heidelberg.   
SB has solely performed, managed, and analysed all the N-body computations presented
in this work.
SB has done all the coding necessary for this work and has prepared the manuscript,
except for a few output manoeuvres which are borrowed from $\nbpp$ (thanks to Rainer Spurzem).

\section*{Data availability}

The simulation data underlying this article will be shared on reasonable
request to the corresponding author. Some of the data are meant for further research
and publications and may not be shared immediately.

\bibliographystyle{mnras}
\bibliography{bibliography/biblio.bib}

\label{lastpage}

\appendix

\section{Implementation of BH spin and BBH merger recoil in updated $\nbseven$}\label{nbspin} 

A preliminary arrangement has been
made to assign spins to BHs, at their birth from a stellar progenitor, based on
stellar-evolutionary models, as described in Sec.~\ref{bhspin}. The Geneva, MESA, and
Fuller BH spin models are implemented in subroutine {\tt Block/kick.f}
where the remnant natal kick (Sec.~\ref{newkick}) is also evaluated and assigned (see Ba20).
An input integer parameter {\tt bhflag} toggles between the various BH-spin models
({\tt bhflag}$=2/3/4$ for Geneva/MESA/Fuller model). The formulae representing
the Geneva and MESA models are functions of carbon-oxygen core mass, $\mco$,
of the progenitor star (see B20) which value (along with fallback fraction,
fallback mass, and ECS indicator; see Ba20) is imported from {\tt Block/hrdiag.f}
(the $\bse/\nbseven$ subroutine that assigns remnant mass; see Ba20)
via a common block.

After computing the magnitude of the dimensionless spin parameter, $a$,
the BH spin magnitude, $S_{\rm BH}$, is evaluated according to Eqn.~\ref{eq:adef}.
$S_{\rm BH}$ is then scaled to ``N-body unit'', $\hat S_{\rm BH}$, by dividing the former by the spin
scaling parameter {\tt SPNFAC}. $\hat S_{\rm BH}$ and $a$ are then copied
to a newly added common block\\
{\tt COMMON /SPIN2/ SPN(NMAX), ASPN(NMAX)}\\
placed in the program-wide header file {\tt common6.h} for
easy accessibility from other subroutines. For maintaining the association
of the spin values with the BH, the values are stored in the above
arrays against the BH's {\tt NAME}, the latter being an integer that
uniquely identifies a member. This is why the existing {\tt SPIN(NMAX)}
array is not used which is continued to be utilized independently and as before by
$\bse$-based routines that treat Roche lobe overflow, CE, and tidal circularization
(except, in {\tt Block/kick.f}, $\hat S_{\rm BH}$ is also copied to
{\tt SPIN(NMAX)} against the newly-formed remnant's array index, once for all).
At present, a fully consistent treatment of the recycling the BH's natal spin,
in the event of mass transfer onto the BH or a merger of the BH with
a stellar member, is unavailable. Note that, at the start of a run,
{\tt SPIN2} common block is initialized, once for all for all stars, in
subroutine {\tt Block/instar.f}. The newly-formed BHs' spins are
then overwritten onto it from {\tt Block/kick.f}, as described above.

In the event of a BH-star merger,
$a=1$ is simply set for the merger-product BH, as discussed in Sec.~\ref{mrgr}, which is done
in the procedures in $\nbseven$ for assembling a new, merged body. Specifically,
the merged object (a BH) assumes the {\tt NAME} of the merging stellar member,
whose spin would typically correspond to $a>1$ (due to
the way stellar spins are assigned in {\tt instar.f} following
$\bse$ prescriptions; \citealt{Hurley_2002}), making the BH maximally spinning
when the BH's spin parameter is later searched for against its {\tt NAME} in {\tt ASPN}
array of {\tt SPIN2} common block. (In the NR subroutines described below, $a=1$
is set if $a>1$.) For the relatively rare case of BH formation during an
interacting-binary phase (a symbiotic, mass transfer, or CE phase), $a=1$ is set for the BH 
by a ``serendipitous bug'' that prevents the newly-formed BH, while being
treated in subroutine {\tt Block/roche.f}, to be accessed by {\tt Block/kick.f} and
hence prevents the reassignment of the BH's spin in the latter routine (see above). Such
a BH will also be found maximally spinning due to its progenitor star's $a>1$. 
(The BH, thereby, also receives zero natal kick but such a BH is typically in
the direct collapse regime, \ie, it would get
zero kick, anyway, even if it would have been formally treated in {\tt Block/kick.f};
see Sec.~\ref{newkick}.) It is checked in the computations presented in
this work (Table~\ref{tab_nbrun}) that the handful of BHs with a history
of matter-interaction with stars inevitably show up with $a=1$,
if they are encountered in the NR routines (see below).

During execution of an in-cluster GR merger in {\tt ARchain/chain.f}, the {\tt NAME}s, masses,
and stellar types of the merging members are collected and stored in
a private (as opposed to code-wide) common block\\
{\tt COMMON/EXTRA3/ NMOBJ1,NMOBJ2,BOBJ1,BOBJ2,KOBJ1,KOBJ2}.\\
This common block is shared with the ``chain termination'' routines
{\tt ARint/chterm.f} and {\tt ARint/chterm2.f}. It is in the latter two
routines where the GR merger recoil kick is evaluated for BBH mergers, along
with the final dimensionless spin parameter, and assigned
to the merged BH (the computed recoil kick is vector-added to the instantaneous
velocity of the BBH's center of mass and the final spin magnitudes are stored
in the {\tt SPIN2} common block against the {\tt NAME} of the merged BH).

This is done through calling a newly added routine {\tt GWREC3} which, in turn,
calls the newly added routine {\tt GWKICK}. Routine {\tt GWKICK} serves as the
``NR engine'' where the NR-based GR recoil \citep[as in][]{vanMeter_2010}
and final spin \citep[as in][]{Rezzolla_2008} formulae are implemented.
Before calling {\tt GWKICK}, {\tt GWREC3} makes use of {\tt EXTRA3} common block
to get the names and masses of the merging bodies. The dimensionless spin magnitudes
of the merging members are then obtained, against the names,
from {\tt SPIN2} common block (see above). The spins are then assigned random orientations
as described in Sec.~\ref{grkick}.
The masses, dimensionless spins, and spin orientations
serve as arguments to {\tt GWKICK} routine.
{\tt GWREC3} also prints detailed information of the merging members and the
merged object.

{\tt GWREC3} is also called from {\tt Block/brake4.f} which routine
treats GR inspiral and merger when these processes are not treated
via $\archain$. In $\nbseven$, {\tt brake4.f} is never used for treating
NS- or BH-containing binaries; it
is used only whilst running in the ``$\nbsix$ mode'', that uses the
classical or KS $\chain$ \citep{Mikkola_1993} instead of $\archain$.

Note that at the moment, only BH spins are assigned and tracked in the ways described above.
The treatment naturally enables second-generation BHs to potentially be present and undergo
second-generation GR mergers inside a cluster. In the following, two such
examples of in-cluster, second-generation BBH mergers are presented from
a computed model with $\mcl(0)=5.0\times10^4\Ms$, $\rh(0)=1.0{\rm~pc}$, $\fbin(0)=0$, $Z=0.001$,
delayed+B16-PPSN/PSN remnant model, and Fuller BH spin model (model 49 of Table~\ref{tab_nbrun}).
Here, relevant text outputs from $\nbseven/\archain$ are shown during BBH inspirals. 
The carry over of the final mass, NR-based final spin, and the identity
of the first-generation BBH merger product (\ie, of the newly-formed
second-generation BH) to the second-generation BBH merger is highlighted with $\ast$ symbols.
The delay times (N-body unit; \citealt{Heggie_1986}) of the merger events are highlighted with
square brackets.

In Example 1, the second-generation BBH merger happens well within the PSN mass gap,
the second-generation BH involved being of $78.3\Ms$. The final outcome of this BBH merger sequence
is a BH of $118.8\Ms$ and dimensionless spin $0.82$ which crosses the cluster's tidal radius
in a few dynamical times (N-body times) and permanently escapes the system, due to
the large recoil ($>300\kmps$) it receives during the second merger. This is
an example of a massive, second-generation BBH merger early in the cluster's evolution
(first and second-generation mergers at 522.2 and 1152.4 N-body times or 51.2 and 112.9 Myr).
In this work, the $\sim$ few percent mass loss due to quadrupole GW energy radiation,
which has a complex dependence on mass ratio and spin orientations \citep[\eg,][]{Sperhake_2015},
is ignored and the mass of the merged BH is simply the sum of the merging BH masses. 

In Example 2, such BBH merger sequence happens at late evolutionary times:
the first-generation merger at $t=43476.8$ (4260.7 Myr)
and the second-generation merger at $t=103684.3$ (10161.1 Myr).
The second-generation merger happened within a compact triple system (as verified by
the runtime data of in-cluster compact subsystems; see \citealt{Banerjee_2018}),
with an ECS-derived NS of mass $1.26\Ms$ as the outer member. After the coalescence,  
an NSBH binary is formed (the {\tt NEW KSREG} statement) which escapes the cluster
after a few dynamical times, due to the GW recoil kick (the {\tt BINARY ESCAPE} statement). 

Example 3 (from model 61 of Table~\ref{tab_nbrun}) is an example for a merger between 
a mass-gained BH ($a=1$) and a zero-spin BH (Fuller BH natal spin model; Sec.~\ref{bhspin}),
leading to a GW170729-like final BH.

\onecolumn
\begin{minipage}[h]{\textwidth}
{\bf Example 1:}\\
\begin{verbatim}
------------------------------------------------------------------------------------------------------
 INSPIRAL    T NP IPN E A TZ TKOZ DW  [522.247]   0   3  0.99971  3.12E-05  5.98E-01  1.00E+04  2.95E-03
 REVERSE INFALL    IBH JBH I* MI MJ TZ    2   1  14  14  7.56E-04  8.10E-04  5.98E-01
 COALESCENCE    T IC IPN NAM E EB PM TZ     522.2467591   0   3    *20286*  28908  0.99970502 -9.80E-03  9.21E-09  5.98E-01
 SWALLOWED STAR/BH    NAMC K* M1 M2  28908  14  37.78  *78.28*
 INFALL CHECK:   T CG     522.25 -2.1E-22 -6.4E-23 -1.9E-22 -8.8E-16 -1.7E-16 -5.0E-16
 CHAIN CHECK    ENERGY ENER0 DE ECH EnGR  -9.8047E-03  0.0000E+00 -9.8047E-03 -9.8033E-03 -1.4395E-06
 END CHAIN    T # N NBH ECC SEMI RX ECH G   522.2468       90   1   1  0.99971  3.12E-05  5.81E-06  0.00E+00  0.00E+00
 BH1,2: m s theta phi a k
    3.78E+01    0.00E+00     271.977     240.010    0.000000    14
    4.05E+01    0.00E+00     132.642      24.660    0.000000    14

 Merged BH: vprp1 vprp2 vpar xeff afin sfin thfin
       18.435255        0.000000        0.000000    0.000000    *0.686259*    1.21E+05       0.000
 COALESCENCE KICK    VF ECDOT VCM VESC      1.24    0.000303   11.0     *13.6*
 TERMINATE ARC    NNB SI SR BCM   133  2.98E-08  2.98E-08  1.57E-03
..............
..............
..............
 INSPIRAL    T NP IPN E A TZ TKOZ DW [1152.357]   0   2  0.80386  1.61E-07  1.00E+00  1.00E+04  1.45E-03
 REVERSE INFALL    IBH JBH I* MI MJ TZ    2   1  14  14  8.10E-04  1.57E-03  1.00E+00
 COALESCENCE    T IC IPN NAM E EB PM TZ    1152.3570814   0   3    *20286*  32673  0.80381349 -3.95E+00  3.15E-08  1.00E+00
 SWALLOWED STAR/BH    NAMC K* M1 M2  32673  14  40.50 *118.78*
 INFALL CHECK:   T CG    1121.16  2.8E-24 -7.7E-25 -4.7E-24 -6.0E-15 -3.9E-17 -7.4E-16
 CHAIN CHECK    ENERGY ENER0 DE ECH EnGR  -3.9466E+00  0.0000E+00 -3.9466E+00 -2.2870E-02 -3.9237E+00
 END CHAIN    T # N NBH ECC SEMI RX ECH G  1152.3571   336794   1   1  0.80381  1.61E-07  1.98E-07  0.00E+00  0.00E+00
 BH1,2: m s theta phi a k
    4.05E+01    0.00E+00     176.869      40.068    0.000000    14
   *7.83E+01*   1.21E+05     337.076      47.201   *0.686259*   14

 Merged BH: vprp1 vprp2 vpar xeff afin sfin thfin
       14.801988      -90.962949      361.319037    0.416548    0.818478    3.33E+05       8.155
 COALESCENCE KICK    VF ECDOT VCM VESC     60.98    0.972884     6.0   *368.6*
 TERMINATE ARC    NNB SI SR BCM   117  6.10E-05  1.22E-04  2.38E-03
..............
..............
..............
 ESCAPE    N = 84392  1265   0  0.7437  -12.665655   1.04  81.38  0.062   0.23  0.00001 84230 *20286*
------------------------------------------------------------------------------------------------------
\end{verbatim}
\end{minipage}
\twocolumn
\newpage
\onecolumn
\begin{minipage}[h]{\textwidth}
{\bf Example 2:}\\
\begin{verbatim}
------------------------------------------------------------------------------------------------------
 INSPIRAL    T NP IPN E A TZ TKOZ DW  [43476.799] 0   2  0.99790  2.89E-06  1.00E+00  1.00E+04  1.56E-03
 COALESCENCE    T IC IPN NAM E EB PM TZ   43476.7996768   0   3   *16743*   34163  0.99789827 -1.28E-02  6.06E-09  1.00E+00
 SWALLOWED STAR/BH    NAMC K* M1 M2  34163  14  13.78  *27.20*
 INFALL CHECK:   T CG   43448.87 -2.6E-23 -2.3E-23 -3.9E-23 -3.5E-16 -7.7E-18 -1.5E-17
 CHAIN CHECK    ENERGY ENER0 DE ECH EnGR  -1.2835E-02  0.0000E+00 -1.2835E-02 -1.7925E-04 -1.2655E-02
 END CHAIN    T # N NBH ECC SEMI RX ECH G 43476.7997    66559   1   1  0.99790  2.88E-06  4.04E-06  0.00E+00  0.00E+00
 BH1,2: m s theta phi a k
    1.34E+01    0.00E+00     246.902      37.005    0.000000    14
    1.38E+01    0.00E+00     223.879       7.597    0.000000    14

 Merged BH: vprp1 vprp2 vpar xeff afin sfin thfin
        6.809817       -0.000000        0.000000   -0.000000    *0.686827*    1.47E+04       0.000
 COALESCENCE KICK    VF ECDOT VCM VESC      0.19   -0.000061     6.2     *1.2*
 TERMINATE ARC    NNB SI SR BCM   255  2.44E-04  2.44E-04  5.44E-04
..............
..............
..............
 ARC SWITCH    T TB-T S1 R DW G *************  4.37E-11  5.82E-11  2.83E-08  1.10E-03  3.85E-07
 EINSTEIN SHIFT    # IPN IX E A DW       1   2  14  0.29411  2.82E-08  1.11E-03
 RELATIVISTIC    ECC AX PM RZ TZ TPOM DW   0.2941  2.82E-08  1.99E-08  8.02E-12  1.61E+00  1.09E-05  1.11E-03
 REVERSE INFALL    IBH JBH I* MI MJ TZ    2   1  14  14  2.72E-04  5.44E-04  1.61E+00
 COALESCENCE    T IC IPN NAM E EB PM TZ  [103684.3414642] 0   2   *16743*   63694  0.29410903 -2.63E+00  1.99E-08  1.61E+00
 SWALLOWED STAR/BH    NAMC K* M1 M2  63694  14  13.60 *40.80*
 INFALL CHECK:   T CG  103684.34  3.1E-26  2.1E-25 -1.1E-24  3.5E-15 -5.2E-16 -1.9E-16
 CHAIN CHECK    ENERGY ENER0 DE ECH EnGR  -2.6260E+00  0.0000E+00 -2.6260E+00 -2.6260E+00  0.0000E+00
 END CHAIN    T # N NBH ECC SEMI RX ECH G **********        1   1   1  0.29411  2.82E-08  3.63E-08  0.00E+00  0.00E+00
 BH1,2: m s theta phi a k
    1.36E+01    0.00E+00      35.571     330.895    0.000000    14
   *2.72E+01*   1.47E+04     111.447     287.338   *0.686827*   14

 Merged BH: vprp1 vprp2 vpar xeff afin sfin thfin
      200.227384       35.752667     -374.437814   -0.167434    0.609463    2.93E+04      27.792
 COALESCENCE KICK    VF ECDOT VCM VESC     27.77    0.415896    14.8   *411.4*
 TERMINATE ARC    NNB SI SR BCM   160  1.86E-09  1.86E-09  8.16E-04

 NEW KSREG   TIME[NB] 1.0368434146E+05 NM1,2,S=    *16743*   (63069)   102405 KW1,2,S=  14  13   0
 IPAIR        1 DTAU  3.83E-03 M1,2[NB]  8.16E-04  2.52E-05 R12[NB]  3.89E-06
 e,a,eb[NB]=   4.8387081E-01  2.6195E-06 -3.93E-03 P[d]=  3.30E+01 H -1.61E+02
 GAMMA  0.00E+00 STEP(ICM)  2.33E-10 NPERT    0 NB(ICM)  159 M1,2[*] *4.08E+01* (1.26E+00)
 RAD1,2,S[*]  5.69E-05  1.40E-05  2.23E+02 RI,VI[NB]=  1.91E-01  3.16E+01
..............
..............
..............
 BINARY ESCAPE    KS =     1  NM =    *16743*  (63069) K* = 14 13  0 -1  M = *40.80* (1.26) EB =   -0.0039  R*/PM =  0.000  V/<V> =  28.89  E =  0.48387081  EI =   0.42009  P = 3.3E+01
------------------------------------------------------------------------------------------------------
\end{verbatim}
\end{minipage}
\twocolumn
\newpage
\onecolumn
\begin{minipage}[h]{\textwidth}
{\bf Example 3:}\\
\begin{verbatim}
------------------------------------------------------------------------------------------------------
 WATCH    # IPN E EN EGR A NAM K* M GP ES R     1190   2   0.9998418  -0.0043159  -0.0001085  4.2151E-05      6313      6355  14  14    6.84E-04    5.32E-04    0.00E+00   -9.67E-05    7.96E-05
 INSPIRAL    T NP IPN E A TZ TKOZ DW   1114.497   0   2  0.99984  4.22E-05  1.00E+00  1.00E+04  2.37E-03
 REVERSE INFALL    IBH JBH I* MI MJ TZ    2   1  14  14  5.32E-04  6.84E-04  1.00E+00
 COALESCENCE    T IC IPN NAM E EB PM TZ    1114.4967764   0   3     6313     6355  0.99984181 -4.32E-03  6.67E-09  1.00E+00
 SWALLOWED STAR/BH    NAMC K* M1 M2   6355  14  39.91  91.19
 INFALL CHECK:   T CG    1114.47  1.2E-22  2.2E-21  1.2E-21  7.3E-18  9.2E-18  1.8E-17
 CHAIN CHECK    ENERGY ENER0 DE ECH EnGR  -4.3161E-03  0.0000E+00 -4.3161E-03 -4.2074E-03 -1.0874E-04
 END CHAIN    T # N NBH ECC SEMI RX ECH G  1114.4968     1193   1   1  0.99984  4.21E-05  8.22E-05  0.00E+00  0.00E+00
 BH1,2: m s theta phi a k
   *3.99E+01*   0.00E+00     209.119      73.663   *0.000000*   14
   *5.13E+01*   8.72E+04      31.900     120.558   *1.000000*   14

 Merged BH: vprp1 vprp2 vpar xeff afin sfin thfin
     -113.815020     -125.059397      548.802952    0.477386   *0.849598*   2.04E+05      11.342
 COALESCENCE KICK    VF ECDOT VCM VESC    283.34    1.617051     2.0   575.1
 TERMINATE ARC    NNB SI SR BCM   194  7.63E-06  6.10E-05  1.22E-03
------------------------------------------------------------------------------------------------------
\end{verbatim}
\end{minipage}
\twocolumn

\section{Merger mass loss, BH-TZO accretion, and $\bse$ user inputs in updated $\nbseven$}\label{nbmrg}

As discussed in Sec.~\ref{mrgr}, in star-star mergers within a cluster, a constant fraction, $\fmrg$, of
the instantaneous secondary mass is eliminated. Also, in a BH-star merger (initially forming a BH-TZO
object), a constant fraction, $\ftz$, of the stellar mass is assumed to be accreted onto the BH. 

The star-star merger mass loss is implemented in the $\nbseven$ routines {\tt Block/mix.f}
and {\tt Block/coal.f}. BH-TZO accretion is also implemented in the same routines. The
values of $\fmrg$ and $\ftz$ are supplied to these routines via common blocks and
are read from input at the start of the run (see below).

Apart from $\nbseven$'s standard parameter and option inputs, all the $\bse$ input parameters
as described in Ba20 (see also \citealt{Hurley_2002}), along
with $\fmrg$ and $\ftz$, are read from a special input file called {\tt input\_bse}.
All the common blocks associated with these input parameters are placed in the
code-wide header file {\tt common6.h} for supplying their input-read values to
the relevant $\bse$ routines. Care is taken that the values of all these
parameters, as well as the arrays in {\tt SPIN2} common block (Sec.~\ref{nbspin}),
are correctly saved in the ``common dump'' file for resuming the runs. 

\section{N-body runs with updated $\nbseven$}\label{runlist}

Table~\ref{tab_nbrun} lists the long-term direct N-body computations
performed in this work with updated $\nbseven$ (Sec.~\ref{newnb}).
These runs are performed on server computers containing {\tt NVIDIA}
{\tt Fermi-}, {\tt Kepler-}, or {\tt Turing-}series GPUs and 4- or 16-thread
CPUs. These calculations, altogether, took about one year to complete
including the times taken for continued code developments, their testing,
the management of the runs,
essential data analyses of the runs, and maintenance of the compute servers.

\onecolumn
\renewcommand*{\arraystretch}{1.5}
\begin{longtable}{>{\stepcounter{rowno}\therowno}rccclcllrcc}
	\caption[Summary of model calculations]
	{Summary of model evolutionary calculations. The columns from left to right give the model
	cluster's (a) ID number, (b) initial mass, $\mcl(0)$, (c) initial half-mass radius, $\rh(0)$,
	(d) metallicity, $Z$, (e) initial fraction of primordial binaries, $\fbin(0)$,
	(f) model evolutionary time, $\tevol$, (g) remnant-mass and PPSN/PSN model (Sec.~\ref{newrem}),
	(h) remnant natal kick model (Sec.~\ref{newkick}), (i) BH natal spin model (Sec.~\ref{bhspin}),
	(j) number of GR mergers within the cluster, $\nmrgin$, (k) number of GR mergers
	after getting ejected from the cluster, $\nmrgout$. $\fbin(0)=0.0$ implies that the model
	initially contains only single stars. When $\fbin(0)>0$, the quoted value represents the
	overall initial binary fraction,
	with the binary fraction for stars with ZAMS mass $\geq\mcrit$ being initially $\fobin=1.0$
	as described in Sec.~\ref{comp}. 
	Unless otherwise indicated, $\mcrit=16.0\Ms$.
	Initially, all stellar members in a cluster are ZAMS stars whose masses are distributed over
	$0.08\Ms-150.0\Ms$ according to the standard IMF. All initial models follow
	a Plummer profile and are unsegregated.
	All model clusters are subjected to a solar-neighbourhood-like
	external galactic field. See the footnotes associated with this table for further details
	and specifications.
	}\label{tab_nbrun}\\
	\hline
	\hline
	\multicolumn{1}{r}{No.} & \mcl(0)/\Ms     & \rh(0)/pc & $Z$ & \fbin(0) & \tevol/Gyr & remnant model & SN kick & BH spin & \nmrgin & \nmrgout \\
	\hline
	\endfirsthead
        
	\multicolumn{7}{c}%
        {{\bfseries \tablename\ \thetable{} -- continued from previous page}} \\
        \hline
	\hline
	\multicolumn{1}{r}{No.} & \mcl(0)/\Ms     & \rh(0)/pc & $Z$ & \fbin(0) & \tevol/Gyr & remnant model & SN kick & BH spin & \nmrgin & \nmrgout \\
	\hline
	\endhead

	\hline \multicolumn{10}{r}{{Continued on next page}} \\ \hline
        \endfoot

        \hline \hline
        \endlastfoot

    & 	$1.0\times10^4$ & 1.0       & 0.001  &  0.10\footnote{The binary fraction is defined
	as $\fbin=2\nbin/N$, $\nbin$ being the total number of binaries and $N$ being
	the total number of members.}   
							&  8.5     &   rapid+B16  &  mom. cons.\footnote{
								mom. cons. $\Rightarrow$ momentum conserving natal kick model,
							col. asym. $\Rightarrow$ collapse-asymmetry-driven natal kick model.} 
							                                        &  Geneva  &  0    & 0     \\
    &   $1.0\times10^4$ & 1.0       & 0.001  &  0.10    &  5.2     &   rapid+B16  &  col. asym. &  Geneva  &  0    & 1     \\
    &   $2.0\times10^4$ & 2.0       & 0.001  &  0.10    &  11.0    &   rapid+B16  &  mom. cons. &  Geneva  &  1    & 0     \\
    &   $2.0\times10^4$ & 2.0       & 0.001  &  0.10    &  8.7     &   rapid+B16  &  col. asym. &  Geneva  &  1    & 0     \\
    &   $2.0\times10^4$ & 2.0       & 0.01   &  0.10    &  4.4     &   rapid+B16  &  col. asym. &  Geneva  &  0    & 0     \\
    &   $2.0\times10^4$ & 2.0       & 0.01   &  0.10    &  5.8     &   rapid+B16  &  mom. cons. &  Geneva  &  1    & 0     \\
    &   $2.0\times10^4$ & 2.0       & 0.02   &  0.10    &  4.4     &   rapid+B16  &  mom. cons. &  Geneva  &  0    & 0     \\
    &   $3.0\times10^4$ & 1.0       & 0.0002 &  0.00    &  11.0    &   rapid+B16  &  mom. cons. &  Geneva  &  3    & 0     \\
    &   $3.0\times10^4$ & 1.0       & 0.01   &  0.00    &  7.2     &   rapid+B16  &  mom. cons. &  Geneva  &  2    & 0     \\
    &   $3.0\times10^4$ & 1.0       & 0.01   &  0.00    &  11.0    &   rapid+B16  &  col. asym. &  Geneva  &  1    & 1     \\
    &   $3.0\times10^4$ & 1.0       & 0.01   &  0.00    &  10.9    & delayed+B16  &  col. asym. &  MESA    &  4    & 0     \\
    &   $3.0\times10^4$ & 1.0       & 0.02   &  0.00    &  11.0    & delayed+B16  &  col. asym. &  MESA    &  0    & 0     \\
    &   $3.0\times10^4$ & 1.0       & 0.02   &  0.00    &  8.2     &   rapid+B16  &  mom. cons. &  Geneva  &  0    & 0     \\
    &   $3.0\times10^4$ & 1.0       & 0.02   &  0.00    &  7.0     &   rapid+B16  &  col. asym. &  Geneva  &  1    & 0     \\
    &   $3.0\times10^4$ & 2.0       & 0.0002 &  0.00    &  11.0    &   rapid+B16  &  mom. cons. &  Geneva  &  1    & 0     \\
    &   $3.0\times10^4$ & 2.0       & 0.01   &  0.00    &  9.7     &   rapid+B16  &  mom. cons. &  Geneva  &  2    & 0     \\
    &   $3.0\times10^4$ & 2.0       & 0.01   &  0.00    &  11.0    &   rapid+B16  &  col. asym. &  Geneva  &  2    & 0     \\
    &   $3.0\times10^4$ & 2.0       & 0.02   &  0.00    &  11.0    &   rapid+B16  &  col. asym. &  Geneva  &  0    & 0     \\
    &   $3.0\times10^4$ & 2.0       & 0.001  &  0.10    &  11.0    &   rapid+B16  &  mom. cons. &  Geneva  &  2    & 0     \\
    &   $3.0\times10^4$ & 2.0       & 0.001  &  0.10    &  11.0    &   rapid+B16  &  col. asym. &  Geneva  &  0    & 0     \\
    &   $3.0\times10^4$ & 2.0       & 0.01   &  0.10    &  11.0    &   rapid+B16  &  mom. cons. &  Geneva  &  0    & 0     \\
    &   $3.0\times10^4$ & 2.0       & 0.01   &  0.10    &  11.0    &   rapid+B16  &  col. asym. &  Geneva  &  1    & 0     \\
    &   $3.0\times10^4$ & 2.0       & 0.02   &  0.10    &  3.0     &   rapid+B16  &  mom. cons. &  Geneva  &  0    & 0     \\
    &   $3.0\times10^4$ & 2.0       & 0.001  &  0.10\footnote{$\mcrit=5.0\Ms$}
	                                                &  11.0    &   rapid+weak &  mom. cons. &  MESA    &  1    & 2     \\
    &   $3.0\times10^4$ & 2.0       & 0.001  &  0.10\footnote{$\mcrit=5.0\Ms$}
							&  9.3     &   rapid+weak\footnote{30\% of the full B10 wind is applied.}
							                          &  mom. cons. &  MESA    &  0    & 1     \\
    &   $3.0\times10^4$ & 3.0       & 0.0002 &  0.00    &  6.6     &   rapid+B16  &  mom. cons. &  Geneva  &  0    & 0     \\
    &   $3.0\times10^4$ & 3.0       & 0.01   &  0.00    &  11.0    &   rapid+B16  &  mom. cons. &  Geneva  &  3    & 0     \\
    &   $3.0\times10^4$ & 3.0       & 0.01   &  0.00    &  11.0    &   rapid+B16  &  col. asym. &  Geneva  &  0    & 0     \\
    &   $3.0\times10^4$ & 3.0       & 0.02   &  0.00    &  10.1    &   rapid+B16  &  mom. cons. &  Geneva  &  2    & 0     \\
    &   $3.0\times10^4$ & 3.0       & 0.02   &  0.00    &  11.0    &   rapid+B16  &  col. asym. &  Geneva  &  0    & 0     \\
    &   $5.0\times10^4$ & 2.0       & 0.0002 &  0.00    &  11.0    &   rapid+B16  &  mom. cons. &  Geneva  &  0    & 1     \\
    &   $5.0\times10^4$ & 2.0       & 0.0002 &  0.00    &  11.0    &   rapid+weak &  mom. cons. &  MESA    &  1    & 0     \\
    &   $5.0\times10^4$ & 2.0       & 0.001  &  0.00    &  11.0    &   rapid+B16  &  mom. cons. &  MESA    &  2    & 0     \\
    &   $5.0\times10^4$ & 2.0       & 0.001  &  0.00    &  11.0    &   rapid+weak &  mom. cons. &  MESA    &  0    & 0     \\
    &   $5.0\times10^4$ & 2.0       & 0.001  &  0.00    &  11.0    &   rapid+B16  &  col. asym. &  MESA    &  2    & 0     \\
    &   $5.0\times10^4$ & 2.0       & 0.005  &  0.00    &  11.0    &   rapid+weak &  mom. cons. &  MESA    &  1    & 0     \\
    &   $5.0\times10^4$ & 2.0       & 0.005  &  0.00    &  11.0    & delayed+B16  &  col. asym. &  MESA    &  4    & 0     \\
    &   $5.0\times10^4$ & 2.0       & 0.01   &  0.00    &  11.0    &   rapid+B16  &  mom. cons. &  Geneva  &  1    & 0     \\
    &   $5.0\times10^4$ & 2.0       & 0.01   &  0.00    &  11.0    &   rapid+B16  &  col. asym. &  Geneva  &  3    & 0     \\
    &   $5.0\times10^4$ & 2.0       & 0.02   &  0.00    &  9.9     &   rapid+B16  &  mom. cons. &  Geneva  &  0    & 0     \\
    &   $5.0\times10^4$ & 2.0       & 0.02   &  0.00    &  11.0    & delayed+B16  &  col. asym. &  MESA    &  3    & 0     \\
    &   $5.0\times10^4$ & 2.0       & 0.0001 &  0.05    &  11.0    &   rapid+B16  &  mom. cons. &  Geneva  &  1    & 1     \\
    &   $5.0\times10^4$ & 2.0       & 0.001  &  0.05    &  10.0    &   rapid+B16  &  mom. cons. &  Geneva  &  4    & 2     \\
    &   $5.0\times10^4$ & 2.0       & 0.001  &  0.05\footnote{$\ftz=0.70$, $\fmrg=0.3$}
                                                        &  11.0    &   rapid+weak &  mom. cons. &  MESA    &  1    & 2     \\
    &   $5.0\times10^4$ & 2.0       & 0.0001 &  0.05\footnote{$\fmrg=0.2$}
                                                        &  11.0    &   rapid+B16  &  mom. cons. &  MESA    &  2    & 2     \\
    &   $5.0\times10^4$ & 2.0       & 0.01   &  0.05\footnote{$\ftz=0.90$, $\fmrg=0.2$}
                                                        &  11.0    & delayed+B16  &  col. asym. &  MESA    &  2    & 0     \\
    &   $5.0\times10^4$ & 1.0       & 0.001  &  0.00    &  11.0    &   rapid+B16  &  mom. cons. &  Geneva  &  1    & 2     \\
    &   $5.0\times10^4$ & 1.0       & 0.001  &  0.00    &  11.0    & delayed+B16  &  col. asym. &  MESA    &  3    & 0     \\
    &   $5.0\times10^4$ & 1.0       & 0.001  &  0.00    &  11.0    & delayed+B16  &  mom. cons. &  Fuller  &  5    & 0     \\
    &   $5.0\times10^4$ & 1.0       & 0.01   &  0.00    &  11.0    & delayed+B16  &  col. asym. &  MESA    &  7    & 0     \\
    &   $5.0\times10^4$ & 1.0       & 0.02   &  0.00    &  11.0    & delayed+B16  &  col. asym. &  MESA    &  4    & 4     \\
    &   $7.5\times10^4$ & 2.0       & 0.001  &  0.00    &  11.0    &   rapid+B16  &  mom. cons. &  Geneva  &  2    & 0     \\
    &   $7.5\times10^4$ & 2.0       & 0.001  &  0.00    &  11.0    &   rapid+weak &  mom. cons. &  MESA    &  3    & 0     \\
    &   $7.5\times10^4$ & 2.0       & 0.001  &  0.00    &  11.0    &   rapid+B16  &  col. asym. &  Geneva  &  3    & 0     \\
    &   $7.5\times10^4$ & 2.0       & 0.005  &  0.00    &  11.0    & delayed+B16  &  col. asym. &  MESA    &  4    & 0     \\
    &   $7.5\times10^4$ & 2.0       & 0.01   &  0.00    &  11.0    &   rapid+B16  &  mom. cons. &  Geneva  &  6    & 0     \\
    &   $7.5\times10^4$ & 2.0       & 0.02   &  0.00    &  11.0    &   rapid+B16  &  mom. cons. &  Geneva  &  6    & 0     \\
    &   $7.5\times10^4$ & 2.0       & 0.02   &  0.00    &  11.0    & delayed+B16  &  col. asym. &  MESA    &  3    & 1     \\
    &   $7.5\times10^4$ & 2.0       & 0.0001 &  0.05\footnote{$\fmrg=0.2$}
                                                        &  11.0    &   rapid+B16  &  mom. cons. &  MESA    &  1    & 2     \\
    &   $7.5\times10^4$ & 2.0       & 0.001  &  0.05\footnote{$\ftz=0.95$, $\fmrg=0.2$}
                                                        &  11.0    &   rapid+B16  &  mom. cons. &  MESA    &  1    & 4     \\
    &   $7.5\times10^4$ & 2.0       & 0.001  &  0.05\footnote{$\ftz=0.95$, $\fmrg=0.2$}
                                                        &   9.8    &   rapid+B16  &  mom. cons. &  Fuller  &  5    & 1     \\
    &   $7.5\times10^4$ & 2.0       & 0.01   &  0.05\footnote{$\ftz=0.95$, $\fmrg=0.2$}
                                                        &  11.0    &   rapid+B16  &  mon. cons. &  Fuller  &  1    & 1     \\
    &   $7.5\times10^4$ & 2.0       & 0.02   &  0.05\footnote{$\fmrg=0.2$}
                                                        &  11.0    & delayed+B16  &  col. asym. &  MESA    &  2    & 0     \\
    &   $1.0\times10^5$ & 2.0       & 0.001  &  0.00    &  11.0    & delayed+B16  &  col. asym. &  Geneva  &  5    & 1     \\
    &   $1.0\times10^5$ & 1.5       & 0.001  &  0.05\footnote{$\ftz=0.95$, $\fmrg=0.2$}
							&   1.0\footnote{ongoing run}
						                   &   rapid+B16  &  mom. cons. &  Fuller  &  4    & 0     \\
\end{longtable}
\twocolumn

\begin{figure*}
\includegraphics[width=8.7cm,angle=0]{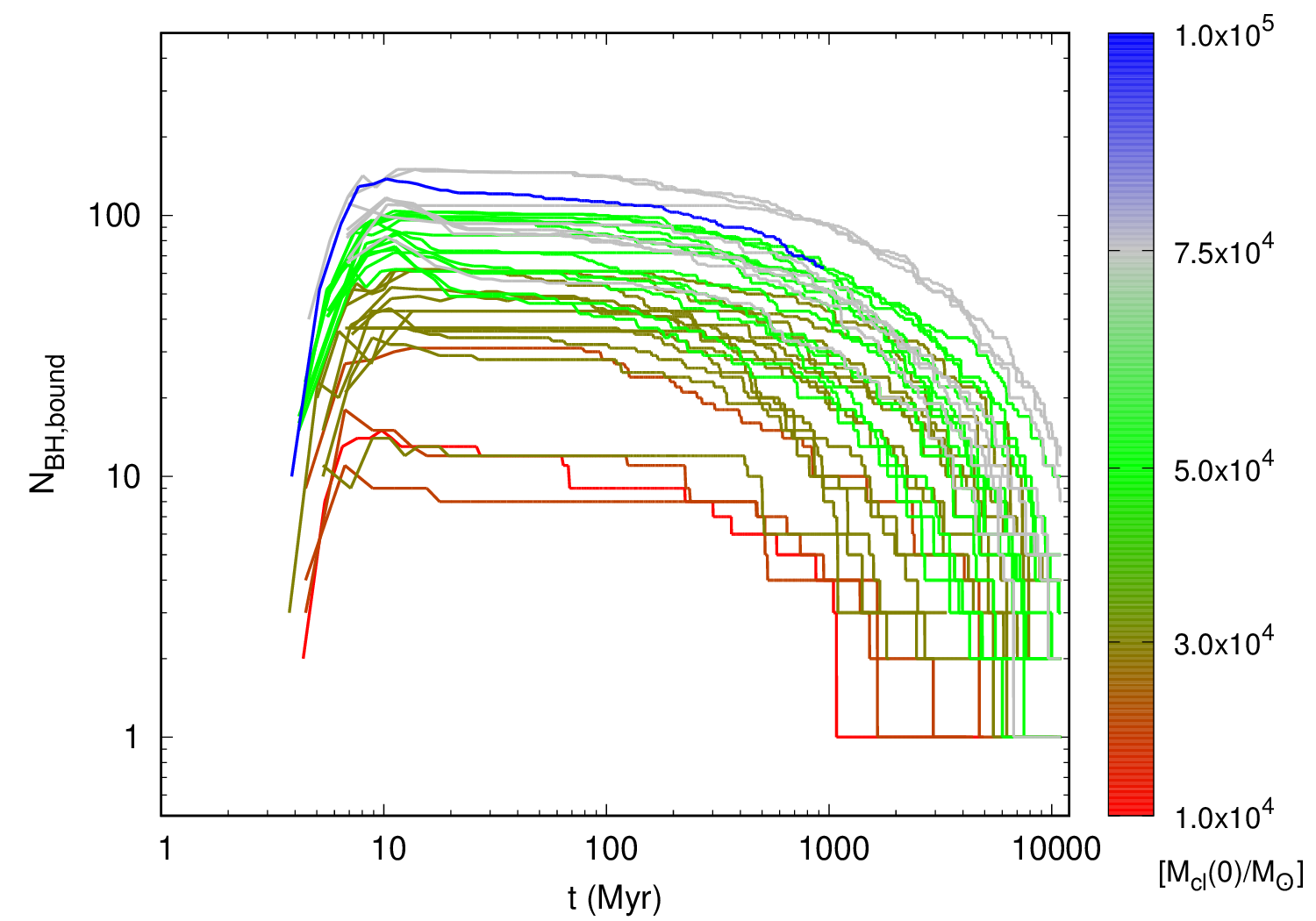}
\includegraphics[width=8.7cm,angle=0]{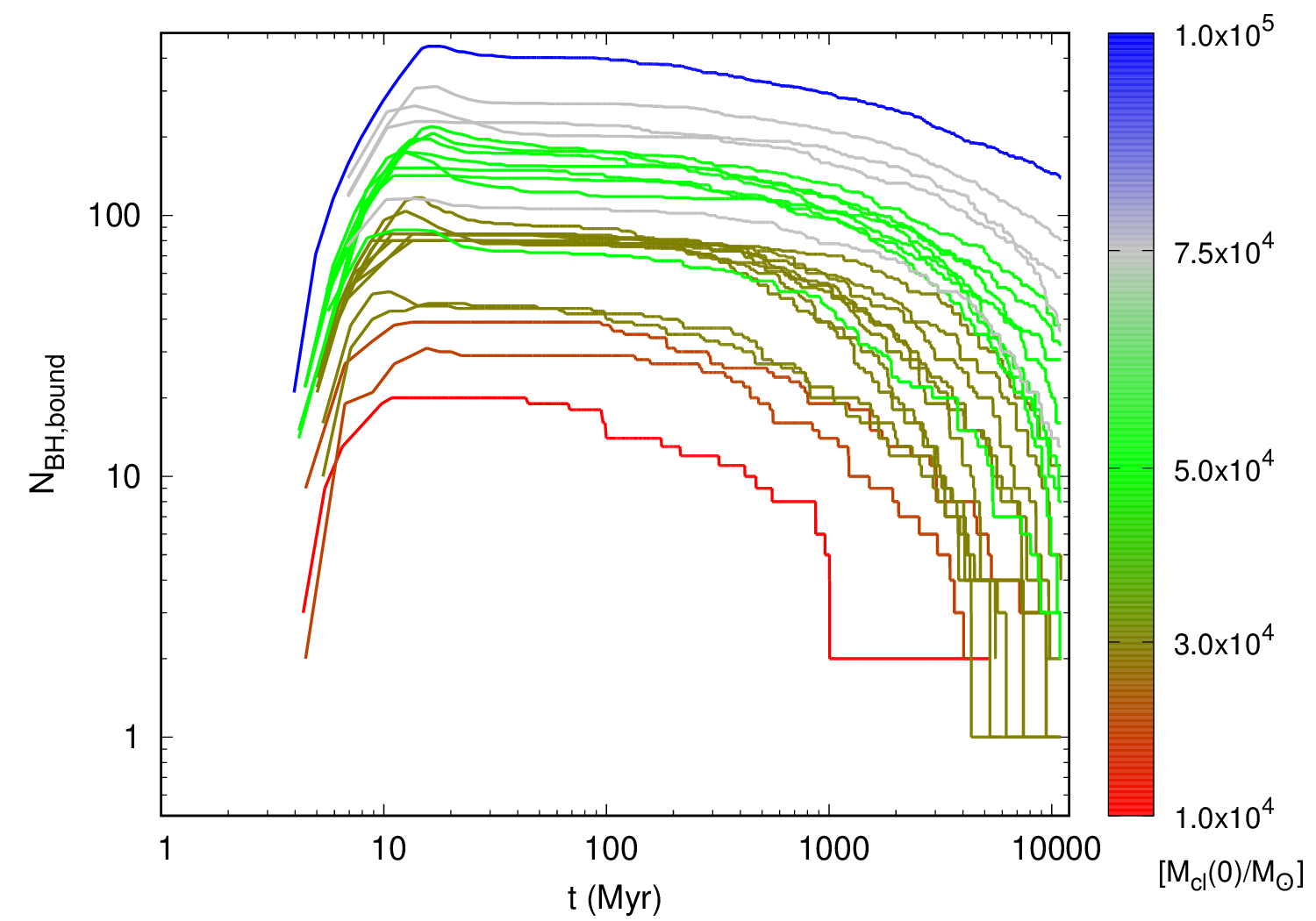}
\caption{Number of BHs, $\nbhbound$, bound to the model clusters of Table~\ref{tab_nbrun}
as a function of the models' evolutionary time, $t$. The left and the right panel
shows $\nbhbound(t)$ for the models with momentum-conserving and collapse-asymmetry-driven
natal kicks, respectively. The lines are colour-coded according to the models' initial
mass, $\mcl(0)$ (colour bar). The initial growth of BH population,
due to the retention of BHs in the clusters at birth (Sec.~\ref{newkick}),
scales, overall, with $\mcl(0)$. All evolutionary models exhibit long-term retention of BHs despite
the decay of their population with time due to dynamical ejections. Since clusters
with collapse-asymmetry-driven natal kick retain larger numbers of BHs at birth (Sec.~\ref{newkick}),
they retain larger numbers of BHs at the end of the computation (right panel).}
\label{fig:nbh}
\end{figure*}

\end{document}